\documentclass[12pt]{article}
\usepackage[dvips]{color,graphicx}
\usepackage{amsmath,amssymb}
\DeclareGraphicsExtensions{.eps}

\newcommand{\gs}{\ensuremath{g_s}} 
\newcommand{\ap}{\ensuremath{\alpha'}} 
\newcommand{\ls}{\ensuremath{l_s}} 

\def\p{\partial}

\newcommand{\tr}{\mathop{\rm Tr}}

\def\expec#1{\langle #1 \rangle}

\newcommand{\cL}{\mathcal{L}}

\newcommand{\cN}{{\mathcal{N}}}
\newcommand{\cO}{{\mathcal{O}}}

\newcommand{\bS}{{\mathbf{S}}}

\newcommand{\tret}{{t_{\mbox{\scriptsize ret}}}}
\newcommand{\tv}{{\tilde{v}}}
\newcommand{\ta}{{\tilde{a}}}

\title{\bf Acceleration, Energy Loss and Screening in Strongly-Coupled Gauge Theories}
\author{Mariano Chernicoff\footnote{e-mail: mariano@nucleares.unam.mx}
~and Alberto G\"uijosa\footnote{e-mail: alberto@nucleares.unam.mx}
\\{\small Departamento de F\'{\i}sica de Altas Energ\'{\i}as,
Instituto de Ciencias Nucleares}\\ {\small Universidad Nacional
Aut\'onoma de M\'exico}\\
{\small Apdo. Postal 70-543, M\'exico D.F. 04510}}
\date{}

\begin{document}
\maketitle
\begin{abstract}
We explore various aspects of the motion of heavy quarks in strongly-coupled gauge theories,
employing the AdS/CFT correspondence. Building on earlier work by Mikhailov, we study the
dispersion relation and energy loss of an accelerating finite-mass quark in $\cN=4$
super-Yang-Mills, both in vacuum and in the presence of a thermal plasma. In the former
case, we notice that the application of an external force modifies the dispersion relation.
In the latter case, we find in particular that when a static heavy quark is accelerated by
an external force, its rate of energy loss is initially insensitive to the plasma, and there
is a delay before this rate approaches the value derived previously from the analysis of
stationary or late-time configurations.

Following up on work by Herzog \emph{et al.}, we also consider the evolution of a quark and
antiquark as they separate from one another after formation, learning how the AdS/CFT setup
distinguishes between the singlet and adjoint configurations, and locating the transition to
the stage where the deceleration of each particle is properly accounted for by a constant
friction coefficient. Additionally, we examine the way in which the energy of a
quark-antiquark pair moving jointly through the plasma scales with the quark mass. We find
that the velocity-dependence of the screening length is drastically modified in the
ultra-relativistic region, and is comparable with that of the
transition distance mentioned above.
\end{abstract}

\vspace{3cm}

\tableofcontents

\section{Introduction and Summary}

\subsection{Brief overview of earlier work}

In the last couple of years, an intense research effort has been
directed toward the use of the AdS/CFT correspondence
\cite{malda,gkpw,magoo} to study parton energy loss in
strongly-coupled thermal plasmas. This endeavor is motivated mostly
  by
the quest to understand the quark-gluon plasma (QGP) \cite{qgprev} produced at RHIC
\cite{rhic} and LHC \cite{alice}, and was stimulated by the pioneering works
\cite{hkkky,gubser,ct,liu}, which were in turn encouraged by the success of the earlier
viscosity calculations \cite{eta,etarev}.\footnote{ Recently it has been argued
\cite{etaqhat} that, at least at weak coupling, there in fact exists a direct link between
viscosity and jet quenching.} The phenomenological
 literature
on energy loss is enormous; for reviews, see, e.g.,
 \cite{baier,energylossrev}.

The drag force experienced by a heavy quark traversing an
$\cN=4$ super-Yang-Mills (SYM) plasma was determined in
\cite{hkkky,gubser}, via its dual description as a  string
moving on an AdS-Schwarzschild background. The
closely related diffusion coefficient
was obtained independently in \cite{ct}. These seminal papers  have been
 generalized and elaborated on in a vast number of
 posterior contributions \cite{dragforcefollowups,dragforceqhat,draggluon,gubserqhat,ctqhat,
 sonnenschein2,liu5,gubsergluon},
including in particular comparisons with
the corresponding weakly-coupled results \cite{weakcoupling},
as well as extensive analyses of the energy-momentum
tensor which paint a detailed and
beautiful picture
of the directionality of energy flow away from the moving quark
\cite{gluonicprofile}.

A second line of development originated from the work \cite{liu},
 whose authors proposed a recipe for the jet-quenching
parameter $\hat{q}$ used in some
phenomenological models of energy loss
\cite{baier}, and
employed it in the context of $\cN=4$ SYM.
Their calculation has been extended in a number of directions
 in \cite{qhatfollowups,dragforceqhat,liu3},
 but also questioned in various ways in \cite{dragqqbar,
draggluon,gubserqhat,argyres2,ctqhat,argyres3}.

A third line was initiated in \cite{liu2,dragqqbar}, which studied mesons moving through an
$\cN=4$ SYM plasma and obtained
 the corresponding quark-antiquark potential and
screening length, using the dual portrayal in terms of a string that moves on an
AdS-Schwarzschild background and has both of its endpoints on the boundary. As was
emphasized in \cite{liu2}, the outcome would be expected to have implications for the issue
of quarkonium suppression in the QGP. Related results were obtained independently in
\cite{sonnenschein}, which determined the spectrum of spinning mesons in the confining
chiral gauge theory dual to the Sakai-Sugimoto model \cite{ss}. Various interesting
extensions and refinements of these calculations have been reported in
\cite{qqbarfollowups,argyres1,sfetsosqqbar,gubserstable,
bky,sonnenschein2,liu4,liu5,iancu2}.

A notable feature is that, in contrast with the quark probes considered
 in \cite{hkkky,gubser,ct,liu}, and the gluon probes studied in \cite{draggluon,gubsergluon},
  mesons do \emph{not} feel a drag force
as they move through the plasma \cite{sonnenschein,liu2,dragqqbar}.
This is because they are color-neutral, and therefore incapable
of setting up the long-range gluonic field profiles that could transport
energy away from them.\footnote{The color field profile set up by
a meson has been explicitly determined at zero temperature in
\cite{cg,linshuryak}.} Indeed,
in \cite{draggluon,liu5}
it has been shown that the other
obvious color-neutral probe, the baryon,
likewise experiences no drag.

Besides the transport and jet quenching properties of the plasma, there have also been
interesting studies of photoemissivity \cite{photon} and deep inelastic scattering
\cite{iancu,iancu2}. Some of the topics we have briefly enumerated here have been reviewed
in more depth in \cite{gubserpitp,adsqgprev}.

\subsection{Motivation and main results}

Naturally,
the initial papers \cite{hkkky,gubser,ct,liu,liu2,dragqqbar} carried out their
calculations under a number of simplifying assumptions. First and
foremost among these is of course the use of an $\cN=4$ SYM plasma
 as a toy model for the real-world QGP. A number of extensions to other
 theories that are in certain ways closer to QCD have already
 been cited above, and a few others can be found in \cite{qcdlike}.
 The issue of how best to compare
 the $\cN=4$ SYM and QCD parameters was discussed in \cite{gubsercompare}.
 Also important is the restriction to an infinite static plasma. Efforts
 to examine the case where the plasma is expanding and/or has a
 finite spatial extent have been made in \cite{expanding}.

 The plasma produced at RHIC or LHC has in addition a finite \emph{temporal} extent,
 so another issue that
 could be significant is the limitation of
 the energy loss computations \cite{hkkky,gubser,ct,liu}
  to the stationary or late-time regime. The actual drag coefficient
  might be expected to differ from the  value obtained
  in these works in a situation where the
  quark moving through the plasma is accelerating, or in the initial
  period  following its production within the thermal medium.
  This point has been emphasized from the phenomenological perspective
  (in the context of collisional energy loss)
  in \cite{peigne1}, where it was argued that, after a quark-antiquark
  pair is produced, there exists
  a significant delay before the rate of energy dissipation coincides
  with the result relevant for the stationary case. Later
  works \cite{peigne2} have called into question the actual duration,
  but not the existence, of this retardation effect.

  The estimates in \cite{peigne1,peigne2} are based on perturbative
  calculations, so it is interesting to inquire into this effect in
  the strongly-coupled systems available to us through
  the AdS/CFT correspondence. It was this question that
  got us started on the investigation that led to
  the present work, which over time
  has expanded toward various other fronts.
  For simplicity, we have carried out our calculations
  in the context of an $\cN=4$ plasma, but we expect most of
  our qualitative conclusions to apply more generally.

  Now, of course,
  the restriction in  \cite{hkkky,gubser}
  to the stationary and asymptotic cases
  was not made gratuitously, but was necessary in order
  to gain analytic control on the problem of energy loss.
  Away from these regimes, the evolution of the quark in
  the thermal plasma, or equivalently, of the string
  on the AdS-Schwarzschild geometry, is complicated, and one must resort
  to a numerical analysis (aspects of which were explored
  already in \cite{hkkky}). As we will examine closely
   in due course, the interpretation of the outcome of such
   an analysis is encumbered by several difficulties,
   chief among which is our ignorance of the thermal
   dispersion relation for the quark.

 In Section \ref{qnoplasmasec}
 we will gain some perspective on these issues by turning off
 the temperature $T$, and analyzing first
 the evolution of an isolated
  quark in vacuum, which is of interest in its own right,
  and easier to interpret because in this context the form of the
   dispersion relation
   is fixed by Lorentz invariance.
   We begin in Section \ref{infinitemasssec} by
   reviewing a remarkable
   paper by Mikhailov \cite{mikhailov},
   who constructed an analytic embedding for the string
   dual to
   an infinitely-massive
    quark in $\cN=4$ SYM that follows an \emph{arbitrary} timelike
    trajectory, and extracted from it
    a rate of energy loss,
    Eq.~(\ref{emikh}), which
    turned out to agree with the Lienard formula from
    classical electrodynamics! In the derivation of his result,
     Mihailov disregarded a total derivative, which we show
     in (\ref{edr}) to give precisely the expected dispersion
     relation for the quark. {}From the results of
     \cite{mikhailov}, then, we are able to draw
     two important lessons: first, that,
     at any given time, the energy of the string
     includes not only the portion intrinsic to the quark,
     but also the part that has been radiated by the quark throughout its
     previous history; second, that the quark's dispersion relation arises
     from a total derivative that ends up being evaluated
     at the string endpoint.

     It is natural to wonder how Mikhailov's results generalize to the
     case where the quark has a finite mass, which
     requires the introduction
     of probe D7-branes on which the string can end
     \cite{kk}. We examine this issue (still
     at $T=0$)
     in Section \ref{finitemasssec}. Interestingly, we discover that
     the resulting quark dispersion relation, Eq.~(\ref{edrf}),
     as well as the rate of energy loss, Eq.~(\ref{emikhf}), depend
     on the external force $F$ exerted on the quark, or equivalently,
     on the string embedding parameter $X'$, which in the gauge
     theory controls the shape of the `gluon cloud' surrounding
     the quark. The dependence is such that the energy and momentum
     of the quark reduce to the familiar expressions
     when $F\to 0$, and on the other hand diverge as the force
     approaches its critical value
     (i.e., the value beyond which $F$
      would be strong enough to nucleate quark-antiquark pairs out
      of the vacuum). We finalize our zero-temperature analysis
      in Section \ref{bhsec}, noting that the
      process of energy loss is accompanied by
      the formation of
      an event horizon (and a stationary limit curve) on the string
      worldsheet, as depicted in Fig.~\ref{bhfig}.

    Armed with the intuition afforded to us
    by Mikhailov's construction, we proceed
    in Section \ref{qplasmasec} to the finite-temperature case.
    In this setting, we are of course
     limited by the fact that the general
    solution to the string equation of motion is not known
    analytically. Nonetheless, we argue that Mikhailov's
    method should admit a $T>0$ generalization, and are able
    to show in Section \ref{constantvsec} that this is indeed
     true for the
    only thermal solution that is thus far available in closed form,
    which corresponds to the stationary quark
    configuration studied in \cite{hkkky,gubser}.
    The resulting dispersion relation, seen in the
    second and third line of Eq.~(\ref{Emikhgubser}), contains
    a novel feature that we argue to apply for all
    finite-temperature configurations,
    including the quark at rest, Eq.~(\ref{Mrest}):
    it receives a contribution not only from the string endpoint
    located on the D7-branes, which is directly
    dual to the quark, but also from the endpoint located at
    the black hole horizon, which, as we explain, encodes the
    initial conditions for the joint quark $+$ plasma system.

    In Section \ref{acceleratedsec} we consider the more general
    case where the quark accelerates within the plasma. After reviewing
    the work done in this context by the authors of \cite{hkkky}, we
    pick up precisely where they left off, integrating
    the string equation of motion numerically for an initially
    static quark that is accelerated by an
    external force over a finite period of time and is thereafter
    released. As shown in Figs.~\ref{vfig},\ref{diffmass},\ref{equalmass},
    we find that
    under such conditions,
    and for values of the mass in the neighborhood of the charm quark,
    there exists a period after release where the quark
    dissipates energy at a rate that is substantially \emph{smaller} than
    the stationary/asymptotic result (\ref{EPlossgubser})
    obtained in \cite{hkkky,gubser}. Additionally, as seen in Fig.~\ref{LW},
    the rate of energy loss in the initial stage where the quark is
    pushed by the external force can be almost completely accounted for
    by the generalized Lienard formula (\ref{emikhf}), describing
     radiation in vacuum.
   In (\ref{Edrplasma}) and the paragraphs
   immediately following, we discuss
    the form of the thermal dispersion relation for the quark
    subject to the above initial conditions, incorporating
    the $F$-dependence expected from our zero-temperature results,
    and explaining in detail
    the relation to the relativistic
    expression (\ref{DRr}) proposed in \cite{hkkky}.
   In Section \ref{bhplasmasec} we then carry over the discussion
   of the worldsheet black hole from Section \ref{bhsec} to the $T>0$ context,
   obtaining a time-dependent analog of the
   Schwarzschild black hole
   encountered for the stationary case in \cite{gubserqhat,ctqhat},
   as schematized in Fig.~\ref{bhplasmafig}.

   Having explored in some depth
   the effect of the acceleration on the rate of energy
   loss for an isolated quark, in Section \ref{qqbarsec}
   we introduce the other element of realism whose importance was
   highlighted by
   the phenomenological studies \cite{peigne1,peigne2}, and examine
   a quark that is produced together with its corresponding antiquark
   at some finite time within the thermal medium.
   The gravity description of this system
    involves a string with \emph{both} of its
   endpoints on the D7-branes, at initially coincident positions.
   As we review in Section \ref{qqbarreviewsec},
   an initial exploration of this type of configuration was
   carried out in \cite{hkkky}. In Section \ref{qqbargeneralizationsec}
   we show how the one-parameter family
   of initial conditions considered in that
   work can be generalized to describe situations with
   vastly different patterns of excitations for the initial gluonic
   fields, and more importantly, to the case where the
   newly-formed quark
   and antiquark transform in the adjoint instead of the singlet
   representation of the color gauge group.
   Curiously, we find that it is only in the adjoint case that
   the initial quark velocity $v_0$ is freely adjustable.
   For the singlet
   case, as in \cite{hkkky} one must necessarily have $v_0=v_m$,
   where $v_m$ is the velocity (\ref{vm}) that
   appeared in different manifestations in the works
   \cite{sonnenschein,liu2,dragqqbar,gubserqhat,ctqhat},
   and on the string theory side of the duality corresponds
   to the proper velocity of light at the location of the string
   endpoint \cite{argyres1}. This identification strongly suggests
   that $v_m$ should be a limiting velocity, as is implicit
   in at least \cite{argyres1,gubserqhat,ctqhat}, and is discussed
   more explicitly in Section \ref{qqbarvmsec} (as well as in
   the very recent works \cite{liu4,argyres3}, which appeared
   while this paper was in preparation).

  In Section \ref{qqbartransitionsec} we study the transition
  of the $q$-$\bar{q}$ pair
   to the asymptotic regime described by a constant friction
    coefficient, which as shown in Fig.~\ref{stage2fig} applies
    uniformly to all different types of initial conditions.
       For singlet configurations,
    the initial evolution of the quark is of course drastically
    affected by the presence of the antiquark.
    In Fig.~\ref{xtransfig} we determine the velocity dependence
    of the transition distance beyond which the quark is effectively
    in the asymptotic regime.
   For adjoint configurations, on the other hand, the interactions
   between the quark and antiquark are suppressed at large $N$,
   and so the two members of the pair evolve independently from
   the start. In this case we find that the transition distance
   is essentially zero.

   A natural question is whether the transition to the asymptotic
   regime occurs right after the quark and antiquark are screened
   from one other by the plasma. This requires a determination
   of the corresponding
   quark-antiquark potential, and more specifically the
   screening length, a problem that we turn to in Section
   \ref{potentialsec}. After reviewing and comparing in Section
   \ref{potentialreviewsec} the results
   obtained for infinitely-massive quarks in \cite{liu2,dragqqbar},
   we generalize to the case of finite mass, first
   at zero temperature in Section \ref{potentialnoplasmasec},
   and then at finite temperature in Section \ref{potentialplasmasec}.
   The resulting potentials are shown in Figs.~\ref{potentialfig},
   \ref{potentialvfig}. As expressed in
   (\ref{Elinearnoplasma}), (\ref{Elinearplasma}),
   (\ref{Elinearplasmav}),
   they are found to be linear instead of divergent when the
   quark and antiquark approach one another, signaling
   the fact that the color sources in this case are no longer
   pointlike.

   In the finite temperature
   case, the potential implies the
   velocity-dependence of the screening length
    $L_{\mbox{\scriptsize max}}$ presented in
   Figs.~\ref{lmaxfig2},\ref{lmaxfig4}. As seen there,
   compared to the infinitely-massive case examined
   in \cite{liu2,dragqqbar}, there is a drastic
   modification of the behavior at high velocities,
   which are now bounded by $v_m$ instead of 1.
    The $v$-dependence near this limit
   can be determined analytically and takes the form
    (\ref{one}), instead of the
    formula (\ref{onequarter})
    obtained in \cite{liu2}.
    Over the whole range $0\le v\le v_m$,
    and for masses similar to the charm quark, the
    behavior can be relatively well
    approximated by (\ref{onethirdvm}),
    which is the obvious generalization
    of the fit (\ref{onethird}) proposed in \cite{dragqqbar}.

  We end the paper by comparing in Section \ref{potentialtransitionsec}
  the screening length against the transition distance determined
  in \ref{qqbartransitionsec}.
  For singlet configurations, we find that the
  magnitude and velocity-dependence of these
  two separations are  comparable,
  as shown in Fig.~\ref{xlmaxfig}.
Notice that this is in spite of the fact that the two relevant string configurations are
rather different. A similar statement can be made for adjoint configurations, where both the
screening and transition lengths are essentially zero. We conclude then that, for
$q$-$\bar{q}$ pairs created within the plasma, the transition to the asymptotic,
constant-drag-coefficient regime takes place immediately after the quark and antiquark lose
contact with one another. That is to say, there is no intermediate stage where the quark and
antiquark decelerate independently from one another at a rate that differs substantially
from the asymptotic result of \cite{hkkky,gubser}.

Our analysis in Section \ref{acceleratedsec} demonstrates that the gluonic fields around a
quark can be disturbed by the application of an external force, to the point of producing a
significant modification of the energy dissipation rate in the period immediately following
release. By analogy with the results we obtained for singlet quark-antiquark configurations
in Section \ref{qqbartransitionsec}, we interpret this to mean that the quark has to escape
far enough from the disturbed region in order to be screened from its effect. In any event,
given that the actual experimental situation resembles the setup of Section \ref{qqbarsec}
much more closely than that of Section \ref{acceleratedsec}, the main overall lesson for QGP
phenomenology would appear to be that, beyond an initial period which is controled by the
screening length, and where the evolution can be modeled relatively well as taking place in
vacuum (and, as such, would be present also in proton-proton collisions), the
stationary/asymptotic rate of energy dissipation determined in \cite{hkkky,gubser,ct} gives
a good approximation to the actual time-dependent dynamics.

It is worth emphasizing that the inferences made in this paper regarding dispersion
relations and energy loss rates are based on the natural split achieved in \cite{mikhailov}
of the total energy of the string. The latter is conserved on the fixed AdS-Schwarzschild
background, but would of course decrease steadily if we take into account the gravitational
(and dilatonic, etc.) radiation given off by the string in the course of its evolution.
Through the GKPW recipe for correlation functions \cite{gkpw}, it is this radiation (or,
more precisely, the full metric perturbation produced by the string), evaluated at the AdS
boundary, that determines the expectation value of the gauge theory energy-momentum tensor.
As has been meticulously studied in \cite{gluonicprofile}, this tensor contains not just the
gross information about the total energy loss rate, but even the fine details about the
directionality of the flow and the relative weight of the various dissipation channels. It
would therefore be very interesting to examine more closely this link between dissipated
energy as encoded on the string and on the gravitational field ultimately generated by it.

\section{Single Quark Evolution: Zero Temperature}
\label{qnoplasmasec}

To analyze the motion of heavy quarks in a strongly-coupled $\cN=4$ $SU(N_c)$ SYM plasma
with coupling $g_{YM}$ and temperature $T$, one must follow the evolution of open strings
that end on a stack of $N_f$ D7-branes \cite{kk} living  on the
(AdS-Schwarzschild)$_5\times\bS^5$ geometry
\begin{eqnarray}\label{metric}
ds^2&=&G_{mn}dx^m dx^n={R^2\over z^2}\left(
-hdt^2+d\vec{x}^2+{dz^2 \over h}\right)+R^2 d\Omega_5~, \\
h&=&1-\frac{z^{4}}{z_h^4}~, \qquad {R^4\over \ls^4}=g_{YM}^2 N_c\equiv\lambda~, \qquad
z_h={1\over \pi T}~,\nonumber
\end{eqnarray}
where $\ls$ denotes the string length. In our work the string will
be taken to lie at a fixed position on the $\bS^5$ (consistent
with the corresponding equations of motion), so the angular
components of the metric will not play any role. The D7-branes
cover the four gauge theory directions $t,\vec{x}$, and extend
along the radial AdS direction up from the boundary at $z=0$ to a
position where they `end' (meaning that the $\bS^3\subset\bS^5$
that they are wrapped on shrinks down to zero size), whose
location $z=z_m$ is related to the quark mass in a way that we
will specify below.

{}From the gauge theory perspective, the introduction of the
D7-branes in the background (\ref{metric}) is equivalent to the
addition of $N_f$ hypermultiplets in the fundamental
representation of the $SU(N_c)$ gauge group, breaking the
supersymmetry down to $\cN=2$. These  are the degrees of freedom
that we refer to as `quarks,' even though they include both spin
$1/2$ and spin $0$ fields. For $N_f\ll N_c$, the backreaction of
the D7-branes on the geometry can be sensibly neglected; in the
field theory this corresponds to working in a `quenched'
approximation which disregards quark loops (as well as the
positive beta function they would generate).

The string dynamics follows as usual from the Nambu-Goto action
\begin{equation}\label{nambugoto}
S_{\mbox{\scriptsize NG}}=-{1\over 2\pi\ap}\int
d^2\sigma\,\sqrt{-\det{g_{ab}}}\equiv {R^2\over 2\pi\ap}\int
d^2\sigma\,\cL_{\mbox{\scriptsize NG}}~,
\end{equation}
where $g_{ab}\equiv\p_a X^m\p_b X^n G_{mn}(X)$ ($a,b=0,1$) denotes
the induced metric on the worldsheet. In the static gauge
$\sigma^0=t$, $\sigma^1=z$, and for motion and deformation of the
string purely along direction $x\equiv x^1$, the non-zero
canonical momentum densities
$\Pi^a_{\mu}\equiv\p\cL_{\mbox{\scriptsize NG}}/\p(\p_a X^{\mu})$
are given by
\begin{eqnarray}\label{momenta}
\Pi^t_t&=&-\frac{h{X^{'}}^{2}+1}{z^2\sqrt{1+h {X^{'}}^{2}-{\dot{X}^2\over h}}}~,\nonumber\\
\Pi^t_x&=&\frac{\dot{X}}{z^2 h\sqrt{1+h{X^{'}}^{2}-{\dot{X}^2\over h}}}~,\\
\Pi^z_t&=&\frac{h\dot{X}X'}{z^2\sqrt{1+h{X^{'}}^{2}-{\dot{X}^2\over h}}}~,\nonumber\\
\Pi^z_x&=&-\frac{h X'}{z^2\sqrt{1+h{X^{'}}^{2}-{\dot{X}^2\over
h}}}~,\nonumber
\end{eqnarray}
where of course $\dot{X}\equiv\p_t X(t,z)$, $X'\equiv\p_z X(t,z)$.
Notice that, due to our normalization of $\cL_{\mbox{\scriptsize
NG}}$, the $\Pi^a_{\mu}$ must be multiplied by
$R^2/2\pi\ap=\sqrt{\lambda}/ 2\pi$ to obtain the physical energy
and momentum densities.

In the present section we will restrict attention to the case of
vanishing temperature ($z_h\to\infty$), in which case we are left
in (\ref{metric}) with a pure AdS geometry, and the D7-brane
parameter $z_m$ is  inversely proportional to the Lagrangian mass
of the quark,
\begin{equation}\label{zmnoplasma}
z_m={\sqrt{\lambda}\over 2\pi m}~.
\end{equation}

A quark that accelerates in vacuum would be expected to emit chromoelectromagnetic
radiation. This problem has been examined from the classical perspective in
\cite{classicalYMrad}, and quantum-mechanically at weak coupling in, e.g., \cite{weakYMrad}.
First steps towards a strong-coupling analysis by means of the AdS/CFT correspondence were
taken in \cite{cg}, which  employed tools developed in \cite{dkk} to study the dilatonic
waves given off by small fluctuations on a radial string in AdS$_5$, and infer from them the
profile of the gluonic field $\expec{\tr F^2(x)}$ in the presence of a quark undergoing
small oscillations. The results of \cite{cg} painted an interesting picture of the
propagation of nonlinear waves in $\cN=4$ SYM, but did not allow a definite identification
of waves with the $1/|\vec{x}|$ falloff associated with radiation. Very recently, this
falloff has been successfully detected in the same setup as \cite{cg} through a calculation
of the energy-momentum tensor $\expec{T_{\mu\nu}}$ \cite{mo}, which appeared while this
paper was in preparation.

\subsection{Infinite mass} \label{infinitemasssec}

The first definite characterization of the radiation rate off an accelerating quark was
found much later than \cite{cg}, and by a completely different route, in an important paper
by Mikhailov \cite{mikhailov}. Remarkably, this author was able to solve the full nonlinear
equation of motion for a string on AdS$_5$, for an \emph{arbitrary} timelike trajectory of
the string endpoint dual to a heavy quark! In terms of the coordinates used in
(\ref{metric}) (where for now $h=1$), his solution is
\begin{equation}\label{mikhsol}
X^{\mu}(\tau,z)=z{dx^{\mu}(\tau)\over d\tau}+x^{\mu}(\tau)~,
\end{equation}
with $\mu=0,1,2,3$, and $x^{\mu}(\tau)$ the worldline of the string endpoint at the AdS
boundary--- or, equivalently, the worldline of the dual, infinitely massive, quark---
parametrized by the proper time $\tau$ defined through
$\eta_{\mu\nu}\;\mathring{}\!\!x^{\mu}\;\mathring{}\!\!x^{\nu}=-1$, where
$\;\mathring{}\!\!x^{\mu}\equiv dx^{\mu}/d\tau$. Equation (\ref{mikhsol}) displays the
string worldsheet as a ruled surface in AdS$_5$, spanned by the straight lines at constant
$\tau$.

Combining (\ref{metric}) and (\ref{mikhsol}), the induced metric on the worldsheet is found
to be
$$
g_{\tau\tau}={R^2\over
z^2}(z^2\,\mathring{}\;\mathring{}\!\!\!x^2-1),\qquad
g_{zz}=0,\qquad g_{z\tau}=-{R^2\over z^2},
$$
implying in particular that the constant-$\tau$ lines are null, a fact that plays an
important role in Mikhailov's construction. In the solution (\ref{mikhsol}), the behavior at
time $t=X^{0}(\tau,z)$ of the string segment located at radial position $z$ is completely
determined by the behavior of the string endpoint at a retarded time $\tret(t,z)$ obtained
by projecting back toward the boundary along the null line at fixed $\tau$. {}From the
$\mu=0$ component of (\ref{mikhsol}), parametrizing the quark worldline by $x^0(\tau)$
instead of $\tau$, and using $d\tau=\sqrt{1-\vec{v}^{\,2}}dx^0$, where $\vec{v}\equiv
d\vec{x}/dx^0$, this amounts to
\begin{equation}\label{tret}
t=z{1\over\sqrt{1-\vec{v}^{\,2}}}+\tret~,
\end{equation}
where the endpoint velocity $\vec{v}$ is meant to be evaluated at $\tret$. In these same
terms, the spatial components of (\ref{mikhsol}) can be formulated as
\begin{equation}\label{xmikh}
\vec{X}(t,z)=z{\vec{v}\over\sqrt{1-\vec{v}^{\,2}}}+\vec{x}(\tret)=(t-\tret)\vec{v}+\vec{x}(\tret)~.
\end{equation}

Working in the static gauge $\sigma^0=t$, $\sigma^1=z$, the total energy of a string that
extends all the way down to the boundary--- i.e., with $z_m=0$, corresponding to an
infinitely massive quark--- follows from  the Nambu-Goto action (\ref{nambugoto}) as
\begin{equation}\label{estring}
E(t)={\sqrt{\lambda}\over 2\pi}\int_0^{\infty} {dz\over z^2}\frac{1+\left({\p\vec{X}\over\p
z}\right)^2} {\sqrt{1-\left({\p\vec{X}\over\p t}\right)^2+\left({\p\vec{X}\over\p
z}\right)^2-\left({\p\vec{X}\over\p t}\right)^2\left({\p\vec{X}\over\p
z}\right)^2+\left({\p\vec{X}\over\p t}\cdot{\p\vec{X}\over\p z}\right)^2}}~.
\end{equation}
Using (\ref{tret}) and (\ref{xmikh}), Mikhailov was able to reexpress this energy (via a
change of integration variable $z\to\tret$) as a local functional of the quark trajectory,
\begin{equation}\label{emikh}
E(t)={\sqrt{\lambda}\over 2\pi}\int^t_{-\infty}d\tret
\frac{\vec{a}^{\,2}-\left[\vec{v}\times\vec{a}\right]^2}{\left(1-\vec{v}^{\,2}\right)^3}
+E_q(\vec{v}(t))~,
\end{equation}
where of course $\vec{a}\equiv d\vec{v}/dx^0$. The second term in the above equation arises
from a total derivative that was not explicitly written down by Mikhailov, but can easily be
worked out to be
\begin{equation}\label{edr}
E_q(\vec{v})={\sqrt{\lambda}\over
2\pi}\left.\left({1\over\sqrt{1-\vec{v}^{\,2}}}{1\over
z}\right)\right|^{z_m=0}_{\infty}=\gamma m~,
\end{equation}
which gives the expected Lorentz-invariant dispersion relation for
the quark. The energy split achieved in (\ref{emikh}) therefore
admits a clear and pleasant physical interpretation: $E_q$ is the
intrinsic energy of the quark at time $t$, and the integral over
$\tret$ encodes the accumulated energy \emph{lost} by the quark
over all times prior to $t$. No less remarkable is the fact that
the rate of energy loss for the quark in this strongly-coupled
non-Abelian theory is found to be in precise agreement with the
standard Lienard formula from classical electrodynamics!

For the momentum of the string, Mikhailov analogously found
\begin{equation}\label{pmikh}
\vec{P}(t)={\sqrt{\lambda}\over 2\pi}\int^t_{-\infty}d\tret
\frac{\vec{a}^{\,2}-\left[\vec{v}\times\vec{a}\right]^2}{\left(1-\vec{v}^{\,2}\right)^3}
\vec{v}+\vec{p}_q(\vec{v}(t))~,
\end{equation}
where the second term can be worked out to be
\begin{equation}\label{pdr}
\vec{p}_q={\sqrt{\lambda}\over
2\pi}\left.\left({\vec{v}\over\sqrt{1-\vec{v}^{\,2}}}{1\over
z}\right)\right|^{z_m=0}_{\infty}=\gamma m\vec{v}~,
\end{equation}
and encodes the momentum intrinsic to the quark. The last two
equations also follow from (\ref{emikh}) and (\ref{edr}) through
Lorentz invariance.

\subsection{Finite mass} \label{finitemasssec}

It is interesting to consider how these results are modified in the case $z_m>0$, where the
mass $m$ of the quark given by (\ref{zmnoplasma}) is large but not infinite. As we have
reviewed above, in Mikhailov's original solution (\ref{mikhsol}) the evolution of the string
at any radial position $z$ follows from knowledge of the trajectory of the endpoint at the
AdS boundary. For a finite-mass quark, we ought to impose boundary conditions on the string
not at $z=0$ but at $z=z_m$: given the worldline $\vec{x}(t)$ of the quark, we must require
that the string worldsheet satisfy $\vec{X}(t,z_m)=\vec{x}(t)$. Moreover, we need only
determine the behavior of the string in the region $z\ge z_m$.

The required physical solution can of course be viewed as merely
the $z\ge z_m$ portion of one particular instance of the general
solution found by Mikhailov. Our task is therefore to reexpress
(\ref{mikhsol}) in terms of the data $\vec{x}(t)$ at the new
boundary $z=z_m$, instead of the (now merely auxiliary) data at
the AdS boundary $z=0$, which we will henceforth distinguish with
a tilde: $\vec{\tilde{x}}(t)$. For simplicity, we will carry out
this translation explicitly only in a setup where the quark moves
purely along direction $x\equiv x^1$, which is all that we will
need for our analysis in subsequent sections.

It follows from (\ref{tret}) and  (\ref{xmikh}) that, at any given point $(t,z)$ on the
string worldsheet,
\begin{eqnarray}\label{differentials}
dt&=&{dz\over\sqrt{1-\tv^2}}+d\tret\left[{\tv\ta z\over(1-\tv^2)^{3/2}}+1\right]~,\\
dX&=&{\tv dz\over\sqrt{1-\tv^2}}+d\tret\left[{\ta z\over\sqrt{1-\tv^2}}+{\tv^2\ta
z\over(1-\tv^2)^{3/2}}+\tv\right]~,\nonumber
\end{eqnarray}
where $\tv,\ta$ denote the velocity and acceleration at the point $(t=\tret,z=0)$ on the AdS
boundary obtained by projecting back from $(t,z)$ along a null trajectory. {}From
(\ref{differentials}) we can deduce that
\begin{equation}\label{xdot}
\left(\p X\over \p t\right)_{\! z} = \frac{\ta z +\tv
(1-\tv^2)^{3/2}}{\tv\ta z +(1-\tv^2)^{3/2}}~.
\end{equation}
Evaluated at the new boundary $z=z_m$, this formula relates the
velocity $v\equiv dx/dt=\p_t X(t,z_m)$ of the actual string
endpoint--- i.e., the velocity of the finite-mass quark--- to the
velocity $\tv$ and acceleration $\ta$ of the `auxiliary endpoint'
at $z=0$. Equation (\ref{xdot}) implies that the quark
acceleration $a\equiv d^2 x/dt^2=\p_t^2 X(t,z_m)$ depends not only
on $\tv$ and $\ta$, but also on the second time derivative of
$\tv$. Because of this, it is not possible to solve for $\tv$ and
$\ta$, the quantities that appear directly in Mikhailov's energy
formula (\ref{emikh}), in terms of $v$ and $a$, the data that we
would naively expect to suffice to characterize the rate of energy
loss of the heavy quark.

On the other hand, from (\ref{tret}) and (\ref{xmikh}) we can infer as well that
\begin{equation}\label{xprime}
 \left(\p X\over \p z\right)_{\! t}
=-\frac{\sqrt{1-\tv^2}\ta z}{\tv\ta z+(1-\tv^2)^{3/2}}~,
\end{equation}
so it is certainly possible to solve for $\tv,\ta$ in terms of $v$ and $\p_z X(t,z_m)$.
Notice that when $z_m\to 0$, we automatically have $\p_z X(t,z_m)\to 0$, which explains why
this parameter was not needed in the description of the infinitely-massive quark. For
$z_m>0$,  the value of $\p_z X(t,z_m)$ encodes how much the string tip tilts away from the
vertical. But what does this mean in gauge-theoretic language? Through a calculation of
$\langle\tr F^2\rangle$ or $\langle T_{\mu\nu}\rangle$ in parallel with that of
\cite{dkk,cg,gluonicprofile}, the slant of the string will have an impact on the shape of
the gluonic field profile in the immediate vicinity of the heavy quark: for $\p_z
X(t,z_m)\neq 0$, this profile is not spherically symmetric. We find it physically more
transparent to express the rate of energy loss in terms not of $\p_z X(t,z_m)$ but of the
string momentum density $\Pi^z_x$ given by (\ref{momenta}), which by use of (\ref{xdot}) and
(\ref{xprime}) can be rewritten as
\begin{equation}\label{pizx}
\Pi^z_x={\ta\over z(1-\tv^2)^{3/2}}~.
\end{equation}
When evaluated at $z=z_m$, this controls the external force
$F\equiv(\sqrt{\lambda}/2\pi)\Pi^z_x(t,z_m)$ acting on the string endpoint, or equivalently,
on the quark.

Inverting (\ref{xdot}) and (\ref{pizx}), we find
\begin{eqnarray}\label{vtildeatilde}
\tv&=&\frac{v- z_m^2 \Pi }{1-z_m^2 v \Pi }~,\\
\ta&=&z_m \Pi \frac{(1-v^2)^{3/2}(1-z_m^4 \Pi^2)}{(1-z_m^2 v \Pi )^3}~,\nonumber
\end{eqnarray}
where we have abbreviated $\Pi\equiv\Pi^z_x$. Using this and (\ref{differentials}) in
(\ref{emikh}), we finally conclude that the total energy of the string at time $t$ is given
by
\begin{equation}\label{emikhf}
E(t)={\sqrt{\lambda}\over 2\pi}\int_{-\infty}^t \!dt\, z_m^2 \Pi
^2\left[\frac{1-v\Pi z_m^2}{1-z_m^4 \Pi^2}\right]+E_q(v(t),F(t))
~.
\end{equation}
As before, the first term represents the accumulated energy lost by the quark at all times
prior to $t$: it is the generalization to the $m<\infty$ case of the Lienard formula
(\ref{emikh}) deduced by Mikhailov.  The second term again denotes a surface term and gives
the modified dispersion relation for the finite-mass quark,
\begin{equation}\label{edrf}
E_q(v,F) ={\sqrt{\lambda}\over 2\pi}\left.\left({1-z_m^2 v \Pi
\over z\sqrt{(1-v^2)(1-z_m^4
\Pi^2)}}\right)\right|^{z_m}_{\infty}=\left({2\pi
m^2-\sqrt{\lambda} v F\over\sqrt{4\pi^2 m^4-\lambda
F^2}}\right)\gamma m~.
\end{equation}

Starting instead from Mikhailov's formula (\ref{pmikh}) for the momentum, we find
\begin{equation}\label{pmikhf}
P(t)={\sqrt{\lambda}\over 2\pi}\int_{-\infty}^t \!dt\, z_m^2
\Pi^2\left[\frac{v-\Pi z_m^2}{1-z_m^4 \Pi^2}\right]+p_q(v(t),F(t))
~,
\end{equation}
where
\begin{equation}\label{pdrf}
p_q(v,F) ={\sqrt{\lambda}\over 2\pi}\left.\left({v-z_m^2  \Pi
\over z\sqrt{(1-v^2)(1-z_m^4
\Pi^2)}}\right)\right|^{z_m}_{\infty}=\left({2\pi m^2
v-\sqrt{\lambda} F\over\sqrt{4\pi^2 m^4-\lambda F^2}}\right)\gamma
m~.
\end{equation}

Notice that $\p E_q /\p p_q =(2\pi m^2 v-\sqrt{\lambda} F)/(2\pi m^2-\sqrt{\lambda} v F)$,
which for $m<\infty$ and $F\neq 0$ differs from the result expected for a pointlike quark,
$\p E/\p p =v$. This reflects the fact that the fundamental source dual to a string that
terminates at $z_m>0$ is indeed \emph{not pointlike}. According to the standard UV/IR
connection \cite{uvir}, it has a linear size of order $z_m$, and it is only because of this
extended nature that, as we saw above, to characterize its state one needs to specify not
only the velocity $v$ but also the parameter $F$ (or $\p_z X(t,z_m)$) that encodes its
shape. The crucial point here is that the source in question should not be thought of as a
bare quark, but as a `dressed' or `constituent' quark, surrounded by a gluonic cloud with
thickness $z_m$ \cite{martinfsq,mateos}. We will see more evidence of this in Section
\ref{potentialsec}.

Another salient feature of the energy and momentum of the quark given by expressions
(\ref{edrf}) and (\ref{pdrf}) is the fact that they both diverge as the value of the
external force approaches
\begin{equation}\label{Fcritnoplasma}
F_{\mbox{\scriptsize crit}}= {2\pi m^2\over \sqrt{\lambda}}~.
\end{equation}
The reason for this is easy to understand on the string theory side. To exert a force $F$ on
the string endpoint, within the D7-branes we must turn on an electric field that has
strength $F_{01}=F$ at $z=z_m$. Working in static gauge, the Born-Infeld Lagrangian on the
D7-branes is then
$$
\sqrt{-\det(g_{ab}+2\pi\ap F_{ab})}\propto\sqrt{-G_{tt}G_{xx}-(2\pi\ap
F_{01})^2}=R^2\sqrt{{1\over z^4}-\left({2\pi\over\sqrt{\lambda}}F_{01}\right)^2}~,
$$
which is real at $z=z_m$ only as long as the electric field is
below the value $F^{\mbox{\scriptsize crit}}_{01}={\sqrt{\lambda}/
2\pi z_m^4}$. Through (\ref{zmnoplasma}), this is seen to coincide
with the value of the critical force (\ref{Fcritnoplasma}). The
physical origin of this bound is the fact that, for
$F_{01}>F^{\mbox{\scriptsize crit}}_{01}$, the creation of open
strings is energetically favored, and so the system is unstable.
According to (\ref{edrf}) and (\ref{pdrf}), then, the energy and
momentum of the constituent quark diverge precisely at the point
where the external force becomes capable of nucleating
quark-antiquark pairs.

\subsection{Late-time behavior and worldsheet black hole}
\label{bhsec}

It is interesting  to consider the evolution of the string in the
case where the quark is accelerated by an external force $F(t)$
over some period of time and is then set free at a time
$t_{\mbox{\scriptsize release}}$. If we put $F=0$ ($\Pi=0$) in
(\ref{vtildeatilde}), then $\ta=0$ (the acceleration of the
auxiliary $z=0$ endpoint of the string vanishes), which in turn
implies through (\ref{xdot}) and (\ref{xprime}) that $\dot{X}=\tv$
and $X'=0$ at all points on the same null (constant $\tret$) line,
independently of the value of $z$. In particular, the lower
($z=z_m$) endpoint of the string travels at speed
$v\equiv\dot{X}(t,z_m)=\tv$, which means that that, for $t\ge
t_{\mbox{\scriptsize release}}$, the quark moves at constant
velocity $v$, as one would expect given the fact that it is in
vacuum.

As time progresses, the null line that departs from the point
$(t_{\mbox{\scriptsize release}},z_m)$, which according to
(\ref{tret}) is given by
\begin{equation}\label{zrelease}
z_{\mbox{\scriptsize
release}}(t)=\sqrt{1-v^2}(t-t_{\mbox{\scriptsize release}})+z_m~,
\end{equation}
reaches further away from the AdS boundary, so there is an
increasing portion of the string ($z_m\le z\le
z_{\mbox{\scriptsize release}}(t)$) that is completely vertical
and moves with the same final velocity $v$. As $t\to\infty$, this
vertical segment describes a quark that moves at constant speed
and is surrounded by a stationary gluonic field profile, related
to that of a static quark through a Lorentz transformation.

At any given time $t\ge t_{\mbox{\scriptsize release}}$, then, the
energy and momentum previously radiated by the quark are stored as
excess energy and momentum on the portion of the string above
$z_{\mbox{\scriptsize release}}(t)$. This suggests the existence
of a geometric region on the worldsheet that `absorbs' the surplus
$E$ and $p$.

 To define this region more precisely, we should note first that, for the
spacetime metric (\ref{metric}) (with $h=1$), null curves on the worldsheet obey
\begin{equation}\label{znulldot}
\dot{z}_{\mbox{\scriptsize
null}}^{(\pm)}(t)=\frac{-X'\dot{X}\pm\sqrt{1+{X^{'}}^{2}-{\dot{X}}^2}}{{X^{'}}^{2}+1}~,
\end{equation}
with the upper or lower sign for the upward- or downward-pointing half of the future light
cone, respectively. Using (\ref{xdot}), (\ref{xprime}) and (\ref{vtildeatilde}), this
translates into
\begin{equation}\label{znulldotplus}
\dot{z}_{\mbox{\scriptsize
null}}^{(+)}(t)=\frac{\sqrt{(1-v^2)(1-z_m^4\Pi^2)}}{1-v z_m^2 \Pi}
\end{equation}
and
\begin{equation}\label{znulldotminus}
\dot{z}^{(-)}_{\mbox{\scriptsize null}}(t)=-\frac{(1-v z^2_m
\Pi)\sqrt{(1-v^2)(1-z^4_m\Pi^2)}\left[1 -(z^2+z_m^2
)z_m^2\Pi^2\right] }{\left[ z z_m\Pi(v-z^2_m\Pi)+(1- v z^2_m
\Pi)\sqrt{1-z^4_m\Pi^2}\right] ^2}~,
\end{equation}
where, as before, $v$ and $\Pi$ refer to the velocity and external force at $z_m$, evaluated
at the retarded time corresponding to the given $(t,z)$.
 As mentioned earlier, Mikhailov's constant-$\tret$ lines (\ref{tret})-(\ref{xmikh}), and
(\ref{zrelease}) in particular, are null, and indeed they can be easily seen to satisfy
(\ref{znulldotplus}).

If we follow the point $z_{\mbox{\scriptsize release}}(t)$ as
$t\to\infty$ (reaching the spacetime horizon for the Poincar\'e
patch,  $z\to\infty$), and then project back along the
\emph{downward} pointing light half-cone $z_{\mbox{\scriptsize
null}}^{(-)}(t)$, we delineate a region ($z\ge
z_{\mbox{\scriptsize null}}^{(-)}(t)$) on the string worldsheet
from which, by construction, no signal can escape to the
asymptotic region corresponding to the final vertical and
stationary segment of the string. In other words, the curve
$z_{\mbox{\scriptsize null}}^{(-)}(t)$ so obtained, which we will
henceforth denote by $z_{\mbox{\scriptsize BH}}(t)$, is the event
horizon of a worldsheet black hole.

A worldsheet black hole figured prominently in the energy loss
analysis of \cite{ctqhat,gubserqhat}, concerning a quark in a
thermal plasma. Those works considered the steady-state
configuration where the quark moves at a constant velocity $v$; as
a consequence, the black hole they encountered was static, with an
event horizon located at the fixed position $z_{\mbox{\scriptsize
BH}}(t)\equiv z_h(1-v^2)^{1/4}$, precisely the radius that played
a crucial role in the drag force calculation of
\cite{hkkky,gubser}.

 In our
non-stationary system, on the other hand, the worldsheet black
hole is dynamical. Let us focus for concreteness on the case where
the quark is static up to a time $t=t_{\mbox{\scriptsize grab}}$,
and is then accelerated until $t=t_{\mbox{\scriptsize release}}$.
The location $z_{\mbox{\scriptsize BH}}(t)$ of the event horizon
will begin descending from $z\to\infty$ even before
$t=t_{\mbox{\scriptsize grab}}$ (in anticipation of the
disturbance produced by the acceleration at the lower endpoint),
reach a minimum value of the radial coordinate, and then move up
again, approaching $z_{\mbox{\scriptsize release}}(t)$ as
$t\to\infty$. Notice from (\ref{znulldotminus}) that, at any given
$z$, the light half-cone $\dot{z}^{(-)}_{\mbox{\scriptsize null}}$
can only point towards increasing $z$ for sufficiently large $\Pi$
(and, for any given $\Pi\neq 0$, it \emph{will} point upward for
sufficiently large $z$). This means that the entire upward portion
of $z_{\mbox{\scriptsize BH}}(t)$ must lie within the region of
maximal disturbance of the worldsheet, i.e., in the diagonal swath
between $z_{\mbox{\scriptsize grab}}(t)\equiv
t-t_{\mbox{\scriptsize grab}}+z_m$ and $z_{\mbox{\scriptsize
release}}(t)$.

\begin{figure}[tbph]
\begin{center}
\vspace*{0.5cm}
 \setlength{\unitlength}{1cm}
\includegraphics[width=11cm,height=6cm]{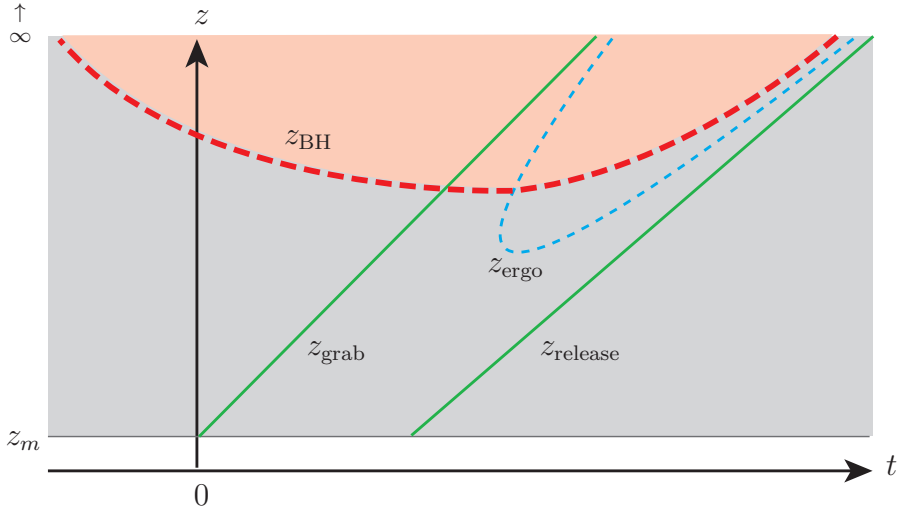}
 \begin{picture}(0,0)
   \put(0,0.1){$t$}
   \put(-9.2,-0.3){$0$}
   \put(-9.2,6.1){$z$}
   \put(-11.7,6){$\uparrow\atop\infty$}
   \put(-11.7,0.5){$z_m$}
   \put(-4.6,1.7){$z_{\mbox{\scriptsize release}}$}
   \put(-7.7,1.7){$z_{\mbox{\scriptsize grab}}$}
   \put(-5.3,2.8){$z_{\mbox{\scriptsize ergo}}$}
   \put(-8.0,4.5){$z_{\mbox{\scriptsize BH}}$}
  \end{picture}
 \end{center}
\caption{Schematic illustration of the string worldsheet (shaded in gray), in the static gauge
$\tau=t$, $\sigma=z$, showing the upward null Mikhailov (fixed $t_{\mbox{\scriptsize tret}}$) lines
 $z_{\mbox{\scriptsize grab}}$ and $z_{\mbox{\scriptsize release}}$ (solid green),
 the stationary limit curve $z_{\mbox{\scriptsize ergo}}$ (dotted blue), and
  the event horizon $z_{\mbox{\scriptsize BH}}$ (thick dotted red) above
  which lies the worldsheet black hole (shaded light red). See text for discussion.}\label{bhfig}
\end{figure}

As usual, determining the exact location of the event horizon is
difficult due to the global character of its definition:  one must
know the entire history of the string and then integrate
(\ref{znulldotminus}), subject to the stated final condition. It
is, however, easy to pinpoint with this same equation the location
on the worldsheet where $\dot{z}^{(-)}_{\mbox{\scriptsize
null}}(t)=0$, which gives a lower bound on the upward portion of
$z_{\mbox{\scriptsize BH}}(t)$ (where, by definition, one has
$\dot{z}^{(-)}_{\mbox{\scriptsize null}}(t)>0$). {}From
(\ref{znulldotminus}), this happens at
\begin{equation}\label{zergo}
z_{\mbox{\scriptsize
ergo}}(t)=\frac{\sqrt{1-z^4_m\Pi^2}}{z_m\Pi}~,
\end{equation}
where, again, the external force $\Pi$ is meant to be evaluated at the retarded time
appropriate for the given $t,z$. At any point along this curve, the downward light half-cone
is horizontal (or, equivalently, $g_{tt}=0$), so timelike trajectories must necessarily
point towards larger $z$. In other words, $z_{\mbox{\scriptsize ergo}}(t)$ is a
stationary-limit curve, and the region between it and the upward portion of the event
horizon is the analog of an ergosphere, a concept whose relevance has been noted previously
in the $T>0$ context in \cite{argyres2}. Setting $\Pi=0$ in (\ref{zergo}) implies
$z_{\mbox{\scriptsize ergo}}\to\infty$, so, unlike the event horizon, the stationary limit
curve is located fully within the diagonal region between $z_{\mbox{\scriptsize grab}}(t)$
and $z_{\mbox{\scriptsize release}}(t)$. It follows from the definitions of the two curves
that $z_{\mbox{\scriptsize BH}}(t)$ crosses $z_{\mbox{\scriptsize ergo}}(t)$ precisely when
the former attains its minimum value, and so the downward portion of the horizon lies below
and to the left of $z_{\mbox{\scriptsize ergo}}(t)$. The situation is summarized in
Fig.~\ref{bhfig}.

It is interesting that at $T=0$ the notion of a worldsheet black hole plays as much of a
role as in previous analyses at finite temperature. The appearance of such causal structure
is seen then to be intrinsically tied to energy dissipation, be it within a thermal plasma
or in vacuum. We will examine the former case in Section \ref{bhplasmasec}. It would be nice
to develop this picture further by exploring the relation between the rate at which energy
crosses the black hole horizon and the modified Lienard formula in (\ref{emikhf}).

\section{Single Quark Evolution: Finite temperature}
\label{qplasmasec}

Having understood the rate of energy and momentum loss and dispersion relation for a heavy
quark that moves in the SYM vacuum, in this section we restore $z_h<\infty$--- and
consequently $h<1$--- in the metric (\ref{metric}), to study the same quantities in the case
where the quark moves through a thermal plasma. In this case, the position $z=z_m\le z_h$
where the D7-branes `end' is related to the Lagrangian mass $m\gg \sqrt{\lambda}T$ of the
quark through \cite{hkkky}
\begin{equation}\label{zm} {1\over z_m}={2\pi
m\over\sqrt{\lambda}}\left[1+{1\over 8}\left(\sqrt{\lambda} T\over 2 m\right)^4-{5\over
128}\left(\sqrt{\lambda} T\over 2 m\right)^8+\cO\left(\left(\sqrt{\lambda} T\over 2
m\right)^{12}\right)\right]~.
\end{equation}

A thorough generalization of Mikhailov's analytic results \cite{mikhailov} to this finite
temperature setup would require finding the exact solution to the Nambu-Goto equation of
motion for the string on the AdS-Schwarzschild background, for any given trajectory of the
string endpoint at $z_m\ge 0$. Sadly, we have not been able to accomplish this feat.
Nevertheless, based on the results discussed in the previous subsection, we expect the total
energy of the string at any given time to again decompose into a surface term that encodes
the intrinsic energy of the quark and an integrated local term that reflects the energy lost
by the quark.

\emph{A priori} it might not be obvious that the rate of energy loss in the presence of the
strongly-coupled non-Abelian plasma should be given by some expression that depends just on
the behavior of the quark at the given instant, and not on its previous history. But in the
AdS/CFT context, this property is strongly suggested by the fact that the energy of the
string is given by a local expression on the worldsheet just as much in the
AdS-Schwarzschild background that is dual to the thermal plasma as in the pure AdS
background that corresponds to the SYM vacuum (which, one should not forget, is in itself a
nonlinear medium). Starting with $E$ written as an integral over $z$, it should again be
possible to project back to the boundary along null trajectories, to obtain a formula that
depends locally on the quark worldline $\vec{x}(t)$. The main difference with the
zero-temperature case would be that the null trajectories are no longer straight lines.

\subsection{Constant velocity} \label{constantvsec}

We have verified that these expectations are borne out in the case of the only
finite-temperature solution that is known analytically: the stationary configuration of
\cite{hkkky,gubser},
\begin{equation}\label{gubsersol}
X(t,z)=v\left[ t-{z_h\over 4}\ln\left(z_h+z\over z_h-z\right)+{z_h\over
2}\tan^{-1}\left({z\over z_h}\right)\right]~,
\end{equation}
which describes a quark moving at constant velocity $v$. Any given point $(t,z)$ on this
worldsheet is connected to a point $(\tret,0)$ on the AdS boundary by a null curve $t(z)$,
defined by
\begin{equation}\label{gubsernull}
\left(dt\over
dz\right)_{\!\tret}=\frac{z_h^4(z_h^2+\sqrt{1-v^2}z^2)}{(z_h^4-z^4)(z^2+z^2_h\sqrt{1-v^2})}~.
\end{equation}
This equation can be integrated to give
\begin{equation}\label{gubsertret}
t=\tret+{z_h\over 4}\ln\left(z_h+z\over z_h-z\right)-{z_h\over 2}\tan^{-1}\left({z\over
z_h}\right)+{z_h\over(1-v^2)^{1/4}}\tan^{-1}\left(z\over z_h(1-v^2)^{1/4}\right)~,
\end{equation}
in terms of which the stationary solution (\ref{gubsersol}) can be written in the form
\begin{equation}\label{gubsermikhsol}
X(\tret,z)={z_h v\over(1-v^2)^{1/4}}\tan^{-1}\left(z\over z_h(1-v^2)^{1/4}\right)+x(\tret)~.
\end{equation}

Knowledge of this solution allows the total energy of the string,
\begin{equation}
E(t)=-\frac{\sqrt{\lambda}}{2\pi}\int_{z_m}^{z_h} dz\,
\Pi^t_t=\frac{\sqrt{\lambda}}{2\pi}\int_{z_m}^{z_h}
dz\,\frac{h{X^{'}}^{2}+1}{z^2\sqrt{1+h{X^{'}}^{2}-{\dot{X}^2\over
h}}}~,
\end{equation}
to be reexpressed as
\begin{multline}\label{Emikhgubser}
E(t)=\frac{\sqrt{\lambda}}{2\pi }\int^{t}_{-\infty} d\tret\frac{v^2}{z_h^2\sqrt{1-v^2}}\\
+\frac{\sqrt{\lambda}}{2\pi}\left[
\frac{1}{z_m\sqrt{1-v^2}}+\frac{v^2}{z_h(1-v^2)^{\frac{3}{4}}}\tan^{-1}
\left(\frac{z_m}{z_h(1-v^2)^{\frac{1}{4}}}
\right)\right.\\
\left.-\frac{1}{z_h\sqrt{1-v^2}}-\frac{v^2}{z_h(1-v^2)^{\frac{3}{4}}}\tan^{-1}
\left(\frac{1}{(1-v^2)^{\frac{1}{4}}} \right) \right]~.
\end{multline}
Similarly, the total momentum of the string,
\begin{equation}
P(t)=\frac{\sqrt{\lambda}}{2\pi}\int_{z_m}^{z_h} dz\,
\Pi^t_x=\frac{\sqrt{\lambda}}{2\pi}\int_{z_m}^{z_h}
dz\,\frac{\dot{X}}{z^2h\sqrt{1+h{X^{'}}^{2}-{\dot{X}^2\over h}}}~,
\end{equation}
can be rewritten in the form
\begin{multline}\label{Pmikhgubser}
P(t)=\frac{\sqrt{\lambda}}{2\pi }\int^{t}_{-\infty} d\tret\frac{v}{\sqrt{1-v^2}}\\
+\frac{\sqrt{\lambda}}{2\pi } \left[\frac{v}{z_m\sqrt{1-v^2}}
+\frac{v}{z_h(1-v^2)^\frac{3}{4}}\tan^{-1}\left(\frac{z_m}{z_h(1-v^2)^\frac{1}{4}} \right)\right.\\
\left. -\frac{v}{z_h\sqrt{1-v^2}}
-\frac{v}{z_h(1-v^2)^\frac{3}{4}}\tan^{-1}\left(\frac{1}{(1-v^2)^\frac{1}{4}}\right)\right]~.
\end{multline}

As expected, the integrated term in the top line of (\ref{Emikhgubser}) and
(\ref{Pmikhgubser}) recovers the result for the stationary rate of energy and momentum loss
obtained in \cite{hkkky,gubser},
\begin{equation}\label{EPlossgubser}
\left({dE_q\over dt}\right)_{\mbox{\scriptsize s}}=-{\pi\over
2}\sqrt{\lambda}T^2\frac{v^2}{\sqrt{1-v^2}}~,\qquad \left({dp_q\over
dt}\right)_{\mbox{\scriptsize s}}=-{\pi\over 2}\sqrt{\lambda}T^2\frac{v}{\sqrt{1-v^2}}~.
\end{equation}
 The terms in the second and third line of (\ref{Emikhgubser}) and (\ref{Pmikhgubser}),
 then, codify the energy $E_q$ and momentum $p_q$ that are intrinsic to the
quark. We can  see that, just like at zero temperature, $\p E_q/\p
p_q$ reduces to $v$ only in the pointlike limit $z_m\to 0$.

An important difference with respect to the $T=0$ case analyzed in
the previous subsection is that here the surface contribution that
determines the quark dispersion relation arises not only from the
lower ($z=z_m$) but also from the upper ($z=z_h$) endpoint of the
string. This is in fact the generic situation in the $T>0$ case,
and holds even for the static radial string, where there is of
course no energy loss term and $E$ is given just by a boundary
contribution that defines the thermal rest mass of the quark,
\begin{equation}\label{Mrest}
M_{\mbox{\scriptsize rest}}={\sqrt{\lambda}\over
2\pi}\left({1\over z_m} - {1\over z_h}\right)~.
\end{equation}

Clearly the second term in (\ref{Mrest}), just like the terms in the third line of
(\ref{Emikhgubser}) and (\ref{Pmikhgubser}), arises from the string endpoint located at the
black hole horizon. It is natural then to wonder to what extent these terms should be
regarded as a contribution to the intrinsic energy of the quark, because their value at any
given time does not depend on the parameters $v$ and $F$ (or $\p_z X(t,z_m)$) that
characterize the state of the \emph{lower} endpoint at the same instant. In fact, since
$z_h$ marks the position of an event horizon, for any finite coordinate time $t$ the value
of the surface contribution at $z_h$ is not influenced by the behavior of the $z<z_h$
portion of the string, but depends only on the string's configuration at $t\to-\infty$. The
same interpretation can be then carried over to the gauge theory: the seemingly extraneous
terms represent a contribution to the energy of the state that depends solely on the initial
configuration of the quark$+$plasma system. Throughout the evolution, causality guarantees
that the behavior of the SYM fields at spatial infinity can only be affected by the initial
configuration at $t\to-\infty$, so we can equivalently think of the surface terms at $z_h$
as encoding information on the asymptotic boundary conditions for the system.  Indeed, for
dynamical processes, the radial location $z=z_h$ in AdS-Schwarzschild corresponds to the
deep IR of the gauge theory.

We conclude then that, to the extent that we wish to compare the energies of configurations
with different initial/boundary conditions, it is important to keep track of the terms
arising from the surface contribution at the horizon, despite the fact that they are
generally independent of the parameters $v$ and $F$ associated with the intrinsic dynamics
of the quark. Notice, in particular, that with this interpretation the negative sign in the
second term of (\ref{Mrest})--- which is at first sight unexpected when viewed as the
leading-order thermal correction to the quark mass \cite{gubserqhat}--- becomes easier to
digest: it reflects the screening effect of the plasma on the long-range gluonic fields set
up by the quark, which implies a \emph{reduction} of the energy stored in the IR, in
comparison with the $T=0$ case.

It is instructive to compare (\ref{Emikhgubser}) with the alternative split achieved in
\cite{hkkky},
\begin{equation}\label{Ehkkky}
E(t)=\frac{\sqrt{\lambda}}{2\pi }\frac{v^2}{z_h^2\sqrt{1-v^2}}{\Delta x\over v}
+\frac{\sqrt{\lambda}}{2\pi}\left({1\over z_m}-{1\over z_h}\right) \frac{1}{\sqrt{1-v^2}}~,
\end{equation}
with $\Delta x\equiv X(t,z_m)-X(t,z_h)$. If one could interpret
this latter quantity as the total distance traversed by the quark
since the beginning of time, then the first term in (\ref{Ehkkky})
would give the overall energy lost by the quark, at the known rate
(\ref{EPlossgubser}), in the total elapsed time $\Delta x/v$. In
view of (\ref{Mrest}), the second term would then imply a standard
relativistic dispersion relation for the quark, with mass
$M_{\mbox{\scriptsize rest}}$.

The problem with this interpretation, however (alluded to already
in \cite{hkkky}), is that it does not properly address the issue
of initial conditions.  The actual distance travelled by the quark
is by definition $X(t,z_m)-X(-\infty,z_m)$, which agrees with
$\Delta x$ only if $X(-\infty,z_m)=X(t,z_h)$.\footnote{When
writing expressions like $X(-\infty,z_m)$, we of course have in
mind evaluating the corresponding quantities at a time that is
fixed but arbitrarily far in the past.} This last equality would
hold if we had started at $t\to-\infty$ with the quark at rest
(i.e., with the string static and completely vertical), but the
energy-loss term in (\ref{Ehkkky}) makes no allowance for an
initial period of acceleration.

In contrast with this, the separation obtained in (\ref{Emikhgubser}) has a clear geometric
origin in the context of the generalization pursued here of Mikhailov's work
\cite{mikhailov} to the finite-temperature case. Within this framework, the portion
$dE(t,z_h)$ of the total string energy $E(t)$ that is contributed by the segment of the
string in the immediate vicinity of the horizon (located at $X(t,z_h)$ in the notation of
(\ref{gubsersol}), or $X(\tret,z_h)$ in the notation of (\ref{gubsermikhsol})) was lost by
the quark at a particular time $\tret$ in the distant past that can be deduced from
(\ref{gubsertret}). Since this is the highest segment of the string that contributes to
$E(t)$, $\tret$ marks the precise instant when we need to begin our accounting of the energy
lost by the quark. At $t=\tret$ the lower endpoint of the string (and, hence, the quark) was
found a distance $d\equiv X(\tret,z_h)-X(\tret,z_m)$ behind the location of the upper
endpoint at $t$, which according to (\ref{gubsermikhsol}) translates into
$$ d={z_h v\over(1-v^2)^{1/4}}\tan^{-1}\left(1\over (1-v^2)^{1/4}\right)
-{z_h v\over(1-v^2)^{1/4}}\tan^{-1}\left(z_m\over z_h(1-v^2)^{1/4}\right)~.
$$
By this logic, the total energy lost by the quark is given by an expression of the same form
as the first term of (\ref{Ehkkky}), but with $\Delta x$ replaced by the actual total
distance $\Delta x + d=v\int_{-\infty}^t d\tret$. And indeed, we see that the integrated
term in the first line of (\ref{Emikhgubser}) is larger than the putative energy loss term
in (\ref{Ehkkky}) precisely by the amount $(d/v)dE_q/dt$, and, correspondingly, the
intrinsic energy of the quark identified in the second and third line of (\ref{Emikhgubser})
is smaller by this same amount than what (\ref{Ehkkky}) would have indicated.\footnote{The
fact that, in going from (\ref{Emikhgubser}) to (\ref{Ehkkky}), part of the total derivative
has been shifted back to the integrated term might give the impression that the split
between the intrinsic energy of the quark and the energy that has already been lost is
inherently ambiguous. The wide latitude available in this case, however, stems from the
steady-state nature of the configuration under scrutiny. In the general case, clearly it is
a very non-trivial property for a particular contribution to the energy $E(t)$ of the string
to be expressible as a functional only of
 the state of the endpoints at the given instant.}

\subsection{Accelerated quark} \label{acceleratedsec}

 Having gained some
intuition from the analysis of a quark moving as in
\cite{hkkky,gubser} at constant speed relative to the
strongly-coupled plasma, let us now turn our attention to the more
general situation where the quark accelerates. In \cite{hkkky}, a
few important steps were taken to have a better understanding of
this case; in the next few paragraphs we will briefly review the
key results.

A quark that undergoes \emph{any} type of (forced or unforced)
motion (and, in the former case, is thereafter released) will be
slowed down by its interaction with the plasma, and eventually
come to rest. The authors of \cite{hkkky} studied the late-time
(and consequently low-velocity, low-acceleration) behavior of such
a quark, by considering small, exponentially damped fluctuations
around the final rest configuration. In dual language, this
involves a determination of the quasi-normal modes on the
worldsheet of a static and purely radial string. {}From their
analysis they were able to numerically deduce, for any given quark
mass parameter $z_m$, the value of the drag coefficient
\begin{equation}\label{mu}
\mu\equiv -{1\over p_q}{dp_q\over dt}~.
\end{equation}
As long as one maintains the restriction to the non-relativistic regime, the above
definition is equivalent to $\mu=-(1/v)dv/dt$, and by construction yields a result that is
independent of $p$ (or $v$). A few representative values  were tabulated in \cite{hkkky}.

Additionally, the authors of \cite{hkkky} gave an analytic
derivation of the low-velocity dispersion relation for the quark,
which they found to take the form
\begin{equation}\label{DRnr}
E_q=M_{\mbox{\scriptsize rest}}+\frac{p_q^2}{2M_{\mbox{\scriptsize
kin}}}+\cO(p_q^4)~,
\end{equation}
where $M_{\mbox{\scriptsize rest}}$ is the thermal rest mass
(\ref{Mrest}), and
\begin{equation}\label{Mkin}
M_{\mbox{\scriptsize kin}}\equiv{\pi\over
2}{\sqrt{\lambda}T^2\over\mu}
\end{equation}
the kinetic  mass of the quark. In the heavy quark limit $m\gg
\sqrt{\lambda}T$ ($z_m\ll z_h$), where (\ref{Mrest}) and
(\ref{zm}) imply that
\begin{equation} \label{mrestm}
M_{\mbox{\scriptsize rest}}=m-{\sqrt{\lambda}T\over 2}+\cO\left(m\left({\sqrt{\lambda}T\over
2m}\right)^4\right)~,
\end{equation}
they found from their quasi-normal mode calculation that
\begin{equation}\label{muqnm}
\mu={\pi\over 2}{\sqrt{\lambda}T^2\over m}\left[1+{\sqrt{\lambda}T\over
2m}+\cO\left(\left({\sqrt{\lambda}T\over 2m}\right)^2\right)\right]~,
\end{equation}
which through (\ref{Mkin}) leads to
\begin{equation}\label{mkinmrest}
M_{\mbox{\scriptsize kin}}=M_{\mbox{\scriptsize rest}}+\cO\left(m\left({\sqrt{\lambda}T\over
2m}\right)^2\right)~.
\end{equation}

 It was noticed in \cite{hkkky} that expressions
(\ref{DRnr}) and (\ref{muqnm}), valid in the low-velocity,
low-acceleration regime, as well as the value of the drag
coefficient (\ref{mu}) deduced from (\ref{EPlossgubser}), valid
for a quark with constant but otherwise arbitrary velocity, are
consistent with a \emph{relativistic} dispersion relation of the
form
\begin{equation}\label{DRr}
 E_q=M_{\mbox{\scriptsize rest}}-M_{\mbox{\scriptsize kin}}+\sqrt{p_q^2+M_{\mbox{\scriptsize kin}}^2}
 =M_{\mbox{\scriptsize rest}}+M_{\mbox{\scriptsize kin}}(\gamma-1)~.
\end{equation}
Additional evidence for this relation was found from the analysis of back-to-back motion of
a quark and antiquark formed within the plasma. The results of \cite{hkkky} in this setting
will be reviewed and extended in Section \ref{qqbarsec}.

In view of the discrepancy between (\ref{DRr}) and the relation
$E_q=\sqrt{p_q^2+M_{\mbox{\scriptsize rest}}^2}$ that one would
naively infer from the second term in (\ref{Ehkkky}), the authors
of \cite{hkkky} emphasized the need for a more detailed study of
the quark's intrinsic dynamics, and proposed a plan of attack.
They observed that if one starts with the quark at rest in the hot
medium, and then accelerates it with an external force, then as
long as energy dissipation is negligible, it is natural to define
the total intrinsic energy of the quark as the initial rest energy
plus the work done by the external agent. Motivated by this
proposal, we have studied the early-time behavior of a quark
initially at rest, which is accelerated by a time-dependent
external force that is turned off after a short period of time,
allowing the quark to move thereafter only under the influence of
the plasma.

As we already mentioned, regrettably, we have  not been able to
find a general exact solution to the Nambu-Goto equation of motion
for the string in the non-stationary case. Nevertheless, it is
certainly possible to find numerical solutions that describe
string configurations dual to the gauge theory setup described
above. Starting at $t=0$, a string extending from a fixed value of
the radial coordinate $z=z_m$ to the horizon at $z=z_h$, initially
at rest and vertical, is accelerated by applying an external force
$F(t)$ to its lower endpoint. After a time $t_{\mbox{\scriptsize
release}}$, this external force is set to zero and the string
moves freely in the curved background.

Using (\ref{metric}) and (\ref{nambugoto}) and imposing the condition that the string moves
only in the $x\equiv x^1$ direction, the equation of motion is
\begin{equation}\label{eomquark}
\frac{\partial}{\partial{z}}\left[\frac{-hX'}{z^2\sqrt{{1+h{X'}^2}
-\frac{{\dot{X}}^2}{h}}}\right]+\frac{\partial}{\partial{t}}
\left[\frac{\dot{X}}{z^2h\sqrt{{1+h{X'}^2}-\frac{{\dot{X}}^2}{h}}}\right]=0~,
\end{equation}
and we must set the initial and boundary conditions to be
\begin{equation}\label{icbc}
X(0,z)=0~, \quad \dot{X}(0,z)=0~, \quad X'(t,z_m)=f(t)~, \quad X(t,z_h)=0~.
\end{equation}
The first two conditions here simply implement the requirement
that the string start out being static and purely radial. In the
third condition, for a given external force
$F(t)\equiv(\sqrt{\lambda}/2\pi)\Pi^z_x(t,z_m)$ acting on the
quark, the function $f(t)$ could be determined using the relation
(\ref{momenta}) between $X'$ and $\Pi^z_x$. In practice we find it
easier, however, to specify $f(t)$ and use the results of the
numerical integration together with (\ref{momenta}) to deduce the
associated $F(t)$. By studying a number of different examples, we
have verified that the instantaneous rate of energy loss after the
quark is released is independent of our choice of $f(t)$ and
$t_{\mbox{\scriptsize release}}$, i.e., the quark does not care
about its  past history. For the trajectories that we will plot
below, we have used $f(t)= bt(t-t_{\mbox{\scriptsize release}})$,
with $t_{\mbox{\scriptsize release}}=0.3/\pi T$ and adjustable
$b$.

The fourth and final condition in (\ref{icbc}) specifies that the string endpoint at the
horizon remain fixed, reflecting the fact that the wavefront for the disturbance produced by
the external agent at the $z=z_m$ endpoint of the string will not reach $z=z_h$ before an
infinite amount of (boundary) time has elapsed. In order to implement this condition in our
numerical integration, we have set  $X(t,z_{\mbox{\scriptsize max}})=0$ at a radial cutoff
$z_{\mbox{\scriptsize max}}=0.999 z_h$, and considered in all cases an integration time
$t_{\mbox{\scriptsize max}}$ smaller than the time $\int_{z_m}^{z_{\mbox{\tiny
max}}}dz/(1-(z/z_h)^4)$ that it takes the wavefront
 to reach the cutoff.

For applications of this formalism to phenomenology, we must choose values of the mass
parameter $z_m$ based on the charm and bottom quark masses, $m\simeq 1.4,4.8$ GeV. The issue
of how best to translate between the SYM and QCD parameters has been discussed in
\cite{gubsercompare}. Taking $\alpha_{QCD}=0.5$ ($g_{QCD}=\sqrt{2\pi}$), $N_c=3$ and
$T_{QCD}=250$ MeV, and employing the ``obvious'' prescription $g_{YM}=g_{QCD}$ and
$T_{SYM}=T_{QCD}$, from (\ref{zm}) we find that $z_m/z_h\sim 0.40$ for charm and
$z_m/z_h\sim 0.11$ for bottom. If, on the other hand, one uses the ``alternative'' scheme
$g_{YM}^2 N\sim 5.5$ (motivated in \cite{gubsercompare} through a rough matching of the
AdS/CFT and lattice quark-antiquark potentials) and $T_{SYM}=3^{-1/4}T_{QCD}\simeq 190$ MeV
(which follows from equating the energy densities of the two theories), then (\ref{zm})
leads to $z_m/z_h\sim 0.16$ for  charm and $z_m/z_h\sim 0.046$ for bottom. In our analysis,
we have covered a significant range of masses, but below we will present the results only
for three representative values in the neighborhood of the charm mass:
$z_m/z_h=0.2,0.3,0.4$.

We have carried out the numerical integration of (\ref{eomquark})
subject to (\ref{icbc}) using the {\tt NDSolve} routine of
Mathematica 5.2. Based on the variation of our results upon
doubling the number of integration steps, we estimate our numerics
to be accurate to better than $1\%$. The integration time shown in
all plots is given in units of 1/$\pi T$, which (for $T_{QCD}=250$
MeV) corresponds to $0.25~\mbox{fm}\!/c$ under the ``obvious'' and
$0.33~\mbox{fm}\!/c$ under the ``alternative'' prescription of
\cite{gubsercompare}. Unfortunately, the numerical integration
degrades rather quickly, so in either scheme our investigation is
limited to intervals that are an order of magnitude below the
experimental timescale of the plasma (typically
$t_{\mbox{\scriptsize breakdown}}\sim 0.9 /\pi T\sim
0.3~\mbox{fm}\!/c$). For the same reason, even though the quark
can be taken to relativistic
 velocities at the point of
maximal acceleration, we can only achieve rather small velocities
at the time of release (for the most part, $v_{\mbox{\scriptsize
release}}<0.1$).

Our results show a qualitative difference between the initial
stage ($0\le t < t_{\mbox{\scriptsize release}}$) where the quark
is accelerated by means of the external force $F(t)$, and the
second stage ($t_{\mbox{\scriptsize release}}\le t
<t_{\mbox{\scriptsize max}}$) where it moves only under the
influence of the plasma. We will begin by discussing the latter
stage, which would appear to be more relevant from the
phenomenological perspective.

\begin{figure}[tbph]
\vspace*{0.5cm}
 \setlength{\unitlength}{1cm}
\includegraphics[width=7cm,height=5cm]{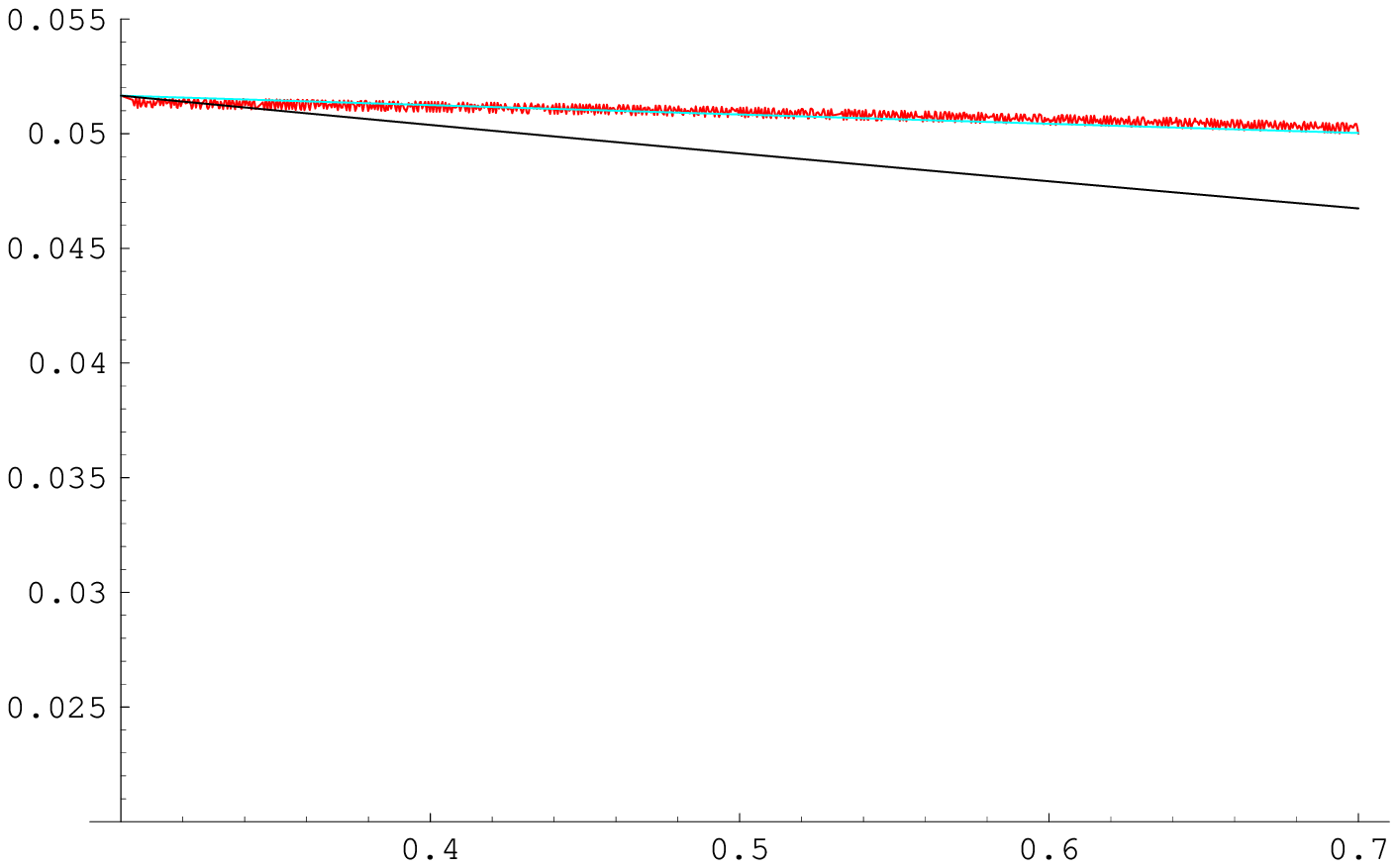}
 \begin{picture}(0,0)
   \put(0,0.3){$t$}
   \put(-6.5,5.1){$v$}
 \end{picture}
\includegraphics[width=7cm,height=5cm]{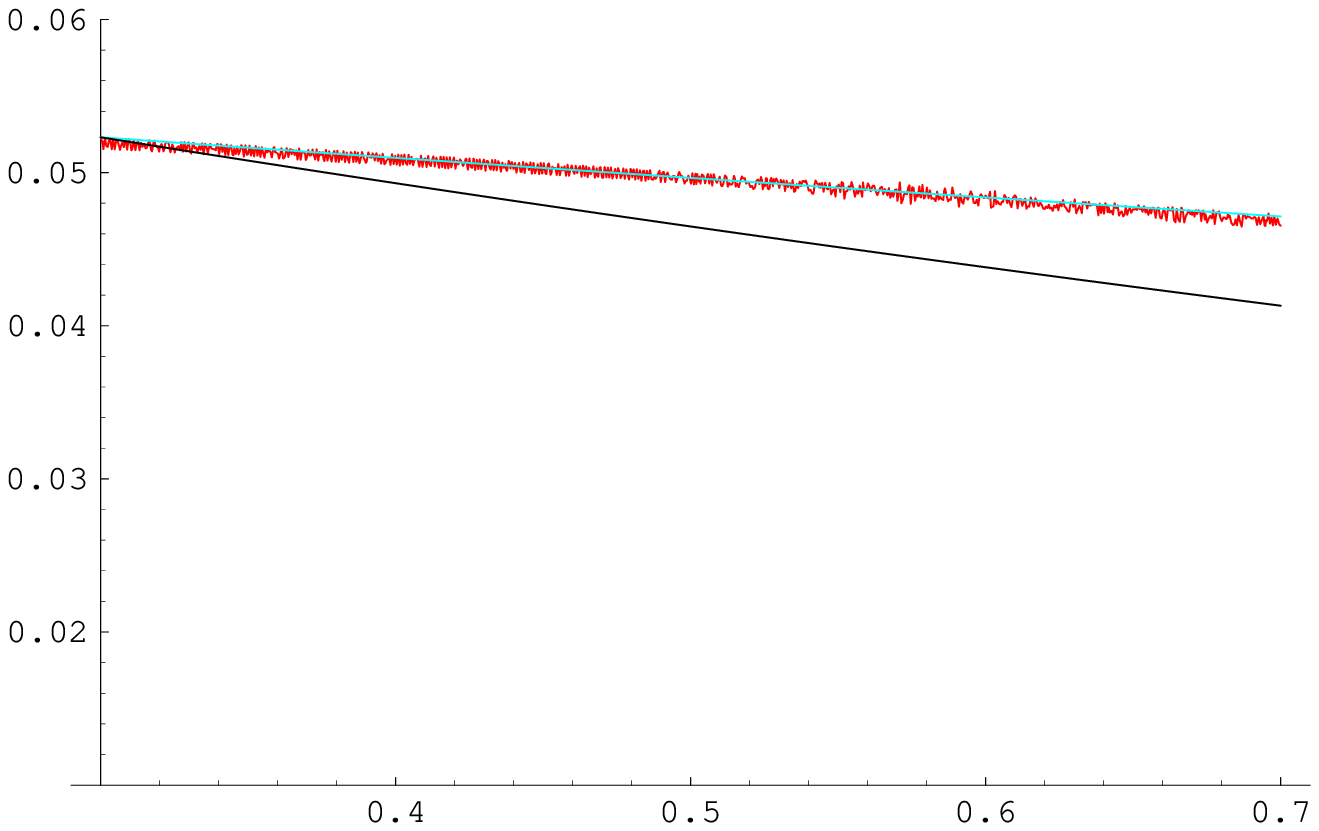}
 \begin{picture}(0,0)
   \put(0,0.3){$t$}
   \put(-6.5,5.1){$v$}
 \end{picture}
\caption{Quark velocity as a function of time from our numerical integration (in red)
compared against (\ref{vhkkkynr}) with the value of $\mu$ deduced in \cite{hkkky} (in
black), and (\ref{vhkkkynr}) with $\mu$ chosen to fit the data (in light blue), for a)
$z_m/z_h=0.2$ and b) $z_m/z_h=0.4$. See text for discussion,}\label{vfig}
\end{figure}

 The most direct way to inquire whether the output of our numerical integration for the
 accelerated quark
 conforms to the the constant-velocity (or late-time) results of  \cite{hkkky,gubser}
 is to compare the corresponding quark trajectories. {} If one assumes the dispersion
 relation (\ref{DRr}), then the equation of motion for a quark subject only to the
 drag force (\ref{mu}) with constant $\mu$ is
 \begin{equation}\label{qeomplasma}
{dv\over dt}=-\mu v(1-v^2)~,
 \end{equation}
 whose solution is \cite{hkkky}
 \begin{equation}\label{vhkkky}
v(t)=\frac{v_{\mbox{\scriptsize
release}}}{\sqrt{v_{\mbox{\scriptsize
release}}^2+(1-v_{\mbox{\scriptsize
release}}^2)e^{2\mu(t-t_{\mbox{\scriptsize release}})}}}~.
 \end{equation}
As we already mentioned, in our numerical results
$v_{\mbox{\scriptsize release}}\ll 1$, so we are only able to test
the non-relativistic version of (\ref{vhkkky}),
\begin{equation}\label{vhkkkynr}
v(t)=v_{\mbox{\scriptsize release}}e^{-\mu(t-t_{\mbox{\scriptsize
release}})}~
\end{equation}
(and, since our integration is limited to small time intervals, we
would in effect see just the linear portion of this function). A
comparison between this analytic prediction and our numerical
results for $t\ge t_{\mbox{\scriptsize release}}$ is given in
Fig.~\ref{vfig}. It is evident from this plot that, in the early
stage of motion covered by our analysis, the quark dissipates
energy at a rate much lower than the late-time result of
\cite{hkkky}. Indeed, for $z_m/z_h=0.2,0.3,0.4$ the asymptotic
friction coefficient is respectively $\mu_{\mbox{\scriptsize
late}}/\pi T=0.25,0.41,0.59$, but our numeric results for $v(t)$
are best approximated by $\mu_{\mbox{\scriptsize early}}/\pi
T=0.08,0.15,0.26$.

We can also attempt to perform the comparison directly at the level of energy loss rates. An
important drawback of working with the numerical solution, however, is that we cannot
achieve a direct splitting of the total energy $E$ of the string, as we did in the
stationary case (as well as in the general case at zero temperature). This means that,
\emph{a priori}, we know neither the correct form of the quark dispersion relation nor the
formula for the rate of energy loss. But, given that their sum remains constant throughout
the evolution, finding a prescription for one of these two quantities would enable us to
compute the other.

The authors of \cite{hkkky} assumed that the energy loss would be
negligible for sufficiently short acceleration intervals (i.e.,
for small $t_{\mbox{\scriptsize release}}$), because the drag
force exerted by the plasma would not have been able to perform a
substantial amount of work. If true, this would allow a direct
empirical determination of the dispersion relation. Unfortunately,
the situation is not so simple, because, as we learned in Section
\ref{qnoplasmasec}, the quark loses energy through radiation even
in the absence of the plasma, and this effect must be taken into
account to establish what fraction of the total string energy $E$
is intrinsically ascribable to the quark.

To attempt to cut this Gordian knot, we should recall, from our
study of the cases where we had analytic control, that the
dispersion relation arises as a surface term, with contributions
from both endpoints of the string. Based on our previous results,
we expect the dispersion relation for the quark in the thermal
medium to take the form
\begin{equation}\label{Edrplasma}
E_{q}(v,F,T)=\frac{\sqrt{\lambda}}{2\pi} \left( \frac{1-z_m^2 v
\Pi}{z_m\sqrt{(1-v^2)(1-z_m^4 \Pi^2)}}-{1\over
z_h}\right)+\cO(z_m^2/z_h^3)~.
\end{equation}
The first term here has been copied from the $T=0$ expression
(\ref{edrf}) (again abbreviating $\Pi\equiv\Pi^z_x$), and is meant
to approximate the contribution from the string endpoint lying on
the $D7$-branes. The second term arises from the endpoint that
reaches the horizon, which we understood above to encode
information about the initial conditions. Knowing that our
starting configuration is static, we can simply read off this
contribution from the second term in the rest energy
(\ref{Mrest}), which we have interpreted as a screening effect.
Including this term is important to ensure that (\ref{Edrplasma})
reduces to the correct result in the case of a quark that is
static and unaccelerated, $E_{q}(0,0,T)=M_{\mbox{\scriptsize
rest}}$.

Clearly equation (\ref{Edrplasma}) is just an approximation, because the first term should
receive thermal corrections. In particular, it is natural to expect the factor of $1-z_m^4
\Pi^2$ in the denominator to be replaced by $h(z_m)-z_m^4\Pi^2$, so that, just like in the
$T=0$ case, the quark energy diverges at the critical value of the electric field, which now
corresponds to $\Pi=\sqrt{h(z_m)}/z_m^2$ \cite{ctqhat}. More generally, since the background
metric (\ref{metric}) knows that $T>0$ only through the factor $h<1$, any corrections to the
$z=z_m$ surface term due to the presence of the plasma should be of order $1\!-\!h$ or $h'$,
and are consequently small in the heavy quark regime $z_m/z_h\ll 1$. For the
phenomenologically interesting values $z_m/z_h=0.2,0.3,0.4$, the error in the dispersion
relation (\ref{Edrplasma}) is estimated to be $\sim 4-16\%$ (much larger than our $\sim 1\%$
numerical error).

Notice that (\ref{Edrplasma}) without any corrections should give the exact dispersion
relation in the infinite-mass limit $z_m\to 0$ that has been the focus of many AdS/CFT
investigations of energy loss (e.g., \cite{gubser,gluonicprofile}). In this limit one is
left with
\begin{equation}\label{Eheaviest}
E_q(v,T)=\frac{\sqrt{\lambda}}{2\pi}
\left(\frac{\gamma}{z_m}-{1\over z_h}\right)~,
\end{equation}
i.e., the $F$-dependence drops out and, just like in the $T=0$
context at the end of Section \ref{qnoplasmasec}, one recovers
pointlike behavior. We can see that, as expected, the first term
in (\ref{Eheaviest}) agrees with the $z_m\to 0$ limit of the
second line of the stationary expression (\ref{Emikhgubser}), but
the second term in (\ref{Eheaviest}) disagrees with  the limit
 of the third line of (\ref{Emikhgubser}), due to the different initial
conditions.

Expression (\ref{Edrplasma}) as a whole looks superficially rather different from the
dispersion relation (\ref{DRr}) proposed in \cite{hkkky}.  In particular, (\ref{DRr})
evidently cannot reproduce the $F$- (or $\Pi$-)dependence seen in (\ref{Edrplasma}), whose
presence is supported by the zero-temperature results of the previous subsection. It is
important to remember, however, that such dependence indeed would not have been visible in
the quasi-normal mode analysis used to derive (\ref{DRnr}) or in the quark-antiquark
evolution that gave part of the support for (\ref{DRr}), because in those calculations the
external force was taken to vanish.

Setting $\Pi=0$, (\ref{Edrplasma}) reduces to
$E_q=(\sqrt{\lambda}/2\pi)(\gamma/z_m-1/z_h)+\cO(z_m^2/z_h^3)$, while (\ref{DRr}) translates
(via (\ref{Mrest}) and (\ref{mkinmrest}))  into
$E_q=(\sqrt{\lambda}/2\pi)(\gamma/z_m-\gamma/z_h)+\cO(z_m/z_h^2)$. The two expressions
differ in the form of the second, $z_m$-independent term, which we have understood to encode
the initial conditions for the gauge system. Given that the initial conditions for all
situations considered in \cite{hkkky} differ from our current setup, there is no reason why
we should expect the corresponding terms to agree. Beyond this, there
 also appears to be a discrepancy in the order of magnitude for the corrections to the
two expressions. We remain puzzled by this apparent mismatch, because as explained below
(\ref{Edrplasma}), we do not see how the zero-temperature relation could receive corrections
higher than order $1-h$ or $h'$.

We interpret the $v$-dependence seen in the surface contribution
at $z_h$ implied by (\ref{DRr}) to be a reflection of the fact
that, for the quasi-normal mode considered in \cite{hkkky}, the
string endpoint at the horizon is indeed moving, unlike what
happens in our case.\footnote{For the quark-antiquark
configurations that were analyzed in \cite{hkkky} and will be
reviewed and generalized in Section \ref{qqbarsec}, there is no
endpoint at the horizon, so we should expect to get only a surface
contribution from $z=z_m$. Independently of that, the evolution in
that case is not sensitive to the value of $M_{\mbox{\scriptsize
kin}}$, which appears on both sides the equation of motion
(\ref{hkkkyforce}), and consequently drops out, leading to
(\ref{qeomplasma}). } In other words, we find that the `kinetic
mass of the quark,' defined as the coefficient of the $v^2/2$ term
in $E_q$, is sensitive to the initial conditions. In particular,
for a quark that is initially static, we have obtained a
dispersion relation of the same relativistic form as (\ref{DRr}),
\begin{equation}\label{ourdr}
E_q=M_{\mbox{\scriptsize rest}}+M_{\mbox{\scriptsize
kin}}(\gamma-1)+\cO\left(m\left({\sqrt{\lambda}T\over 2m}\right)^3\right)~,
\end{equation}
 but with
$M_{\mbox{\scriptsize kin}}=\sqrt{\lambda}/2\pi z_m=m+ \cO(m({\sqrt{\lambda}T/ 2m})^4)$.

We are finally in position to address energy loss. For the type of process we consider, the
total energy of the string at time $t$ is given by
$$E(t)=M_{\mbox{\scriptsize rest}}+E_{input}(t)~,$$
where the second term is the work performed by the external force,
which can be computed by integrating the energy current
$-\Pi^z_t(t,z_m)$ shown in (\ref{momenta}). The energy lost by the
heavy quark can then be obtained as
\begin{equation}\label{Eloss}
E_{\mbox{\scriptsize lost}}(t)\equiv
E(t)-E_{q}(v(t),F(t),T)=E_{input}(t)-E_{\mbox{\scriptsize
kin}}(t)~,
\end{equation}
where the last term is the kinetic energy
\begin{equation}\label{Ek}
E_{\mbox{\scriptsize kin}}\equiv E_{q}(v,F,T)-M_{\mbox{\scriptsize
rest}}=\frac{\sqrt{\lambda}}{2\pi}\frac{1}{z_m}\left(
\frac{1-z^2_mv\Pi}{\sqrt{(1-v^2)(1-z^4_m\Pi^2)}}-1\right)~.
\end{equation}
Of course, in the $t\ge t_{\mbox{\scriptsize release}}$ ($\Pi=0$)
stage, with the small velocities that we achieve we are only
sensitive to the non-relativistic (quadratic in $v$) terms in this
equation.

\begin{figure}[tbph]
\vspace*{0.5cm}
 \setlength{\unitlength}{1cm}
\includegraphics[width=7cm,height=5cm]{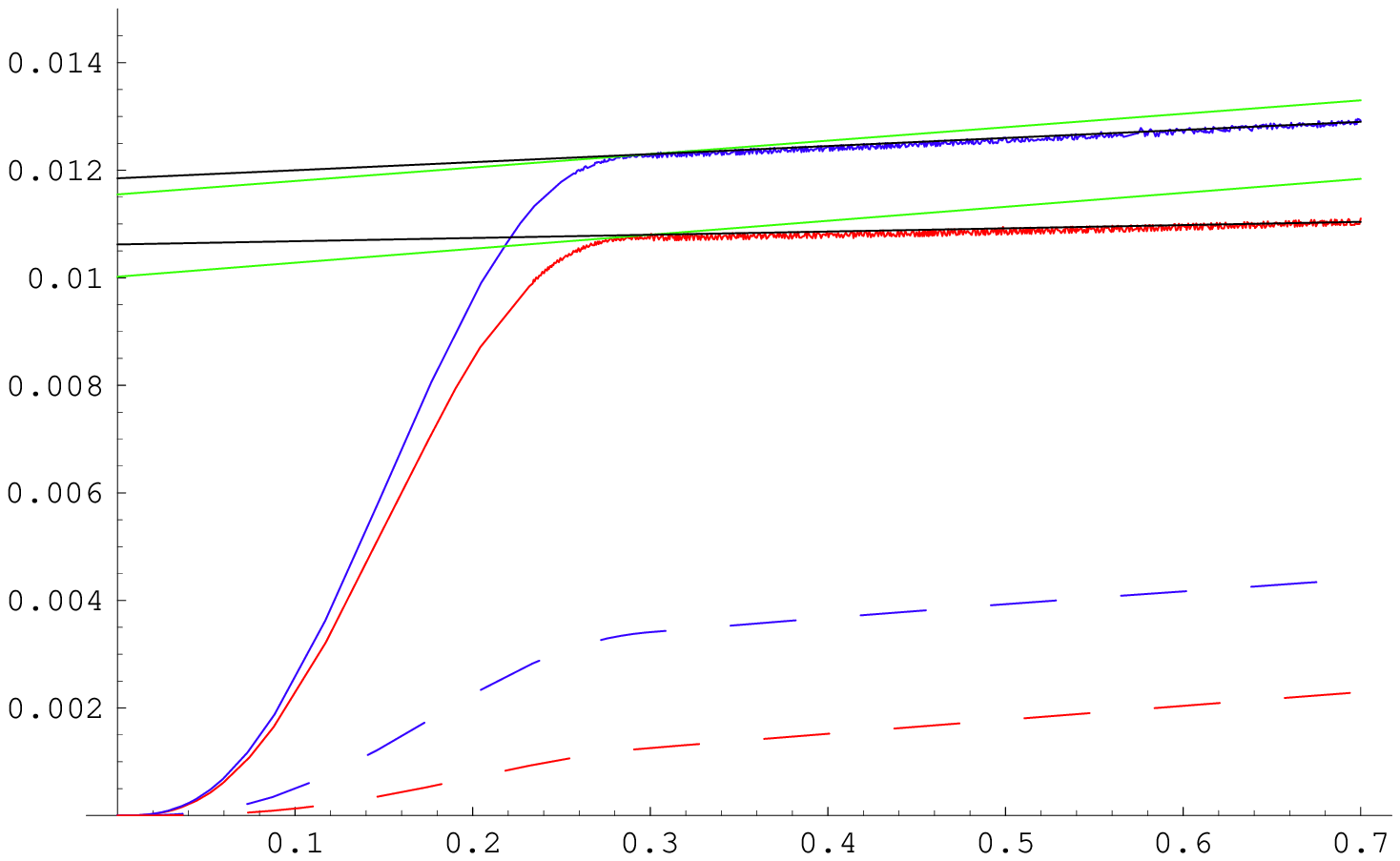}
 \begin{picture}(0,0)
   \put(0,0.3){$t$}
   \put(-6.5,5.1){$E$}
 \end{picture}
\includegraphics[width=7cm,height=5cm]{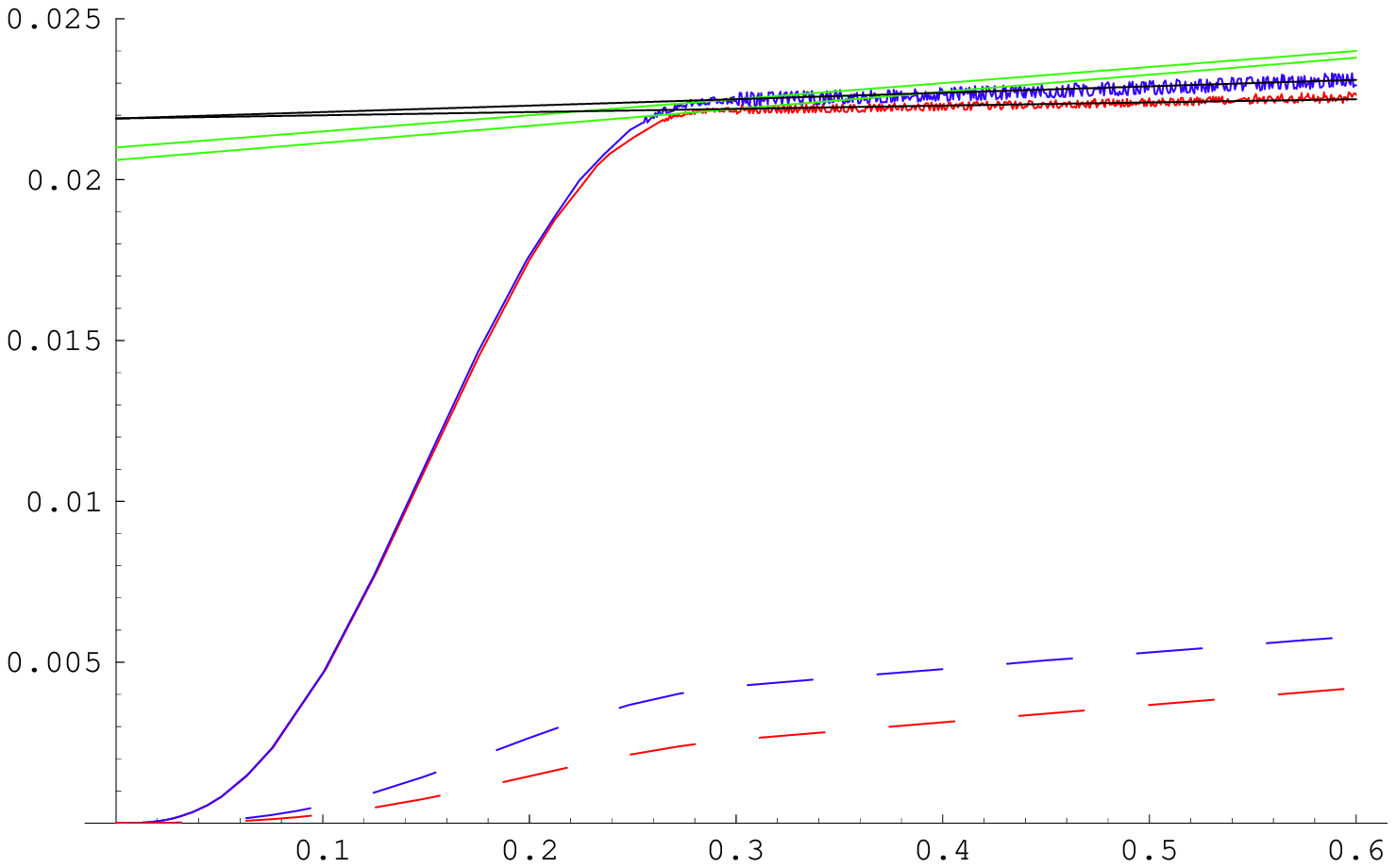}
 \begin{picture}(0,0)
   \put(0,0.3){$t$}
   \put(-6.5,5.1){$E$}
 \end{picture}
\caption{a) Accumulated energy loss (in units of
$\sqrt{\lambda}T/2$) as a function of time (in units of $1/\pi T$)
for a) $z_m=0.2$ (red) and $z_m=0.4$ (blue) with
$v_{\mbox{\scriptsize release}}=0.051$, and b) $z_m=0.2$ (red) and
$z_m=0.3$ (blue) with $v_{\mbox{\scriptsize release}}=0.073$. For
comparison, the dashed curves of the same colors give the energy
loss that would follow from the stationary rate
(\ref{EPlossgubser}) obtained in \cite{hkkky,gubser}. The green
lines represent the
 rate  (\ref{EPlossgubser})
evaluated with $v=v_{\mbox{\scriptsize release}}$, which can be
contrasted against the slope of the numerical curves, shown in
black.}\label{diffmass}
\end{figure}

In Fig.~\ref{diffmass}, we compare the energy loss at for
different masses but equal velocity at the time
$t_{\mbox{\scriptsize release}}$ when the external force is set to
zero. The curves have essentially constant slope for
$t>t_{\mbox{\scriptsize release}}$, meaning that the rate of
energy loss is nearly constant in the limited time window that we
have access to. The lines in the figure contrast the instantaneous
rate of energy loss in this interval, obtained by a numerical fit
and denoted henceforward by $(\p_t E_q)_n$, against the
corresponding stationary result of \cite{hkkky,gubser}, which we
will denote by $(\p_t E_q)_s$. For $z_m=0.4$, $(\p_t E_q)_n$ is
close to two times bigger than $(\p_t E_q)_s$, and as the value of
$z_m$ decreases (corresponding to heavier quarks), the difference
between the two rates grows bigger. In particular, for $z_m=0.2$,
$(\p_t E_q)_s$ is at least three times bigger than $(\p_t E_q)_n$.

\begin{figure}[tbph]
\vspace*{0.5cm}
 \setlength{\unitlength}{1cm}
\includegraphics[width=7cm,height=5cm]{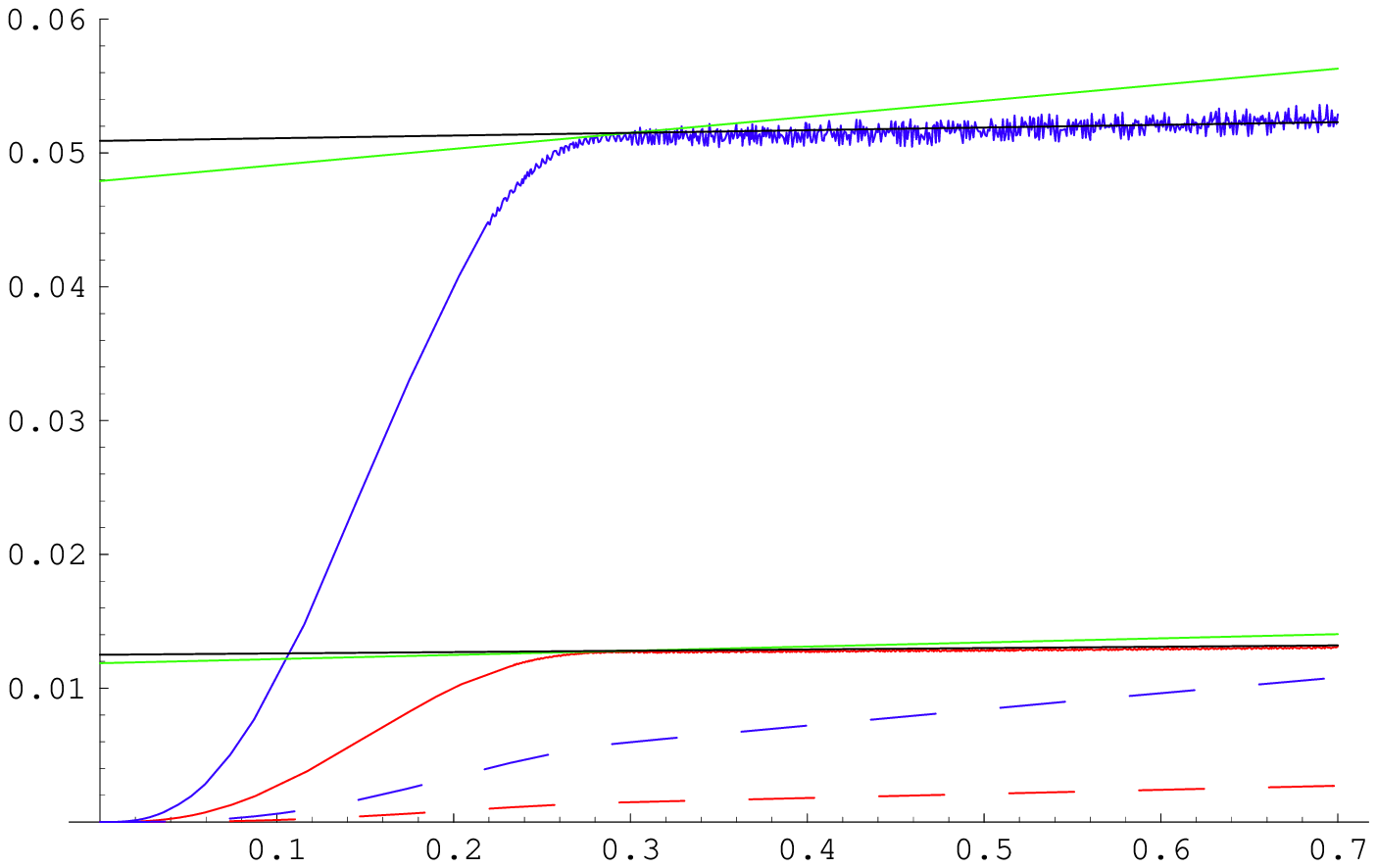}
 \begin{picture}(0,0)
   \put(0,0.3){$t$}
   \put(-6.5,5.1){$E$}
 \end{picture}
\includegraphics[width=7cm,height=5cm]{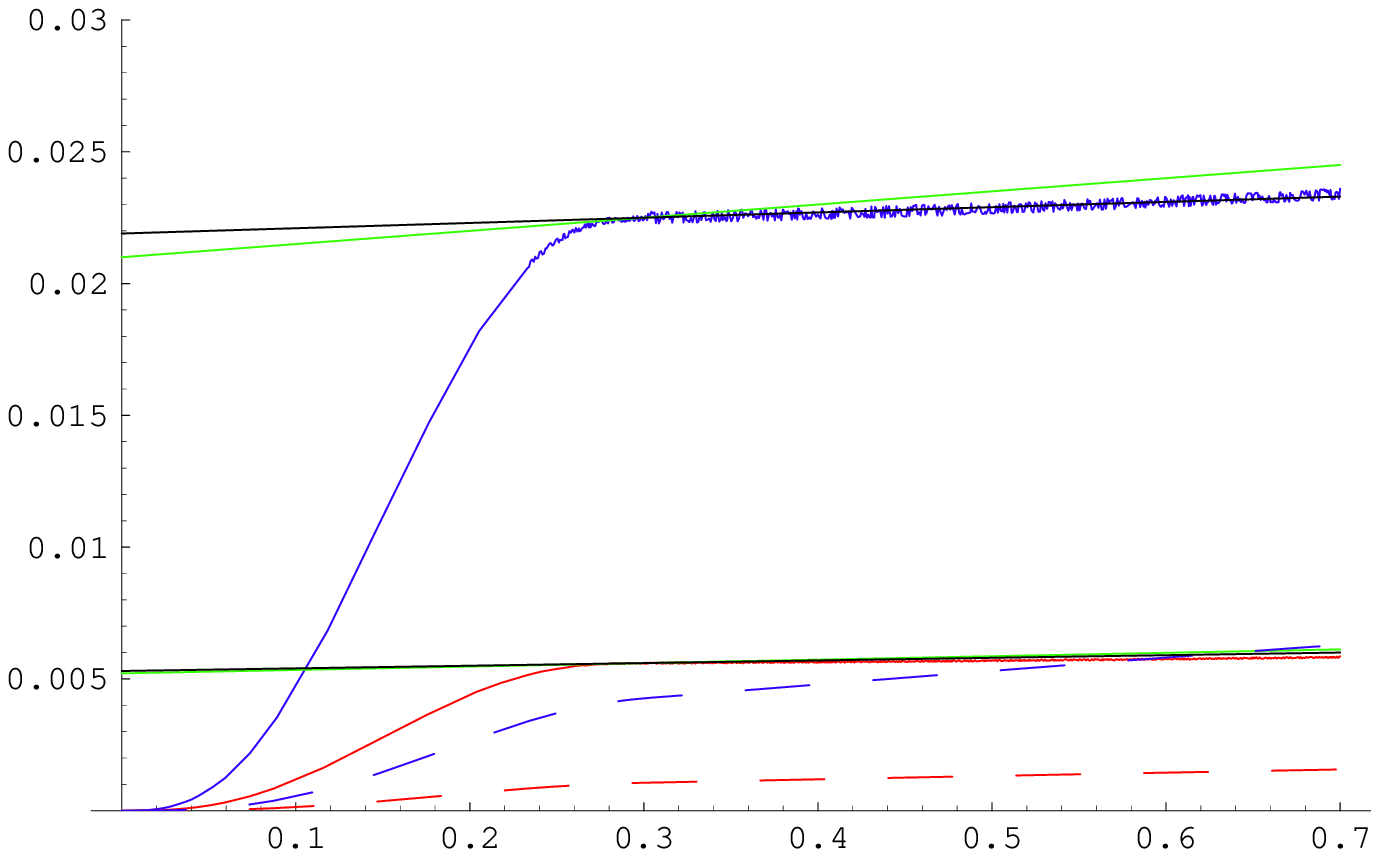}
 \begin{picture}(0,0)
   \put(0,0.3){$t$}
   \put(-6.5,5.1){$E$}
 \end{picture}
\caption{a) Accumulated energy loss (in units of
$\sqrt{\lambda}T/2$) as a function of time (in units of $1/\pi T$)
for a) $z_m=0.2$ with $v=0.056$ (red) and $v=0.111$ (blue) and b)
$z_m=0.3$ with $v=0.036$ (red) and $v=0.073$ (blue). The dashed
curves of the same colors give the  energy loss obtained with the
stationary rate (\ref{EPlossgubser}). The green lines represent
this rate evaluated at $t=t_{\mbox{\scriptsize release}}$ with
velocity $v$, which is to be contrasted against the slope of the
numerical curves, shown in black.} \label{equalmass}
\end{figure}

A similar comparison is shown is Fig.~\ref{equalmass}, but with a fixed mass value and
different release velocities. In all cases, we have found again that the rate of energy loss
(\ref{EPlossgubser}), valid in the stationary regime, is above our numerical result. The
discrepancy observed is significantly larger than our estimated margin of error. Our results
therefore provide clear evidence that, in the mass range of primary phenomenological
interest, there exist conditions under which the rate of energy loss for a heavy quark that
moves only under the influence of the plasma can be substantially smaller than the rate
obtained in \cite{hkkky} for the steady-state or late-time configuration. We will return to
this point in Section \ref{potentialtransitionsec}.

\begin{figure}[tbph]
\vspace*{0.5cm} \setlength{\unitlength}{1cm}
\includegraphics[width=7cm,height=5cm]{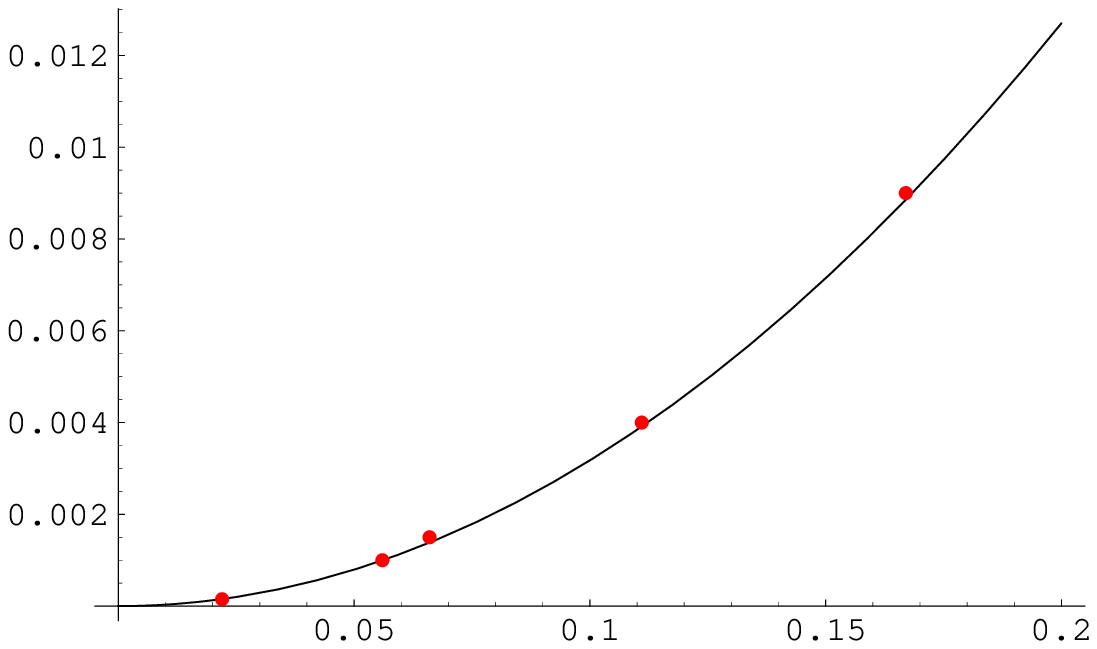}
 \begin{picture}(0,0)
   \put(0,0.3){$v$}
   \put(-6.5,5.1){$(\p_t E_q)_n$}
\end{picture}
\includegraphics[width=7cm,height=5cm]{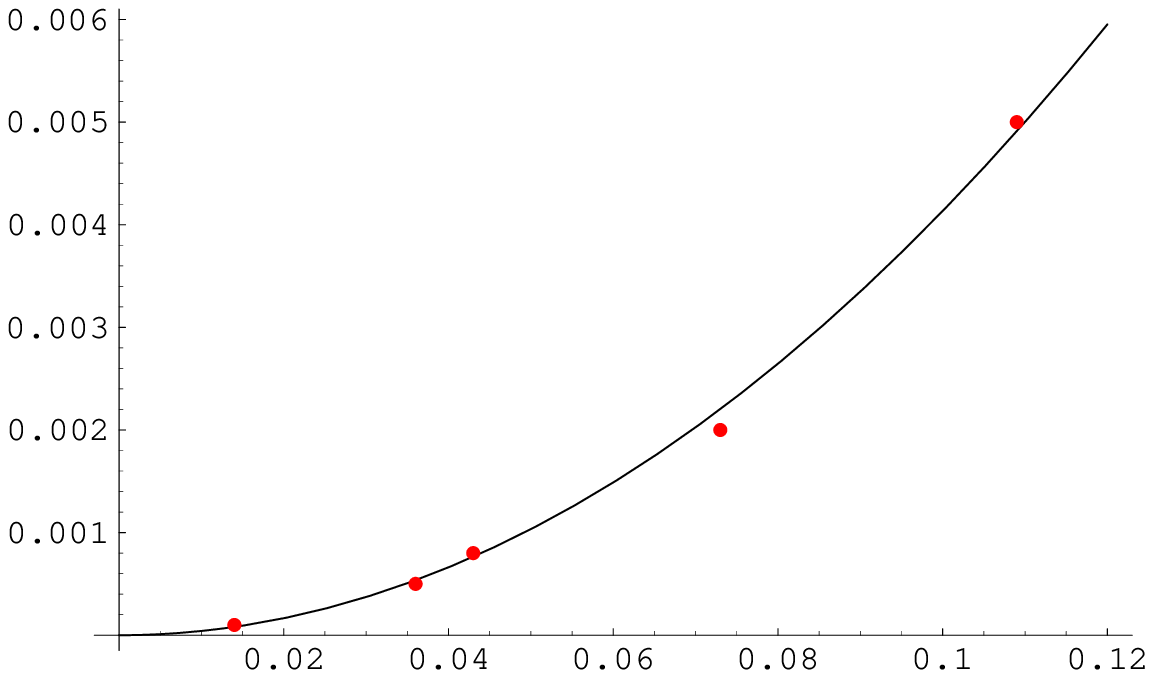}
 \begin{picture}(0,0)
   \put(0,0.3){$v$}
   \put(-6.5,5.1){$(\p_t E_q)_n$}
 \end{picture}
\caption{ Rate of energy loss (points, in units of
$\sqrt{\lambda}\pi T^2/2$) as a function of velocity for a)
$z_m=0.2$ and b) $z_m=0.3$, together with the corresponding
quadratic fits $(\p_t E_q)_n(0.2,v)=0.31v^2$ and $(\p_t
E_q)_n(0.3,v)=0.41v^2$. } \label{parabola}
\end{figure}

As one would expect, the numerical rate of energy loss depends on
$z_m$ and $v$.  As shown in Fig.~\ref{parabola}, for fixed mass
values, it varies quadratically with the velocity of the quark. As
we had mentioned before, for the small velocities that we can
attain, we must deal only with the non-relativistic approximation
of the dispersion relation (\ref{Ek}). This implies that the
accumulated energy loss (\ref{Eloss}) varies quadratically with
the quark velocity $v$, and so the constant slope that we are
reading off of the $t_{\mbox{\scriptsize release}}\le t\le
t_{\mbox{\scriptsize max}}$ portion of the numerical curves in
Figs.~\ref{diffmass} and \ref{equalmass} should be proportional to
$v\Delta v$, where $\Delta v$ denotes the small change in velocity
in the given time interval. The parabolic behavior seen in
Fig.~\ref{parabola}, then, tells us that $\Delta v\propto v$. In
other words, in the leading non-relativistic version of
(\ref{qeomplasma}), the friction coefficient $\mu$ is  independent
of the quark velocity, as one would expect.

Unfortunately, due to limitations with the numerical integration, we have not been able to
characterize the dependence of the dissipation rate on the mass of the quark, beyond the
statement that $(\p_t E_q)_n$ increases roughly linearly with increasing $z_m/z_h$
(decreasing $m$). In particular, above $z_m/z_h\sim 0.75$,
 the numeric rate becomes indistinguishable (within our margin of error) from the stationary
result (\ref{EPlossgubser}).

\begin{figure}[tbph]
\vspace*{0.5cm}
 \setlength{\unitlength}{1cm}
\includegraphics[width=7cm,height=5cm]{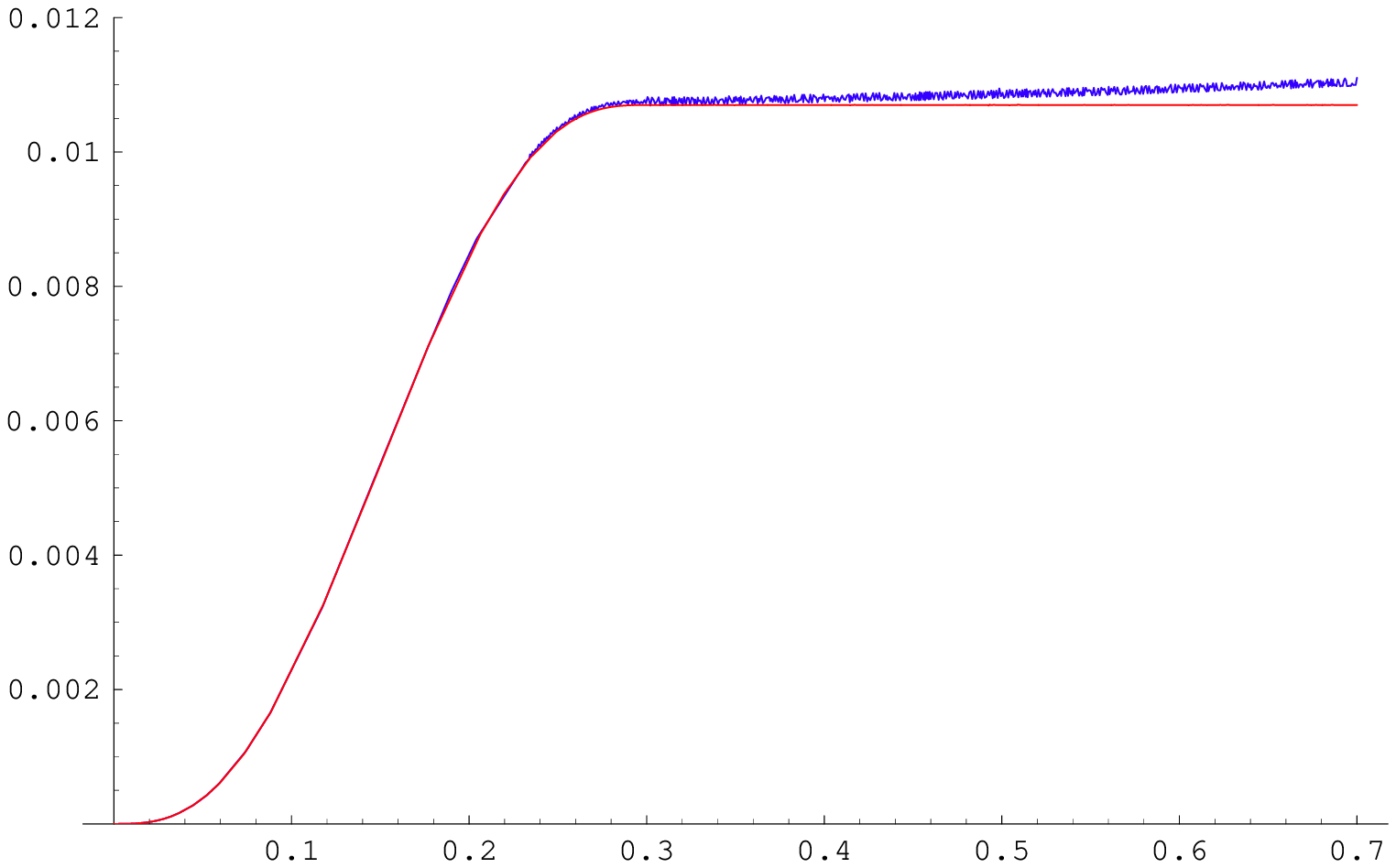}
 \begin{picture}(0,0)
   \put(0,0.3){$t$}
   \put(-6.5,5.1){$E$}
 \end{picture}
\includegraphics[width=7cm,height=5cm]{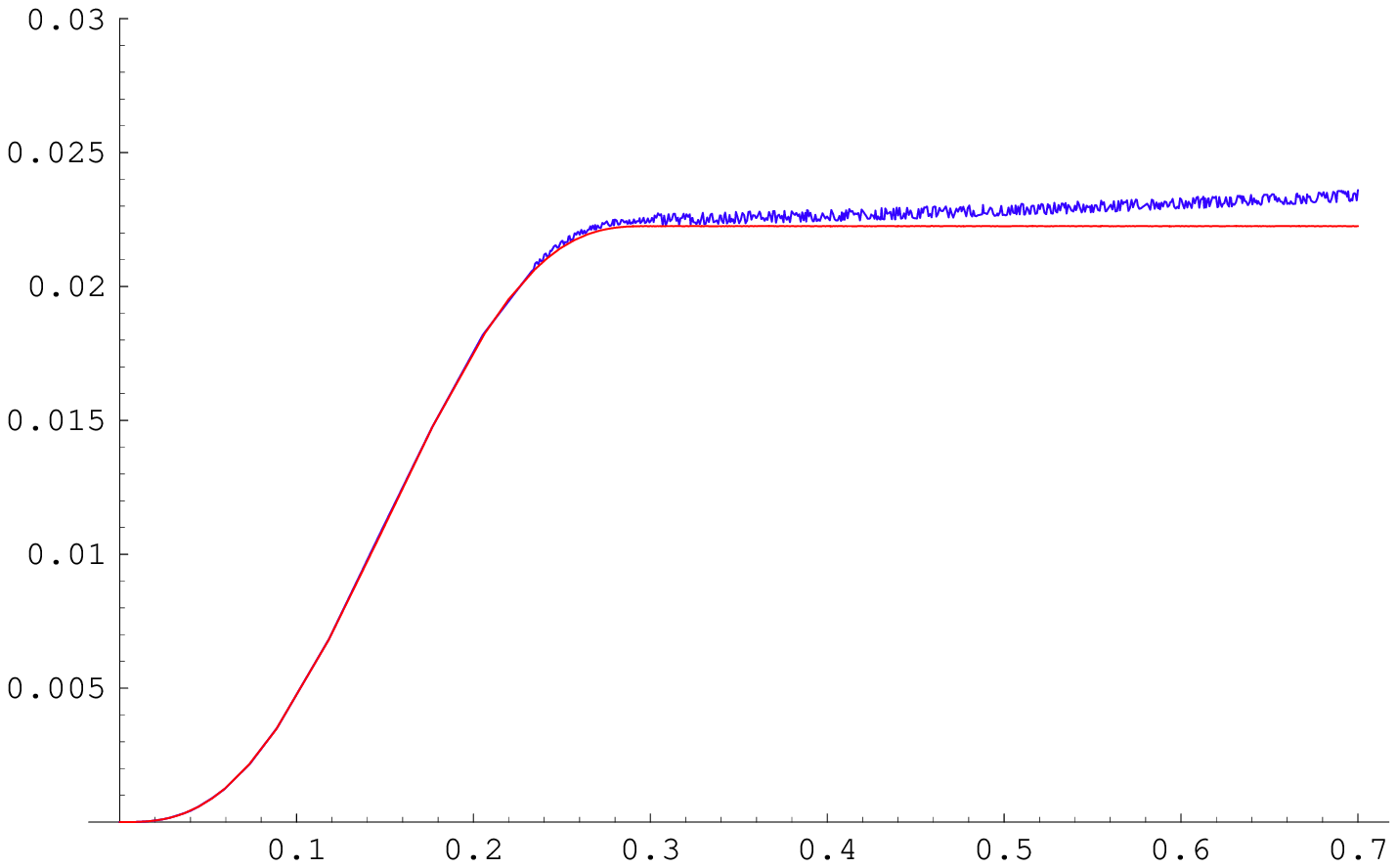}
 \begin{picture}(0,0)
   \put(0,0.3){$t$}
   \put(-6.5,5.1){$E$}
 \end{picture}
\caption{Comparison between the accumulated energy loss (in units
of $\sqrt{\lambda}T/2$) versus time (in units of $1/\pi T$) using
(\ref{Eloss}) in blue and the modified Lienard formula
 (\ref{emikhf}) in red for a) $z_m=0.2$  and b)  $z_m=0.3$.
 See text for discussion.} \label{LW}
\end{figure}

Let us now consider the initial stage $0\le t<t_{\mbox{\scriptsize
release}}$. {}From the beginning portion of the curves in
Figs.~\ref{diffmass} and \ref{equalmass}, we can see that the
situation when the quark is subjected to an external force is
opposite to what we described above for unforced motion: the rate
at which energy is dissipated can be substantially \emph{larger}
than (\ref{EPlossgubser}), suggesting that in this segment of the
quark trajectory the mechanism of energy loss is qualitatively
different.  In particular, the energy lost in this region is by no
means negligible, which is precisely the obstacle that prevents us
from directly inferring the quark's dispersion relation from our
numerical results, as envisioned in \cite{hkkky}. We had
anticipated this already in the discussion above, based on our
results for the $T=0$ case in Section \ref{qnoplasmasec}.

When, using the numerical data of the evolution within the plasma, we compare as in Fig.~\ref{LW} the energy lost by the heavy quark, Eq.~(\ref{Eloss}), against the modified Lienard
formula (\ref{emikhf}), we find that for $t<t_{\mbox{\scriptsize release}}$ the two curves are virtually
indistinguishable from one another when $z_m/z_h=0.2$, $0.3$, and, to a lesser extent,
$0.4$. In fact, this approximate agreement continues to hold (albeit somewhat
reduced)
even if in (\ref{emikhf}) we plug in the data of the quark trajectory at zero temperature. In other words, for quark masses in the phenomenologically interesting range, the
quark behaves initially as if there were no plasma, and loses energy through radiation, at a
rate equal to the modified Lienard formula given by the first term in (\ref{emikhf}). This
is of course as one would expect from the gauge theory perspective, for a heavy quark should
indeed be insensitive to the plasma for very early times.

On the gravity side, the issue is that, for these relatively low values of $z_m/z_h$, the
factor of $h$ is so close to unity that the modifications induced by the black hole horizon
on $E_{\mbox{\scriptsize input}}$ and $E_{\mbox{\scriptsize lost}}$ only become appreciable
when a sufficiently large time has elapsed.
We should emphasize that it is \emph{not} possible to reproduce the numerical results for
$v(t)$ and $E_{\mbox{\scriptsize input}}(t)=E_{\mbox{\scriptsize
kin}}(t)+E_{\mbox{\scriptsize lost}}(t)$ in this initial stage using the dispersion relation
(\ref{DRr}) and dissipation rate (\ref{EPlossgubser}) derived in \cite{hkkky} (the values of
the latter are indicated by the dotted curves in Figs.~\ref{diffmass}-\ref{equalmass}).

Beyond $t_{\mbox{\scriptsize release}}$ the curves in Fig.~\ref{LW}  separate, showing the
influence of the hot medium, although somewhat diminished compared to the stationary result
(\ref{EPlossgubser}), as we know from Figs.~\ref{diffmass}-\ref{equalmass}. {}Since it was
shown in \cite{hkkky} that (\ref{EPlossgubser}) will hold at asymptotically late times, we
expect the rate of energy loss seen in Figs.~\ref{vfig}-\ref{LW} to increase as the system
evolves further. Regrettably, with our very limited integration time we are not able to
track the evolution far enough to locate the characteristic transition time to the
asymptotic behavior. We will return to this issue from a different perspective in Section
\ref{qqbartransitionsec}.

\subsection{Late-time behavior and worldsheet black hole}
\label{bhplasmasec}

 It is
interesting to visualize the evolution of the system beyond the
limited time interval covered by our numerical data, in parallel
with our discussion for the zero-temperature case in Section
\ref{bhsec}. The quark, initially static, and accelerated by an
external force $F(t)$ between $t=t_{\mbox{\scriptsize grab}}$
(originally $t=0$) and $t=t_{\mbox{\scriptsize release}}$, will
thereafter decelerate under the influence of the plasma,
approaching rest at some location $x_{\infty}$ as $t\to\infty$. In
the dual gravity description, this means that the final string
embedding, just like the initial ($t\le t_{\mbox{\scriptsize
grab}}$) one, must include a static vertical segment extending all
the way from the D7-branes at $z=z_m$ to the black hole horizon at
$z=z_h$, to represent the quark at rest.

\begin{figure}[htbp]
\begin{center}
\vspace*{0.2cm}
 \setlength{\unitlength}{1cm}
\includegraphics[width=11cm,height=6cm]{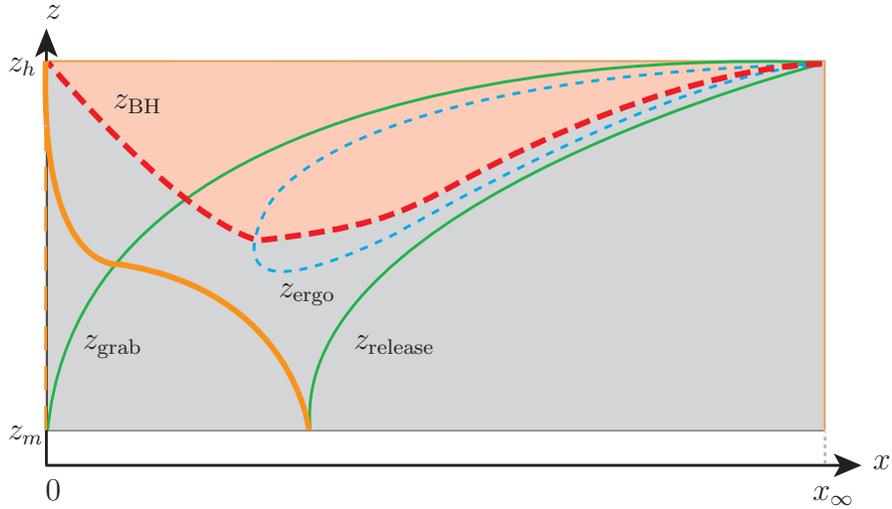}
 \begin{picture}(0,0)
   \put(0,0.1){$x$}
   \put(-0.8,-0.3){$x_{\infty}$}
   \put(-11,-0.3){$0$}
   \put(-11,6.1){$z$}
   \put(-11.5,5.4){$z_h$}
   \put(-11.5,0.5){$z_m$}
   \put(-6.9,1.7){$z_{\mbox{\scriptsize release}}$}
   \put(-10.5,1.7){$z_{\mbox{\scriptsize grab}}$}
   \put(-7.9,2.4){$z_{\mbox{\scriptsize ergo}}$}
   \put(-10.1,4.9){$z_{\mbox{\scriptsize BH}}$}
  \end{picture}
 \end{center}
\caption{Schematic illustration of the string worldsheet (the rectangle shaded
 in gray), projected for convenience
 onto the spacetime $x$-$z$ plane. To aid the visualization of the evolution,
 snapshots of the string are given for three different instants: at (any time up to) $t=t_{\mbox{\scriptsize grab}}$ (thin dotted orange), when the string is at rest and vertical at $x=0$; at $t=t_{\mbox{\scriptsize release}}$ (thick solid orange), when it has already
 been partially deformed by the application of the external force $F(t)$; and at $t\to\infty$ (thin solid orange), when it has adopted a $\neg$ shape, and its vertical segment has come to rest at $x=x_{\infty}$. The diagram additionally shows the upward null (fixed $t_{\mbox{\scriptsize tret}}$) curves
 $z_{\mbox{\scriptsize grab}}$ and $z_{\mbox{\scriptsize release}}$ (solid green),
 the stationary limit curve $z_{\mbox{\scriptsize ergo}}$ (dotted blue), and
  the event horizon $z_{\mbox{\scriptsize BH}}$ (thick dotted red) above
  which lies the worldsheet black hole (shaded light red). See text for discussion.}\label{bhplasmafig}
\end{figure}

As time progresses, the lower ($z=z_m$) string endpoint traces out the trajectory of the
quark, moving from $x=0$ to $x=x_{\infty}$. The upper ($z=z_h$) endpoint, on the other hand,
remains at $x=0$ for all finite times, because the wavefront generated on the string by the
acceleration of the bottom tip will reach the spacetime horizon only at $t\to\infty$. Given
these boundary conditions, on the $x$-$z$ plane the string will clearly evolve from purely
vertical to $\neg$-shaped, with a horizontal segment at $z=z_h$, extending from $x=0$ to
$x=x_{\infty}$. All of the energy (and momentum) lost by the quark throughout its evolution
ends up in this top portion of the string, which encodes the IR region of the gauge theory.
The region of no escape is bounded as in Section \ref{bhsec} by a worldsheet horizon
 that can be identified by standing on the $(x_{\infty},z_h)$ corner of the string
and projecting back along the downward-pointing light half-cone
$\dot{z}^{(-)}_{\mbox{\scriptsize null}}(t)$, now defined by
\begin{equation}\label{znulldotplasma}
\dot{z}_{\mbox{\scriptsize null}}^{(\pm)}(t)=\frac{
X'\dot{X}\pm\sqrt{1+h
X^{'2}-\frac{{\dot{X}}^2}{h}}}{X^{'2}+{1\over h}}~.
\end{equation}
 The resulting curve $z_{\mbox{\scriptsize BH}}(t)$ descends from $z\to\infty$ before $t=t_{\mbox{\scriptsize grab}}$, reaches a
 minimum value of the radial
coordinate, and then moves up again, finally approaching the `corner' as $t\to\infty$. The
upward segment lies fully within the maximally-disturbed region of the worldsheet, located
between the \emph{upward} null curves $z_{\mbox{\scriptsize grab}}(t)$ and
$z_{\mbox{\scriptsize release}}(t)$ obtained by integrating (\ref{znulldotplasma}) with the
upper choice of sign and with initial condition $z_{\mbox{\scriptsize
grab}}(t_{\mbox{\scriptsize grab}})=z_m$ or $z_{\mbox{\scriptsize
release}}(t_{\mbox{\scriptsize release}})=z_m$, respectively. A lower bound on the location
of the upward portion of $z_{\mbox{\scriptsize BH}}(t)$ is given by the (upward segment of
the) stationary limit curve $z_{\mbox{\scriptsize ergo}}(t)$, defined as the locus where
$g_{tt}=0$, i.e., $\dot{X}^2=h$. The situation is summarized in Fig.~\ref{bhplasmafig}.

If, instead of releasing the quark, we pull it with constant velocity $v$ for an arbitrarily
long period of time (i.e., if $t_{\mbox{\scriptsize release}}\to\infty$), then we approach
the steady-state configuration of \cite{hkkky,gubser}, and in so doing stabilize the
worldsheet horizon (as well as the stationary-limit curve) at $z=z_v\equiv (1-v^2)^{1/4}$,
just like in \cite{ctqhat,gubserqhat} (and \cite{argyres2}).

The behavior of the string long after $t_{\mbox{\scriptsize release}}$ was determined
quantitatively in \cite{hkkky}, and is of the form
$X(z,t)=x_{\infty}-A(z)e^{-\mu_{\mbox{\scriptsize late}}t}$, with the friction coefficient
$\mu_{\mbox{\scriptsize late}}$ given by the lowest quasi-normal frequency. It was shown
there that the imposition of a purely ingoing boundary condition near the spacetime horizon
$z=z_h$ forces the string to deviate from the vertical by a divergent amount
($A(z)\to\infty$ as $z\to z_h$). This is consistent with the development of a `corner', just
as we have argued above for the case where the quark is initially static.

Just as we found at $T=0$, we see here that energy dissipation seems to be irrevocably tied
to the appearance of a worldsheet black hole. Since our numerical results show that the
initial energy loss is controlled by a friction coefficient $\mu_{\mbox{\scriptsize
early}}<\mu_{\mbox{\scriptsize late}}$, it would appear like in the unforced case the
asymptotic rate sets in only when the string is sufficiently close to its final $\neg$
shape. It would be interesting to establish a more detailed connection between the
instantaneous rate of energy dissipation from the quark and the rate at which energy crosses
the worldsheet horizon, generalizing the results of \cite{ctqhat,gubserqhat} for the
stationary configuration.


\section{Quark-Antiquark Evolution}
\label{qqbarsec}

A string with both of its endpoints on the D7-branes describes a quark-antiquark pair. The
situation that is closest to modeling the dual process of primary phenomenological
relevance, where a heavy quark and antiquark are created within the plasma at time $t=0$ and
then separate from one another, is such that the string endpoints start out with coincident
positions but different velocities. In a first exploration of this system, it is interesting
enough to consider the simple case where  the string endpoints are taken to separate
back-to-back with the same initial speed $v_0$, meaning that the pair's center of mass frame
coincides with the plasma rest frame. We will let $x$ denote the direction of motion.

\subsection{Review of earlier results}
\label{qqbarreviewsec}

A numerical study of this problem was carried out in \cite{hkkky}. It was found that the
quark and antiquark trajectories can be more efficiently followed to later times if instead
of describing the string embedding $X^{\mu}(\tau,\sigma)$ in the obvious static gauge
$\tau=t,\sigma=z$, one astutely chooses worldsheet coordinates for which the constant $\tau$
slices manage to reach larger values of $x$ near $z=z_m$ while staying away from the horizon
at $z=z_h$. This is most easily implemented by working with the Polyakov (rather than
Nambu-Goto) action
\begin{equation}\label{polyakov}
S_P=-{1\over 4\pi\ap}\int_{-\infty}^{\infty} d\tau
\int_0^{\pi}d\sigma\,\sqrt{-g}g^{ab}G_{\mu\nu}\p_a X^{\mu}\p_b X^{\nu}\equiv~{R^2\over
2\pi\ap}\int d^2\sigma\,\cL_P,
\end{equation}
with $g_{ab}$ the intrinsic metric on the string worldsheet, and making the non-standard
gauge choice $g_{\tau\tau}=-s$, $g_{\sigma\sigma}=1/s$, $g_{\sigma\tau}=0$. The `stretching
factor' $s=s(\sigma,\tau)$ implicitly defines our choice of worldsheet coordinates, and is
meant to be adjusted by hand to keep the numerical integration away from the horizon for the
longest possible time. For a given trajectory, it is the function $z(\tau,\sigma)$ that
determines the proximity to the horizon, so this goal is achieved by setting
$s=s(z(\tau,\sigma))$ (at the level of the equations of motion). Following \cite{hkkky} we
will use $s\propto (z_h-z)^p$ in the examples below.

In this setting, the evolution of the string is controlled by the equations of motion
\begin{eqnarray}\label{eom}
\p_{\tau}\left(h\dot{t}\over s z^2\right)-\p_{\sigma}\left(s h t'\over z^2\right)&=&0~,\\
\p_{\tau}\left(\dot{x}\over s z^2\right)-\p_{\sigma}\left(s  x'\over z^2\right)&=&0~,\nonumber\\
\p_{\tau}\left(\dot{z}\over s h z^2\right)-\p_{\sigma}\left(s z'\over h z^2\right)&=&{1\over
2s}\left[(\dot{z}^2-s^2 z'^2)\p_z\left(1\over h z^2\right)-(\dot{t}^2-s^2
t'^2)\p_z\left(h\over
z^2\right)\right.\nonumber\\{}&{}&\quad\quad\quad\quad\quad\quad\quad\quad\quad
\quad\quad\quad\left.-(\dot{x}^2-s^2 x'^2)\left(2\over z^3\right)\right]~,\nonumber
\end{eqnarray}
(where  $\dot{}\equiv\p_{\tau}$~, $'\equiv\p_{\sigma}$~,) supplemented with the constraints
\begin{eqnarray}\label{constraints}
-h \dot{t}t'+\dot{x}x'+h^{-1}\dot{z}z'&=&0~,\\
-h(\dot{t}^2+s^2 t'^2)+(\dot{x}^2+s^2 x'^2)+h^{-1}(\dot{z}^2+s^2 z'^2)&=&0~,\nonumber
\end{eqnarray}
which as always amount to the statement that the intrinsic metric on the worldsheet must be
proportional to the induced metric (thereby establishing the classical equivalence with the
Nambu-Goto formalism). Given initial data that satisfy (\ref{constraints}), the requirement
that the constraints continue to hold throughout the evolution gives an important
consistency check on the numerical integration. The results we will report throughout this
section were obtained with Mathematica 5.2's {\tt NDSolve} integration routine, with the
constraints typically satisfied at the $10^{-5}$ or $10^{-6}$ level.

In terms of the momentum densities $\Pi^a_{\mu}\equiv\p\cL_P/\p(\p_a X^{\mu})$, we see that,
as usual, the first two equations in (\ref{eom}) express the conservation of the Noether
currents $\Pi^{a}_t$ and $\Pi^{a}_x$, respectively associated with invariance under
translations in $t$ and $x$. To describe a quark-antiquark pair that is not acted upon by
any agent other than the plasma, we must choose the standard Neumann/Dirichlet boundary
conditions
\begin{equation}\label{bc}
 t'(\tau,0)=t'(\tau,\pi)=0~,\quad x'(\tau,0)=x'(\tau,\pi)=0~,\quad
z(\tau,0)=z(\tau,\pi)=z_m\quad\forall\,\tau~.
\end{equation}
The total energy
\begin{equation}\label{energy}
E={R^2\over 2\pi\ap}\int_0^{\pi} d\sigma\,(-\Pi^{\tau}_t)={\sqrt{\lambda}\over
2\pi}\int_0^{\pi}d\sigma\,{h\dot{t}\over s z^2}
\end{equation}
and $x$-momentum
\begin{equation}\label{momentum}
P={R^2\over 2\pi\ap}\int_0^{\pi} d\sigma\,\Pi^{\tau}_x={\sqrt{\lambda}\over
2\pi}\int_0^{\pi}d\sigma\,{h\dot{x}\over s z^2}
\end{equation}
of the string are then conserved.

For the problem at hand, the authors of \cite{hkkky} identified a one-parameter family of
initial conditions that correctly satisfy the constraints (\ref{constraints}) and are
compatible with the boundary conditions (\ref{bc}). Working from now on in units where
$z_h=1/\pi T=1$, these conditions take the form
\begin{eqnarray}\label{hkkkyic}
t(0,\sigma)=0~,&\quad&\dot{t}(0,\sigma)=A~,\\
x(0,\sigma)=0~,&\quad&\dot{x}(0,\sigma)=A\sqrt{1-z_m^4}\cos\sigma~,\nonumber\\
z(0,\sigma)=z_m~,&\quad&\dot{z}(0,\sigma)=A[1-z_m^4]\sin\sigma~,\nonumber
\end{eqnarray}
and describe a string that is pointlike at $t=0$ and grows for $t>0$ as a result of its
non-zero initial velocity
\begin{equation} \label{hkkkyv0}
v_x(\sigma)={\dot{x}\over\dot{t}}=\sqrt{1-z_m^4}\cos\sigma~,\quad
v_z(\sigma)={\dot{z}\over\dot{t}}=[1-z_m^4]\sin\sigma~.
\end{equation}
The parameter $A$ (which has been rescaled here by a factor of $\sqrt{1-z_m^4}$ with respect
to \cite{hkkky}) controls the energy (\ref{energy}) of the configuration,
\begin{equation}\label{pointenergy}
E={\sqrt{\lambda}\over 2\pi}\int_0^{\pi}d\sigma\,\left({h\dot{t}\over s
z^2}\right)_{\tau=0}={\sqrt{\lambda}\over 2}{(1-z_m^4)A\over z_m^2 s(z_m)}~.
\end{equation}
As expected, the total $x$-momentum vanishes.

Using the initial conditions (\ref{hkkkyic}), it was found in \cite{hkkky} that, depending
on the value of $E$, the subsequent behavior of the string endpoints can be of two different
types. When the energy of the pair is large enough (essentially, $E>2M_{\mbox{\scriptsize
rest}}$) the quark and antiquark move apart and are able to escape from one another's
influence, so they simply slow down monotonically until they are finally (at $t=\infty$)
brought to rest by the plasma. For low $E$, on the other hand, the mutual attraction of the
quark and antiquark manages to stop them and make them reverse direction, after which they
undergo a number (larger than one quarter) of oscillations before dissipating all of their
energy to the plasma. During consecutive half-cycles the body of the string is alternately
above and below the $z=z_m$ line, as a result of which the corresponding motion is
asymmetric.\footnote{Configurations where the string lies below its endpoints have been
considered previously in \cite{argyres2,argyres3}. In such circumstance, one suspects that
the string would prefer to shrink by sliding its endpoints along the D7-branes, moving them
closer to the boundary and to each other. To examine this question, however, one must
remember: first, that the boundary conditions for the string are purely Neumann or Dirichlet
only when the D7-brane embedding is described in the original Cartesian coordinates, and are
actually mixed in the spherical coordinates that are naturally employed after taking the
AdS/CFT limit; second, that the string coordinates on the $\bS^5$ describe the internal
$SU(4)$ degrees of freedom of the quark, which one may or may not wish to fix externally.}

\subsection{Generalized initial conditions}
\label{qqbargeneralizationsec}

Intuitively, it should be possible to generalize the initial velocity profiles
(\ref{hkkkyv0}) proposed in \cite{hkkky} to more general functions $v_x(\sigma)$,
$v_z(\sigma)$, which amounts to stipulating that $\dot{x}(0,\sigma)=A v_x(\sigma)$,
$\dot{z}(0,\sigma)=A v_z(\sigma)$. Based on the symmetry of our problem, for simplicity we
restrict attention to functions $v_x(\sigma)$ that are odd on the interval $[0,\pi]$.
Compatibility with the Neumann boundary condition in (\ref{bc}) requires that $v'_x(0)=0$,
the Hamiltonian constraint in (\ref{constraints}) determines $v_z(\sigma)$ in terms of
$v_x(\sigma)$, and the Dirichlet boundary condition in (\ref{bc}) demands that $v_z(0)=0$.
Altogether, then, we find that we are allowed to choose
\begin{eqnarray}\label{pointic}
t(0,\sigma)=0~,&\quad&\dot{t}(0,\sigma)=A~,\\
x(0,\sigma)=0~,&\quad&\dot{x}(0,\sigma)=A v_x(\sigma)~,\nonumber\\
z(0,\sigma)=z_m~,&\quad&\dot{z}(0,\sigma)=A\sqrt{1-z_m^4}\sqrt{1-z_m^4-v_x(\sigma)^2}~,\nonumber
\end{eqnarray}
with $v_x(0)=\sqrt{1-z_m^4}$ and $v'_x(0)=0$, which constitutes an infinite-parameter
generalization of the initial conditions (\ref{hkkkyic}).

\begin{figure}[htb]
\vspace*{0.5cm} \setlength{\unitlength}{1cm}
\includegraphics[width=6cm,height=4cm]{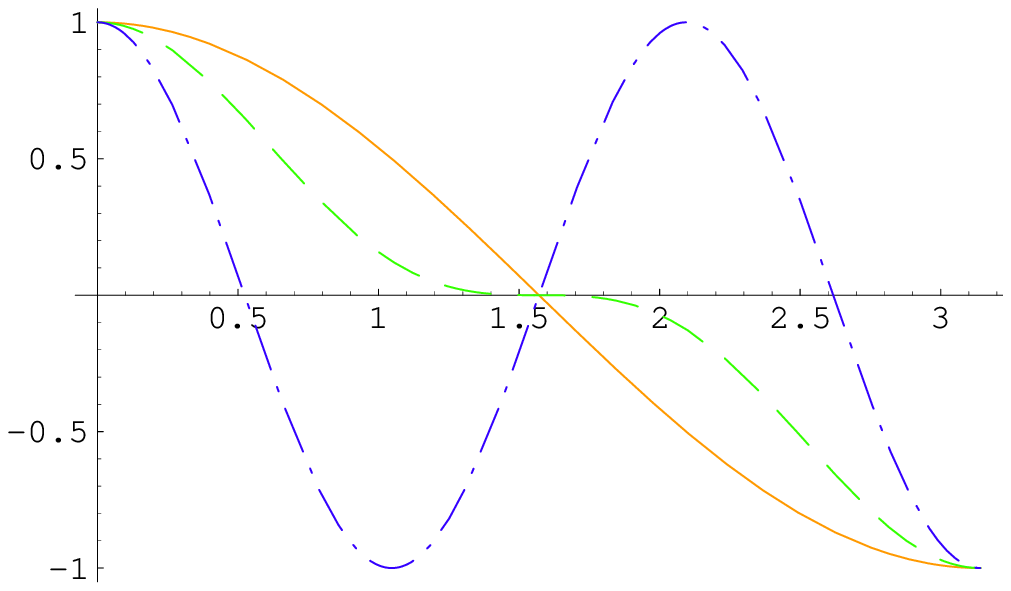}
 \begin{picture}(0,0)
   \put(0,1.9){$\sigma$}
   \put(-5.5,4.1){$v_x(\sigma)$}
 \end{picture}\hspace{1cm}
 \includegraphics[width=6cm,height=4cm]{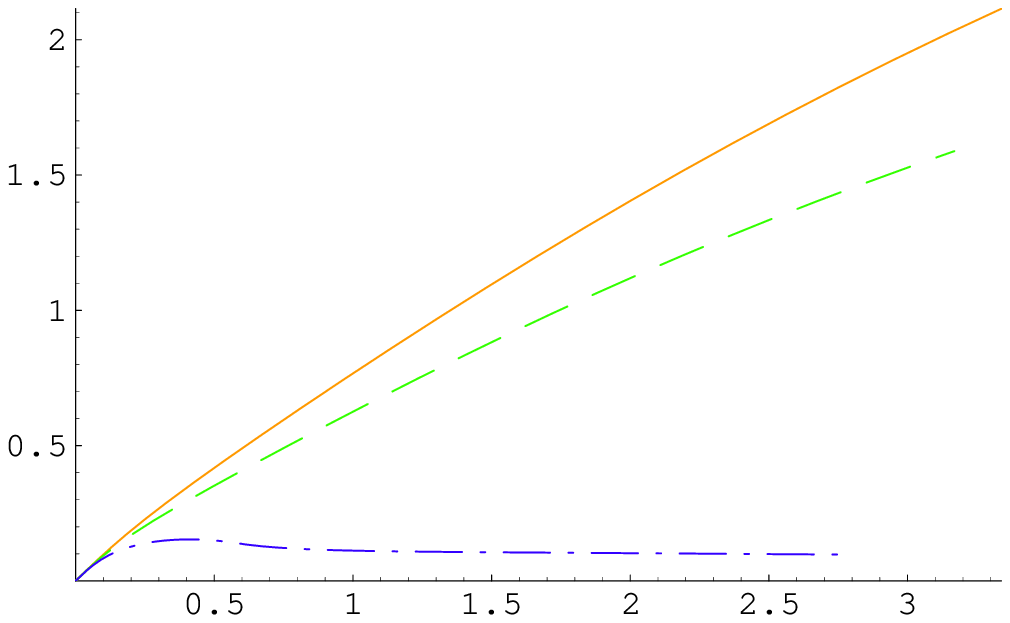}
 \begin{picture}(0,0)
   \put(0,0.3){$t$}
   \put(-5.5,4.1){$x$}
 \end{picture}
\caption{(a) Three different initial velocity profiles for the string: $\cos\sigma$ (solid),
$\cos^3\sigma$ (dashed), and $\cos3\sigma$ (dashdotted). (b) The corresponding trajectories
for the $\sigma=0$ string endpoint, for quark mass parameter $z_m=0.2$. Even though all 3
configurations have the same energy $E/2M_{\mbox{\scriptsize rest}}=2.45$ and initial quark
velocity $v_0=0.9992$, the evolution of the quark is clearly rather different in each case.
} \label{vsfig}
\end{figure}

The energy of all of these configurations is  given by the same formula (\ref{pointenergy}).
Since the string is initially a point, it might seem peculiar that one can physically
distinguish among various possible velocities for its different `internal points'--- one
might suspect that the seemingly different initial profiles (\ref{pointic}) are all related
to each other by gauge transformations. That this is not the case can be seen most
straightforwardly  by evolving  a few of these initial configurations forward in time: as
shown in Fig.~\ref{vsfig}, the resulting spacetime trajectories are found to be distinct.
Notice in particular that in the dashdotted trajectory the quark actually reverses direction
before coming to rest, which shows that the boundary between oscillatory and non-oscillatory
behavior depends strongly on the way in which the gluonic field (dual to the string) is
excited. So the moral of the story is that, while the string is initially a point in
spacetime, it is most definitely \emph{not} a point in phase space, and it is this fact that
allows the existence of a truly infinite-dimensional family of initial conditions.

The question of the gauge-dependence of our description does
however serve to highlight a useful point: instead of sampling
different initial conditions by varying the functional form of
$v_x(\sigma)$ in (\ref{pointic}) (for a fixed choice of the
worldsheet coordinate $\sigma$), we can keep the form of
$v_x(\sigma)$ fixed and change the meaning of $\sigma$. One way to
do the latter is to modify our choice of the initial stretching
factor $s(0,\sigma)$. It is easy to see that this indeed leads to
different spacetime trajectories, even if the initial energies
(\ref{pointenergy}) are appropriately matched. Conversely, when
tweaking $s(\tau,\sigma)$ to find the choice that optimizes the
numerical integration for a given physical string configuration,
one must keep the initial stretching factor $s(0,\sigma)$ fixed,
to ensure that the initial conditions for the evolution do not
change. For our calculations we found it convenient to use
$s(\tau,\sigma)=(1-z(\tau,\sigma))^p/(1-z_m)^{p-1}$, with
adjustable $p$.

It is natural to wonder what the interpretation of the different initial conditions
(\ref{pointic}) is in the dual SYM language. The answer is provided to us by the standard
recipe for correlation functions \cite{gkpw}: different time-dependent string profiles
correspond to different time-dependent configurations of the gluonic fields \cite{dkk,cg}.
Just like the string embedding is not completely characterized by giving the location of its
endpoints, the initial state of the gauge theory is not uniquely characterized by specifying
the position of the quark and antiquark. This is of course true already in zero-temperature
QED, but in that case the linear character of the equations of motion makes it easy to
identify, for a given field configuration, the portion that is directly ascribable to the
 sources of interest.

Notice from (\ref{pointic}) that the initial velocity of the quark, $v_0\equiv
v_x(0)=\sqrt{1-z_m^4}$, is \emph{not} a free parameter of the system, but is uniquely fixed
by the choice of $z_m$, or equivalently, by the quark's Lagrangian mass $m$, according to
(\ref{zm}). The reason for this is easy to understand on the string theory side. At $t=0$,
the pointlike string happens to obey not only $\dot{z}=0$, in compliance with (\ref{bc}),
but also $z'=0$, according to (\ref{pointic}), as would befit an endpoint that is free in
all spacetime directions. It is well-known that the endpoints of such a string must move at
the speed of light (see, e.g., \cite{brinkhenneaux}), and, given that the endpoints are
located at $z=z_m$, we see from (\ref{metric}) that, indeed, a \emph{coordinate} velocity
$v_x=\sqrt{1-z_m^4}$ corresponds precisely to a \emph{proper} velocity $V_x\equiv
v_x/\sqrt{-G_{tt}}=1$, a fact that was first pointed out in \cite{argyres1}.\footnote{The
same reasoning in fact applies to `all points' on the string: their proper initial
velocities
$$
V_x(\sigma)\equiv{v_x(\sigma)\over\sqrt{1-z_m^4}}~,\quad V_z(\sigma)\equiv{v_z(\sigma)\over
1-z_m^4}={\sqrt{1-z_m^4-v_x(\sigma)^2}\over\sqrt{1-z_m^4}}~,
$$
clearly satisfy $V_x^2+V_z^2=1$, independently of the choice of $v_x(\sigma)$.} In the gauge
theory, this identification confers then a special status to the mass-dependent velocity
\begin{equation}\label{vm}
v_m\equiv\sqrt{1-z_m^4}~,
\end{equation}
whose meaning will be discussed further in the next subsection.

{}At least from the gauge theory perspective, one would expect to be able to find
configurations in which the initial quark velocity $v_0$ \emph{is} freely adjustable. Given
the discussion of the previous paragraph, we see that on the AdS side this can be achieved
with initially coincident string endpoints only if we choose $z'\neq 0$. To satisfy the
first constraint in (\ref{constraints}) we must then set $\dot{z}=0$. Picking for the string
a velocity profile $v_x(\sigma)$ that is an arbitrary odd function on the interval
$[0,\pi]$, the complete second set of allowed initial conditions is then
\begin{eqnarray}\label{lineic}
t(0,\sigma)=0~,&\quad&\dot{t}(0,\sigma)=A~,\\
x(0,\sigma)=0~,&\quad&\dot{x}(0,\sigma)=Av_x(\sigma)~,\nonumber\\
z(0,\sigma)=\zeta(\sigma)~,&\quad&\dot{z}(0,\sigma)=0~,\nonumber
\end{eqnarray}
with $\zeta(\sigma)$ an even function on the $[0,\pi]$ interval, which satisfies
$\zeta(0)=z_m$ for compatibility with the Dirichlet boundary condition in (\ref{bc}), and
\begin{equation}\label{zetadiffeq}
\zeta'(\sigma)=\pm {A\sqrt{1-\zeta(\sigma)^4}\over
s(\zeta(\sigma))}\sqrt{1-\zeta(\sigma)^4-v_x(\sigma)^2}~,
\end{equation}
to comply with the Hamiltonian constraint in (\ref{constraints}). By construction, the
initial quark velocity can now be chosen arbitrarily, as long as $v_0\leq v_m$ in order for
the right-hand side of (\ref{zetadiffeq}) to be real at $\sigma=0,\pi$.  The energy
(\ref{energy}) of the configuration is now given by
\begin{equation}\label{lineenergy}
E={\sqrt{\lambda}\over 2\pi}\int_0^{\pi}d\sigma\,\left({h\dot{t}\over s
z^2}\right)_{\tau=0}={\sqrt{\lambda}A\over
2\pi}\int_0^{\pi}d\sigma\,\frac{1-\zeta^4}{s(\zeta)\zeta^2}~.
\end{equation}
The linear $x$ momentum is still zero.

\begin{figure}[htb]
\begin{center}
\vspace*{0.5cm} \setlength{\unitlength}{1cm}
\includegraphics[width=6cm,height=4cm]{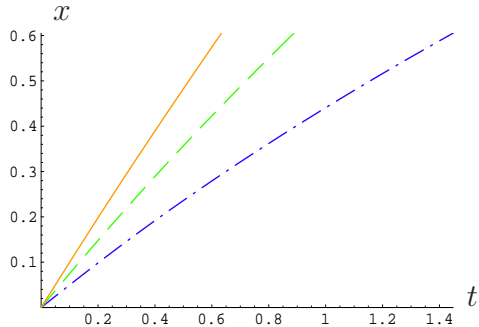}
 \begin{picture}(0,0)
   \put(0,0.3){$t$}
   \put(-5.5 ,4.1){$x$}
 \end{picture}
\caption{Quark trajectories for adjoint $q$-$\bar{q}$
configuration, for $z_m=0.2$ and $v_0=v_m$ (solid), $v_0=0.75v_m$
(dashed), $v_0=0.5v_m$ (dashdotted). The corresponding energies
are $E/2M_{\mbox{\scriptsize
rest}}=2.86,1.38,1.13$. In this case
no oscillating configurations are found.}\label{linefig}
\end{center}
\end{figure}

Conditions (\ref{lineic}) describe a linelike string that extends purely along the radial
AdS coordinate, stretching upward from $z=z_m$ up to some turning point $z=\zeta(\pi/2)$ and
then returning back down to $z=z_m$. The function $\zeta(\sigma)$ will be smooth at the
turning point only if $\zeta'(\pi/2)=0$, which combined with (\ref{zetadiffeq}) and the
requirement $v_x(\pi/2)=0$ implies that $\zeta(\pi/2)=1$, i.e., the string turns around at
the horizon. For this to occur, given an initial velocity profile $v_x(\sigma)$, the value
of $A$ must be tuned in order for the numerical solution of (\ref{zetadiffeq}) to reach
$\zeta=1$ precisely at $\sigma=\pi/2$. So for these initial conditions, where $v_0$ is a
free parameter, $A$ is not. Having determined $\zeta(\sigma)$, one can proceed as before to
the numerical integration of the equations of motion (\ref{eom}). A few representative quark
trajectories are shown in Fig.~\ref{linefig}, with $v_x(\sigma)=v_0\cos\sigma$ and
$s=(1-z)^{1/2}$ (chosen to simplify the equation of motion and increase the stability of the
numerical integration needed to obtain the initial profile $\zeta(\sigma)$).

The existence of the two qualitatively distinct sets of initial conditions for the string
describing the creation of a quark-antiquark pair has a direct field-theoretic
interpretation: the product of a fundamental $q$ and an antifundamental $\bar{q}$ can lead
to a $q$-$\bar{q}$ pair either in the singlet or the adjoint representation of the $SU(N_c)$
gauge group, and  each of the above string configurations is dual to one of these. Indeed,
the linelike string (\ref{lineic}) is precisely the system considered in
\cite{gubserpitp,gubserstable,gp,draggluon} to model a color source in the adjoint
representation. Due to its extended nature, it sets up long-range supergravity fields that
translate through the standard recipe of \cite{gkpw} into a long-range gluonic field
profile, indicative of a source with net color charge. The completely pointlike string
(\ref{pointic}), on the other hand, sets up no long-range chromoelectromagnetic field and so
describes the singlet. Given this correspondence, it is interesting that the AdS/CFT duality
predicts that (at large $N_c$ and large $\lambda$) the initial quark velocity is freely
adjustable in the adjoint, but not the singlet, configuration.

\subsection{Limiting velocity}
\label{qqbarvmsec}

In the previous subsection we have learned that the initial velocity of the quark and
antiquark at the moment of the pair's creation is bounded above by the speed $v_m$ defined
in (\ref{vm}). The reason for this is easy to understand on the string theory side of the
duality, since, as we have noted above, the coordinate velocity $v_x=v_m$ corresponds to a
proper velocity $V_x$ equal to that of light at the position of the string endpoints,
$z=z_m$ \cite{argyres1}. The interesting feature is that it is $v_x$, and not $V_x$, that
corresponds to the gauge theory velocity.

We would naturally expect the restriction to subluminal velocities, $v\le v_m$, to apply to
more general string configurations. In complete analogy with the point particle case, the
easiest way to deduce this restriction is to go back to the Nambu-Goto action
(\ref{nambugoto}), and observe that the requirement that it be real (i.e., that the string
worldsheet be timelike) imposes a bound on physically realizable embeddings. Indeed, working
for simplicity in the static gauge $\tau=t,\sigma=z$, it is easy to see that the Nambu-Goto
square root is real only as long as the embedding function $\vec{X}(z,t)$ satisfies
\begin{equation}\label{ngbound1}
\left({\p\vec{X}\over \p t}\right)^2\le h\frac{1+h\left({\p\vec{X}\over \p z}\right)^2}
{1+h\left({\p\vec{X}\over \p z}\right)^2\sin\alpha }~,
\end{equation}
where $\alpha\equiv\angle(\p \vec{X}/\p t,\p \vec{X}/\p z)$. For a string that moves and stretches
along a single direction $x$, as we have considered up to now, $\alpha=0$ or $\pi$ and this
reduces to
\begin{equation}\label{ngbound2}
\left({\p X\over \p t}\right)^2\le h\left(1+h\left({\p X\over \p z}\right)^2\right)~.
\end{equation}

It might seem peculiar that, as long as the string segment under consideration is not
vertical ($\p X/\p z\neq 0$), the bound (\ref{ngbound2}) allows the proper velocity of the
segment, $(1/\sqrt{h})\p X/\p t$, to exceed the speed of light (by an amount that becomes
arbitrarily large in the limit $|\p X/\p z|\to\infty$). One should note, however, that this is a
gauge-dependent statement, because unless the string segment is vertical, $x$ is not
entirely transverse to it, and the longitudinal component of the string motion is of course
unphysical (in particular, motion along $x$ is entirely unphysical in the limit
$|\p X/\p z|\to\infty$, where the string becomes horizontal). {}From (\ref{ngbound1}) we can see
that, for purely transverse motion, $\alpha=\pm\pi/2$ (as in the meson configurations we
will consider in Section \ref{potentialsec}), the proper velocity is indeed required to be
less than unity, i.e., the coordinate (or, equivalently, gauge theory) velocity is bounded
by $\sqrt{1-z^4}$.

The argument of the preceding paragraph applies to a generic point in the interior of the
string, but the situation is different for the endpoints, where motion along the body of the
string \emph{is} physical. As we know, a string endpoint on the D7-branes is dual to a quark
(or antiquark), so the velocity $\vec{v}$ of the latter must necessarily respect the bound
 (\ref{ngbound1}), or, for collinear motion, (\ref{ngbound2}). Evaluating this last equation
 at $z=z_m$ and parametrizing as in Sections \ref{qnoplasmasec} and \ref{qplasmasec} the
$\p X/\p z$-dependence in terms of the momentum density $\Pi\equiv\Pi^z_x$ given by
(\ref{momenta}), which controls the external force $F=
(\sqrt{\lambda}/2\pi)\Pi$ applied to the quark, we deduce that $v$ is bounded by
\begin{equation}\label{vmpi}
v\le \frac{h_m}{\sqrt{(h_m-z_m^4\Pi^2)(1+z_m^4\Pi^2)}} =\frac{v_m^2}{\sqrt{(v_m^2-\lambda
F^2/4\pi^2 m^4 []^4)(1+\lambda F^2/4\pi^2
m^4 []^4)}}~,
\end{equation}
where $h_m\equiv h(z_m)$ and $[]$ denotes the expression within brackets in (\ref{zm}).

For the case we have considered in the previous two subsections, where the string endpoint
is free, corresponding to a quark that evolves only under the influence of the plasma,
(\ref{vmpi}) is precisely the statement that $v\le v_m$. As expected, we see that this bound
(or, more generally, the analogous bound deduced from (\ref{ngbound1})) applies not only to
quark-antiquark configurations, but also to isolated quarks.\footnote{This observation has
also been made very recently in \cite{argyres3}. Simultaneously, $v_m$
 has been
 shown to emerge as a limiting
velocity directly from the microscopic meson dispersion relation \cite{liu4}.
 Both of these works appeared while the present paper was in
preparation.}

For the case where the quark is externally forced, on the other hand, the bound (\ref{vmpi})
becomes less restrictive: as $F$ increases, the quantity on the
right grows monotonically, and in fact diverges when the force approaches
the critical value $F_{\mbox{\scriptsize crit}}=(\sqrt{\lambda}/2\pi)h_m/z_m^4$ mentioned
already in Section \ref{acceleratedsec} (this is the same as the divergence
seen in (\ref{ngbound1}) and (\ref{ngbound2}) when  $\p X/\p z\to \infty$). {}From this perspective alone, then,
it would seem possible to take the quark to velocities larger than $v_m$ while
exerting a force on it (even though, after release, one would again
have $v\le v_m$).

 It is not guaranteed, however, that this possibility is realized in practice.
In the forced stationary  case considered in \cite{hkkky,gubser}, for instance,
it is found that $v> v_m$ would necessarily require $F>F_{\mbox{\scriptsize crit}}$, and is consequently unattainable \cite{ctqhat}. More generally, the question of whether $v_m$ is limiting
or not
is dynamical in nature, and essentially depends on the form of the thermal dispersion
relation for the quark. In the discussion following (\ref{Edrplasma}) we noted that, based on our
zero-temperature results of Section \ref{qnoplasmasec}, the quark's intrinsic energy $E_q$ (and
momentum $p_q$) should diverge as $F\to F_{\mbox{\scriptsize crit}}$. Given the connection
between the $1-v^2$ factor in the denominator of  (\ref{Edrplasma}) and the Lorentz invariance of the metric
(\ref{metric}) at $T=0$, it is natural to expect it to be replaced by $h_m-v^2=v_m^2-v^2$
at finite temperature, which would imply that $E_q\to\infty$ (and $p_q\to\infty$) as
$v\to v_m$,
meaning that $v>v_m$ is physically unattainable. The results
we will obtain in
Section \ref{potentialplasmasec} appear to support this expectation, but it
would certainly be nice to be able to show this directly.\footnote{For mesons this
was done recently in \cite{liu4}, which appeared while this paper was in preparation.}

\subsection{Transition to asymptotic regime}
\label{qqbartransitionsec}

For the non-oscillatory string trajectories with the one-parameter
family of initial conditions (\ref{hkkkyic}), it was found in
\cite{hkkky} that the late-time quark motion is independent of the
energy $E$ of the configuration, and coincides with the behavior
expected from a particle with relativistic dispersion relation
$p\propto v/\sqrt{1-v^2}$, subject to a damping force
\begin{equation}\label{hkkkyforce}
{dp\over dt}=-\mu p~,
\end{equation}
with $\mu$ a $p$-independent friction coefficient that was
tabulated in \cite{hkkky} for various values of $z_m$, and which
is consistent with the drag force (\ref{EPlossgubser}) obtained in
\cite{hkkky,gubser,ct} for the stationary configuration.

\begin{figure}[htb]
\vspace*{0.2cm} \setlength{\unitlength}{1cm}
\begin{center}
\includegraphics[width=9cm,height=6cm]{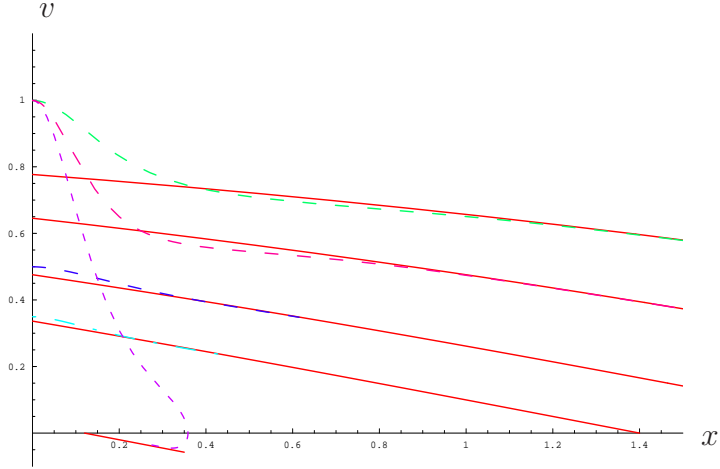}
 \begin{picture}(0,0)
   \put(0.1,0.4){$x$}
   \put(-8.7,6.1){$v$}
 \end{picture}\hspace{1cm}
 \end{center}
 \vspace*{-0.5cm}
\caption{Quark evolution (velocity as a function of traveled distance,
in units of $1/\pi T$)
 for five different initial conditions. To explore the neighborhood
 of the charm quark, the
  mass parameter has been chosen as $z_m=0.2$, corresponding
  to a limiting velocity $v_m=0.9992$. The dotted curves show the results
 of our numerical integration, contrasted against fits in solid red that use
  the analytic expression (\ref{constmuvx}), with the value $\mu=0.25$ obtained
  in \cite{hkkky} and an optimal choice of the stopping distance
  $x_{\infty}$. The three dotted curves starting
  at the same point describe singlet configurations (and therefore have $v_0=v_m$,
  as explained in the main text). The green, magenta and purple curves
  correspond respectively to total
  energies $E/2M_{\mbox{\scriptsize rest}}=2.45,2.45,1.01$
  and initial string velocity profiles
 $v_0\cos\sigma$,
$v_0\cos^3\sigma$ and $v_0\cos\sigma$ (leading to $x_{\infty}=4.15,3.07, 0.12$).
Notice in particular that the purple curve describes a
situation where the quark and antiquarks turn around and come to rest while approaching
one another.
The two
remaining curves arise from adjoint configurations with different energies and initial quark
velocities: the case $E/2M_{\mbox{\scriptsize rest}}=1.12$ and $v_0=0.5 v_m$ (leading to
$x_{\infty}=2.07$) is shown in dark blue, while $E/2M_{\mbox{\scriptsize rest}}=1.05$ and
$v_0=0.35 v_m$ ($x_{\infty}=1.4$) is shown in light blue.
} \label{stage2fig}
\end{figure}

The
agreement between the analytic and late-time numeric results is
most cleanly seen if instead of comparing graphs of $x(t)$ or
$v(t)$ for the quark (where one would need to look at
$t\to\infty$), one examines the plots of $v(x)$,\footnote{We thank
Antonio Garc\'\i a for suggesting this.} where the analytic
behavior for constant $\mu$ takes the simple form
\begin{equation}\label{constmuvx}
v(x)=\tanh[\mu(x_{\infty} - x)]~,
\end{equation}
which is linear with slope $-\mu$ near the final rest point
$x=x_{\infty}$ (whose value is meant to be adjusted to fit the
data). These plots are shown in Fig.~\ref{stage2fig}, where it is
seen that the late-time behavior is well-described by
(\ref{constmuvx}) also for oscillating trajectories and for the
more general singlet configuration (\ref{pointic}), as well as for
the adjoint configuration (\ref{lineic}).
Notice that the agreement holds
 in the late-time regime where (\ref{constmuvx}) reduces to a linear
 expression, but, as in the cases studied by the authors of
  \cite{hkkky},
it extends beyond the range covered by their general
quasi-normal mode analysis, because the velocity of the
 quark is not necessarily small.

In Fig.~\ref{stage2fig}, we see
 that there is an initial period
where the behavior differs
 from the late-time frictional evolution (\ref{constmuvx}).
 This difference is clearly more significant for
 the singlet than the adjoint case. The \emph{time} that
 must elapse before the asymptotic behavior sets
 in becomes arbitrarily large for singlet
 configurations that are close to being oscillatory.
 In these cases, essentially all of the energy of the quark
 is lost not through the constant-$\mu$ frictional force due to
 the plasma, but as a result of the chromoelectromagnetic
 force exerted by the antiquark.

 On the string theory side of the duality,
 the issue is that, as the initially pointlike string grows
  and falls toward the black hole,
  it takes some time before it is close enough to
  $z=1$ to be deformed into (two juxtaposed copies of) the
  asymptotic $\neg$ shape discussed in Section \ref{bhplasmasec}.
  This picture seems rather close to the phenomenological
  discussion given
  in \cite{peigne1} (in the context of collisional
 energy loss):
  when the singlet quark-antiquark pair is formed
 within the plasma, there is a delay before
 the interaction between the newly created sources and the plasma
 can set up the long range gluonic field profile that is responsible
 for the late-time dissipation.

To examine in more detail the transition to the late-time behavior
(\ref{constmuvx}),  for a variety of trajectories we have
determined the point $(x_f,v_f)$ beyond which the numeric $v(x)$
curve agrees with the analytic curve (\ref{constmuvx}) to the
accuracy indicated by the fraction $f$. Even though, judging by
the effect of halving the grid spacing, our numerical results
seem to be accurate to at least 1\%, the precision with which we
can determine $(x_f,v_f)$ is limited by the uncertainty in the
 fitting parameter $x_{\infty}$, which we estimate to be of order 5-10\%.
 It does not make
 much sense therefore to consider a value of $f$ smaller than this.

\begin{figure}[htb]
  \vspace*{0.5cm}
 \begin{center}
\setlength{\unitlength}{1cm}
\includegraphics[width=6cm,height=4cm]{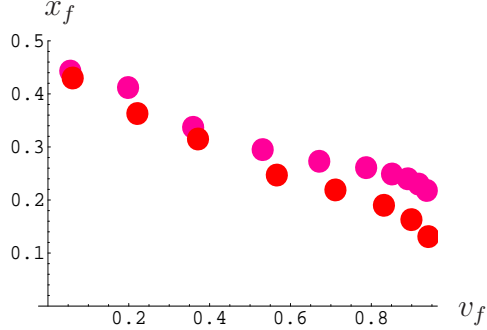}
 \begin{picture}(0,0)
   \put(0,0.2){$v_f$}
   \put(-5.5,4.2){$x_{f}$}
 \end{picture}\hspace{1cm}
 \caption{Transition distance $x_f$ (in units of $1/\pi T$)
as a function of the transition velocity $v_f$, for $f=0.1$ (red) and $f=0.05$ (magenta). The size of the dots
gives a rough indication of the margin of error.} \label{xtransfig}
\end{center}
\end{figure}

 A representative sample of
 our results for
 $f=0.1$ and $f=0.05$, in the case of singlet configurations
 is shown in Fig.~\ref{xtransfig}. The two sets
 of data have somewhat different functional forms,
 but are consistent with one another within the rather large margin of
 error. The general tendency is for $x_f$ to approach zero as
 $v_f\to v_m$.

 For adjoint configurations, we find that the numeric $v(x)$ curves
 are within 5-10\% of (\ref{constmuvx}) already at the start of the evolution, so,
 to be consistent with our pre-established criterion, in this case we
 must identify the transition length as essentially $x_f=0$. It is still
 worth noting, however, that there is an initial period where
 the functional form of the numeric and analytic plots is different, as can be seen in Fig.~\ref{stage2fig}. This difference is negligible
 for small initial velocities, and becomes more pronounced
 (both in magnitude and in duration) as $v_0$ increases.
 The
 transition length identifiable at the 3-4\% level thus follows
 a trend opposite to the one for the singlet case. Nonetheless,
 up to the velocities $v_f\sim 0.5$
  that we have been able
 to explore (corresponding to $v_0\sim v_m$), it remains smaller than
 the singlet $x_f$ shown in Fig.~\ref{xtransfig}.

 A question worth considering is whether the transition to the regime
 where the quark experiences a constant drag coefficient occurs
 right after the quark and antiquark are screened from each other
 by the plasma, or if there is an intermediate regime where the
 quark moves independently from the antiquark but nevertheless
 feels
 a drag force that differs from the stationary result of \cite{hkkky,gubser,ct},
 as we found when applying an external force in Section
 \ref{acceleratedsec}. To answer this question, we will determine the
relevant screening length in the next section.

It would also be very interesting to develop an interpretation of energy loss for the
evolving quark-antiquark pair in parallel with the picture for an isolated quark proposed by
Mikhailov \cite{mikhailov} and developed further in
Sections~\ref{qnoplasmasec},\ref{qplasmasec}. An effort in this direction was in fact made
in the work \cite{sinzahed}, but regrettably we do not understand its use of trajectories
that lie outside of the string worldsheet. We suspect that a treatment based on null curves
\emph{on} the worldsheet should be possible. Of course, progress is again hampered by the
lack of an analytic solution describing the back-to-back $q$-$\bar{q}$ evolution considered
in this section (or even its counterpart at zero temperature). Another potentially confusing
issue is the fact that, in contrast with the isolated quark case, there are now two string
endpoints at $z=z_m$, and \emph{a priori} it would be possible for null trajectories to
`bounce' repeatedly between them. One should however bear in mind that, at least beyond some
finite interval of time, this would be prevented by the formation of a worldsheet horizon
analogous to the one discussed in Sections~\ref{bhsec},\ref{bhplasmasec}.

 \section{Quark-Antiquark Potential}
 \label{potentialsec}

\subsection{Review of earlier results}
\label{potentialreviewsec}
 The potential $E(L)$ for an infinitely massive static quark and
antiquark in a strongly-coupled $\cN=4$ SYM plasma was determined
in \cite{theisen},\footnote{For the corresponding weak coupling calculation
in QCD, see, e.g., \cite{lprt} and references therein.} using
the dual description of the pair in terms a string with both of its endpoints on the AdS
boundary $z=0$. As expected, for small quark-antiquark separation $L$, the effects of the
plasma are negligible and the potential matches the zero-temperature result
\cite{maldawilson}
\begin{equation}\label{coulomb}
E(L)=-\frac{4{\pi}^2\sqrt{g^2_{YM}N}} {{\Gamma(\frac{1}{4})}^4L}~.
\end{equation}
As the separation grows, however, the effects of the plasma
progressively screen the quark and antiquark from one another, and
as a consequence raise the system's energy above the Coulombic
behavior. The $\cap$-shaped string embedding that is employed for
the calculation exists only up to a maximal separation
$L_{\mbox{\scriptsize max}}=0.865/\pi T$, and its energy exceeds
that of the disconnected solution beyond the somewhat smaller distance
$L_*=0.755/\pi T$. In the $N_c,\lambda\to\infty$ limit where the
string-theoretic calculation is easy to carry out, the potential
has a kink at $L_*$, and beyond this point the string path
integral is dominated instead by a configuration with high
worldsheet curvature which for large $L$ describes graviton
exchange between two disconnected strings, leading to a tail
$E(L)\sim -T\exp(-L/L_{\mbox{\scriptsize gap}})$, with
$L_{\mbox{\scriptsize gap}}=0.428/\pi T$
\cite{bky}.

This computation can easily be extended to the case where the
$q$-$\bar{q}$ pair moves with velocity $v$ with respect to the
plasma, to obtain the energy $E(L,v)$ of the pair
\cite{dragqqbar,liu3}. For any $v$, the potential reduces to
(\ref{coulomb}) at small $L$. The main overall effect of
increasing the velocity is to move the non-Coulombic portion of
the $E(L,v)$ curve down and to the left, leading in particular to
the identification of the two screening lengths
$L_*=L_{\mbox{\scriptsize max}}$ for $v\ge 0.447$.

The function
$L_{\mbox{\scriptsize max}}(v)$ was determined numerically in
\cite{liu2,dragqqbar} (and a related length was plotted in
 \cite{sonnenschein}).\footnote{The
works \cite{liu2,liu3} additionally explored the dependence of
$L_{\mbox{\scriptsize max}}$ on the angle $\theta$ between the
direction of motion and the dipole axis (which was fixed at
$\theta=\pi/2$ in \cite{dragqqbar}), finding it to be weak.} Over
the entire range $0\le v\le 1$ its behavior may be approximated
as \cite{dragqqbar}
\begin{equation} \label{onethird}
L_{\mbox{\scriptsize max}}(v)\approx{0.865\over \pi
T}(1-v^2)^{1/3}~,
\end{equation}
while in the ultra-relativistic limit, it can be shown analytically that \cite{liu2}
\begin{equation}\label{onequarteranal}
L_{\mbox{\scriptsize max}}(v)\to {1\over \pi T}{
3^{-3/4}4\pi^{3/2}\over\Gamma(1/4)^2}(1-v^2)^{1/4}\simeq
{0.743\over \pi T}(1-v^2)^{1/4}\quad\mbox{for $v\to 1$.}
\end{equation}
The full curve $L_{\mbox{\scriptsize max}}(v)$ does not deviate
far from this asymptotic form, so a decent
approximation to it is obtained by replacing $0.743\to 0.865$ in
(\ref{onequarter}), to reproduce the correct value at $v=0$ (at
the expense of introducing a 16\% error as $v\to 1$) \cite{liu2}:
\begin{equation} \label{onequarter}
L_{\mbox{\scriptsize max}}(v)\approx{0.865\over \pi
T}(1-v^2)^{1/4}~.
\end{equation}

\begin{figure}[htb]
\vspace*{0.5cm}
\begin{center}
\setlength{\unitlength}{1cm}
\includegraphics[width=6cm,height=4cm]{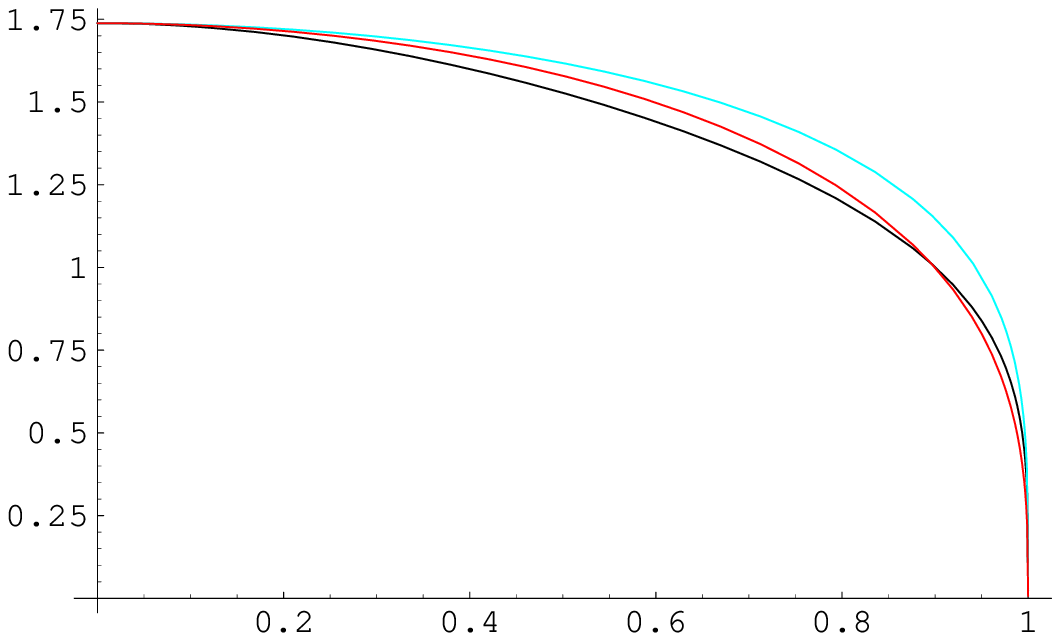}
 \begin{picture}(0,0)
   \put(0.2,0.3){$v$}
   \put(-6.1 ,4.2){$L_{\mbox{\scriptsize max}}$}
 \end{picture}
\caption{Screening length $L_{\mbox{\scriptsize max}}$ (in units
of $1/2\pi T$) as a
function of velocity (in black) compared against the
approximations (\ref{onethird}) (in red) and (\ref{onequarter}) (in
blue).}\label{onethirdfig}
\end{center}
\end{figure}

A comparison between the two approximations (\ref{onethird})
 and (\ref{onequarter}) is shown
 in Fig.~\ref{onethirdfig}: overall, the
 exponent $1/3$
is better than $1/4$ in the sense that
it implies a smaller squared deviation from the numerical results,
even though $1/4$ leads
to a smaller \emph{percentage} error in the range $v>0.991$
($\gamma>7.3$).
An attempt to better parametrize the deviation away from the
ultra-relativistic behavior was made in \cite{sfetsosqqbar}.
In any case, one should bear in mind that the region of principal
interest at RHIC is not really $v\to 1$, but
 $\gamma v\sim 1$ (see, e.g., \cite{mt}).

It is interesting to consider how these results generalize to the case where the quark and
antiquark have a large but finite Lagrangian mass $m$, which as reviewed in Section
\ref{qnoplasmasec} corresponds to letting the flavor D7-branes extend up to a position $z_m>0$ given
by (\ref{zm}). As explained in Section \ref{acceleratedsec}, from the
phenomenological perspective we are mostly interested in values of the mass
parameter $z_m$ in the range $z_m/z_h\sim 0.2-0.4$, which corresponds to the charm mass.

\subsection{Finite mass at zero temperature}
\label{potentialnoplasmasec}

In a static
 configuration, the equation of motion for the string amounts to the statement
that the momentum density $\Pi^z_x$ given by (\ref{momenta}) is independent of $x$. In the pure
AdS geometry (i.e., in the metric (\ref{metric}) with $T=0$) the resulting embedding is
\cite{maldawilson}
\begin{equation}\label{Xnoplasma}
X(z)=\pm z_{\mbox{\scriptsize max}}\int^{z_{\mbox{\tiny
max}}/z}_1 \frac{d\zeta}{\zeta^2\sqrt{\zeta^4-1}}~,
\end{equation}
where $z_{\mbox{\scriptsize max}}\in [z_m,\infty)$ denotes the
turning point of the $\cap$-shaped string and the plus or minus
sign applies respectively to its right and left half. {}From this
it follows in particular that the distance between the string
endpoints at $z=z_m$ (i.e., the $q$-$\bar{q}$ separation) is
\begin{equation}\label{Lnoplasma}
L=2 z_{\mbox{\scriptsize max}}\int^{z_{\mbox{\tiny
max}}/z_m}_1
\frac{d\zeta}{\zeta^2\sqrt{\zeta^4-1}}=2z_{\mbox{\scriptsize
max}}\left[-{1\over 4} B\left( \left({z_m\over
z_{\mbox{\scriptsize max}}}\right)^4;{3\over 4},{1\over
2}\right)+{\sqrt{\pi}\Gamma({3\over 4})\over\Gamma({1\over
4})}\right]~,
\end{equation}
where $B$ denotes the incomplete Euler beta function. The energy of the $\cap$-shaped
string, renormalized by subtracting the energy of two purely radial strings, follows from
(\ref{nambugoto}) as
\begin{equation}\label{Enoplasma}
E={\sqrt{\lambda}\over\pi z_{\mbox{\scriptsize
max}}}\left\{\int^{z_{\mbox{\tiny max}}/z_m}_1
d\zeta\,\left[{\zeta^2\over\sqrt{\zeta^4-1}}-1\right]-1\right\}~.
\end{equation}
The corresponding results for the case of a moving $q$-$\bar{q}$ pair are of course
determined in terms of these by Lorentz invariance.

{}Using (\ref{Lnoplasma}) and (\ref{Enoplasma}), it is easy to see
that for $L\gg \sqrt{\lambda}/m$ (i.e., $z_{\mbox{\scriptsize
max}}\gg z_m$)
 the quark-antiquark potential $E(L)$ approaches the same Coulombic form (\ref{coulomb})
 as in the infinitely massive case ($z_m=0$). On the other hand,
 in the short distance limit $L\to 0$ we have $z_{\mbox{\scriptsize max}}\to z_m$ and so the potential is
 linear,
 \begin{equation}\label{Elinearnoplasma}
E(L)=2 m\left[-1+{\pi\over\sqrt{\lambda}}m L + \mathcal{O}\left(\left({m
L\over\sqrt{\lambda}}\right)^2\right)\right]~.
 \end{equation}

These results are consistent with the form of the gluonic field around an isolated
finite-mass quark \cite{martinfsq},
\begin{eqnarray}\label{Fsquarednoplasma}
\expec{\cO_{F^2}(\vec{x})}_q&=&{\sqrt{\lambda}\over
16\pi^2|\vec{x}|^4}\left[1-\frac{1+{5\over 2}\left({2\pi
m|\vec{x}|\over\sqrt{\lambda}}\right)^2}{\left(1+\left({2\pi
m|\vec{x}|\over\sqrt{\lambda}}\right)^2\right)^{5/2}}\right]\\
{}&=&{\sqrt{\lambda}\over 128\pi^2}\left[-\left({2\pi m\over\sqrt{\lambda}}\right)^4+{7\over
4|\vec{x}|^4}\left({2\pi m|\vec{x}|\over\sqrt{\lambda}}\right)^6+\ldots\right]\nonumber
\end{eqnarray}
(where $\cO_{F^2}=\tr\{F^2+\ldots\}/4 g_{YM}^2$ denotes the operator dual to the dilaton
field \cite{ktr}), which is Coulombic at long distances and non-singular at the location of the quark.

In both (\ref{Elinearnoplasma}) and (\ref{Fsquarednoplasma}) we are seeing the effect of a
color charge distribution that cloaks the fundamental sources, or in other words, of the
gluon (and scalar, etc.) cloud that surrounds the would-be bare quark to turn it into a
`dressed' or `composite' quark, in line with the inferences we made in Section
\ref{finitemasssec}. Notice from (\ref{Fsquarednoplasma}) that the size of this cloud is not
the Compton
 wavelength of the quark, $1/m$
 (indicating that it is not made of virtual quarks and antiquarks),
 but the much larger scale $\sqrt{\lambda}/m$,
which is the characteristic size of the deeply bound
microscopic mesons of the theory
\cite{martinmeson,strassler,martinfsq}---
 so perhaps it would be more appropriate to refer to this
charge distribution as a `meson
cloud.'

\subsection{Finite mass at finite temperature}
\label{potentialplasmasec}

We will consider first the case where the quark-antiquark pair is static with respect to the
plasma. Repeating the calculation of the previous subsection for $T>0$, the separation
between the quark and antiquark is found to be
\begin{equation}\label{Lplasma}
L=2 z_{\mbox{\scriptsize max}}\int^{z_{\mbox{\tiny
max}}/z_m}_1 d\zeta\,\sqrt{1-\zeta^4_h\over
\zeta^4-\zeta^4_h}{1\over\sqrt{\zeta^4-1}}~,
\end{equation}
with $\zeta_h\equiv z_{\mbox{\scriptsize max}}/z_h$, and the
renormalized energy of the pair is given by
\begin{equation}\label{Eplasma}
E={\sqrt{\lambda}\over\pi z_{\mbox{\scriptsize
max}}}\left\{\int^{z_{\mbox{\tiny max}}/z_m}_1
d\zeta\,\left[\sqrt{\zeta^4-\zeta^4_h\over
\zeta^4-1}-1\right]-1\right\}~.
\end{equation}

{}From (\ref{Lplasma}) and (\ref{Eplasma}), the short-distance behavior of the potential is
now
 \begin{equation}\label{Elinearplasma}
E(L)= {\sqrt{\lambda}\over \pi}\left[-\left({1\over z_m}-{1\over z_h}\right)+{L \over
2z^2_m}\sqrt{1-\left({z_m\over z_h}\right)^4} + \mathcal{O}\left(\left({L\over
z_m}\right)^2\right)\right]~,
 \end{equation}
with $z_m$ related to the quark Lagrangian mass $m$ through (\ref{zm}). By comparing with
numerical plots (as in Fig.~\ref{potentialfig} below), this linear expression can be seen to give a good approximation of the
actual potential at least up to $L\sim z_m/2$.

The constant term in (\ref{Elinearplasma}) is clearly nothing but (minus) the thermal rest
mass (\ref{Mrest}) of the isolated quark and antiquark, which was subtracted in our choice
of renormalization. For any $L$, the force $F_{\mbox{\scriptsize pot}}(L)\equiv -dE/dL$
deduced from the potential balances the external force
\begin{equation}\label{Fext}
F_{\mbox{\scriptsize ext}}(L)\equiv {dp\over dt}=
{\sqrt{\lambda}\over 2\pi}\left.\Pi^z_x\right|_{z=z_m}=\mp{\sqrt{\lambda}\over 2\pi
z_{\mbox{\scriptsize max}}^2}\sqrt{1-\left({z_{\mbox{\scriptsize
max}}\over z_h}\right)^4}
\end{equation}
required to hold the string endpoint in place, and the coefficient
of the linear term in (\ref{Elinearplasma}) correctly encodes the
$L\to 0$ ($z_{\mbox{\scriptsize max}}\to z_m$) limit of this
force. Using (\ref{metric}) and (\ref{vm}), this can in turn be
rewritten in the form
\begin{equation}\label{Fcrit}
F_{\mbox{\scriptsize ext}}(0)=\mp {\pi\over 2}\sqrt{\lambda}T^2{v^2_m\over\sqrt{1-v_m^2}}~,
\end{equation}
which coincides with the drag force (\ref{EPlossgubser}) computed with a
trailing string in \cite{hkkky,gubser}, evaluated at the velocity
$v=v_m$. This agreement has a simple interpretation. As explained
in \cite{ctqhat}, the drag force calculation of
\cite{hkkky,gubser} has a limited range of validity ($z_m\le
z_h(1-v^2)^{1/4}$ for a given $v$, or equivalently, $v\le v_m$ for
a given $z_m$), arising from the condition that the electric field
$F_{0x}$ needed to keep the string endpoint moving at constant
speed does not exceed its critical value at $z_m$,
$F_{0x}^{\mbox{\scriptsize crit}}=(\sqrt{\lambda}/2\pi)h(z_m)(z_h/z_m)^2$.
This critical electric field is thus equivalent to the maximal attainable
 drag force, $F_{\mbox{\scriptsize drag}}(v_m)$. The fact that it
  agrees with the right-hand side of
(\ref{Fcrit}), then, is not surprising, because, by definition,
$F_{\mbox{\scriptsize ext}}(0)$ encodes the force needed to pull apart the endpoints
of a zero-size string at $z=z_m$, i.e., the value of
$F_{0x}^{\mbox{\scriptsize crit}}$. In gauge theory language,
this is the force required to nucleate a quark-antiquark pair.

For $L> z_m/2$, the potential is screened and deviates significantly from the linear form
(\ref{Elinearplasma}). Its behavior is shown in Fig.~\ref{potentialfig}. As can be seen
there, the location of the screening lengths $L_*$ and $L_{\mbox{\scriptsize max}}$ shift to
the left with increasing $z_m$ (e.g., for $z_m/z_h=0,0.2,0.4,0.75$ we respectively find $\pi
T L_*=0.755,0.746,0.707,0.511$ and $\pi T L_{\mbox{\scriptsize
max}}=0.865,0.861,0.828,0.616$).

\begin{figure}[htb]
\vspace*{0.5cm}
\begin{center}
\setlength{\unitlength}{1cm}
\includegraphics[width=7cm,height=4cm]{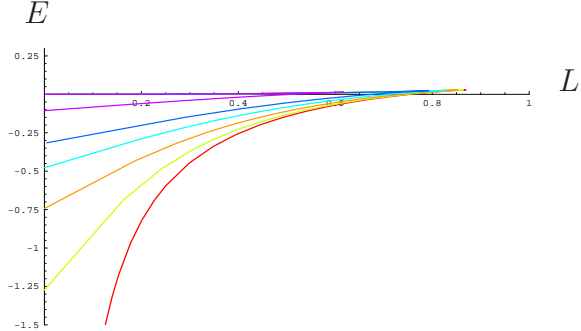}
 \begin{picture}(0,0)
   \put(0.2,3.3){$L$}
   \put(-6.9 ,4.2){$E$}
 \end{picture}
\caption{Quark-antiquark potential (in units of $T\sqrt{\lambda}/4$)
 in the static finite-mass case, for $z_m/z_h=0.001$
(red), 0.2 (light green), 0.3 (orange), 0.4 (light blue), 0.5 (blue) and 0.75 (purple).
The separation $L$ is given in units of $1/2\pi T$. See the main
text for discussion.}\label{potentialfig}
\end{center}
\end{figure}

Let us now go on to consider the case where the quark-antiquark pair moves through the
plasma. For simplicity, as in \cite{dragqqbar} we will restrict ourselves to the case where
the motion is perpendicular to the dipole axis (other angles have been studied
in \cite{liu2,sfetsosqqbar,liu3}).
Orienting the former along $x$ and the
latter along $y$, the expression for the  $q$-$\bar{q}$ separation  can be copied directly
from Eq.~(37) of \cite{dragqqbar}. It is given by
\begin{equation}\label{L}
L(f_y,v)=\frac{f_y}{2{\pi}T}\int^{h_m}_{h_{\mbox{\tiny min}}}
{\frac{dh}{(1-h)^{\frac{1}{4}}\sqrt{(h-v^2)h-(1-h)h{f_y^2}}}}~,
\end{equation}
where
$$f_y\equiv z_h^2 \Pi^z_y= {2\over \pi \sqrt{\lambda} T^2}F_y$$
is a rescaled
version of the force $F_y$ needed to hold the quark and antiquark in place, $h(z)$ is the
function appearing in the metric (\ref{metric}),
\begin{equation} \label{hmin}
h_{\mbox{\scriptsize min}}\equiv h(z_{\mbox{\scriptsize max}})=\frac{v^2+f_y^2}{1+f_y^2}~,
\end{equation}
and we have taken into account the finiteness of the quark Lagrangian mass $m$ by changing
the upper limit of the integration range from 1 to $h_m\equiv h(z_m)$. The form
of expression (\ref{L})
reflects the fact that, in spite of its motion with respect to  the black hole,
the string remains completely vertical, in contrast with the trailing configuration
studied in \cite{hkkky,gubser}.  In the gauge theory, this translates into the
interesting property that, unlike isolated quarks, mesons  \emph{feel no drag}
\cite{sonnenschein,liu2,dragqqbar}.

Similarly, the energy
of the $q$-$\bar{q}$ pair in its rest frame\footnote{For a discussion of subtleties in the
calculation of the pair energy in the rest frame of the plasma, see \cite{dragqqbar}.} can
be read from Eq.~(38) of \cite{dragqqbar} as
\begin{equation}\label{E}
E(f_y,v)=\frac{T\sqrt{\lambda}}{4}\left[ \int^{h_m}_{h_{\mbox{\tiny
min}}}{\frac{dh(h-v^2)\gamma}{(1-h)^{\frac{5}{4}} \sqrt{(h-v^2)h-(1-h)h{f_y^2}}}}
-\int^{h_m}_{0}{\frac{dh}{(1-h)^{\frac{5}{4}}}}\right]~.
\end{equation}
The second term in this expression represents the energy of two disjoint strings
trailing their boundary endpoints as in \cite{hkkky,gubser},
dual to an unbound quark and antiquark (which \emph{do} experience
a drag force). This contribution is subtracted
in order for $E=0$ to correspond to the energy beyond which the system would become
unbound.

\begin{figure}[htb]
\vspace*{0.5cm} \setlength{\unitlength}{1cm}
\includegraphics[width=6cm,height=4cm]{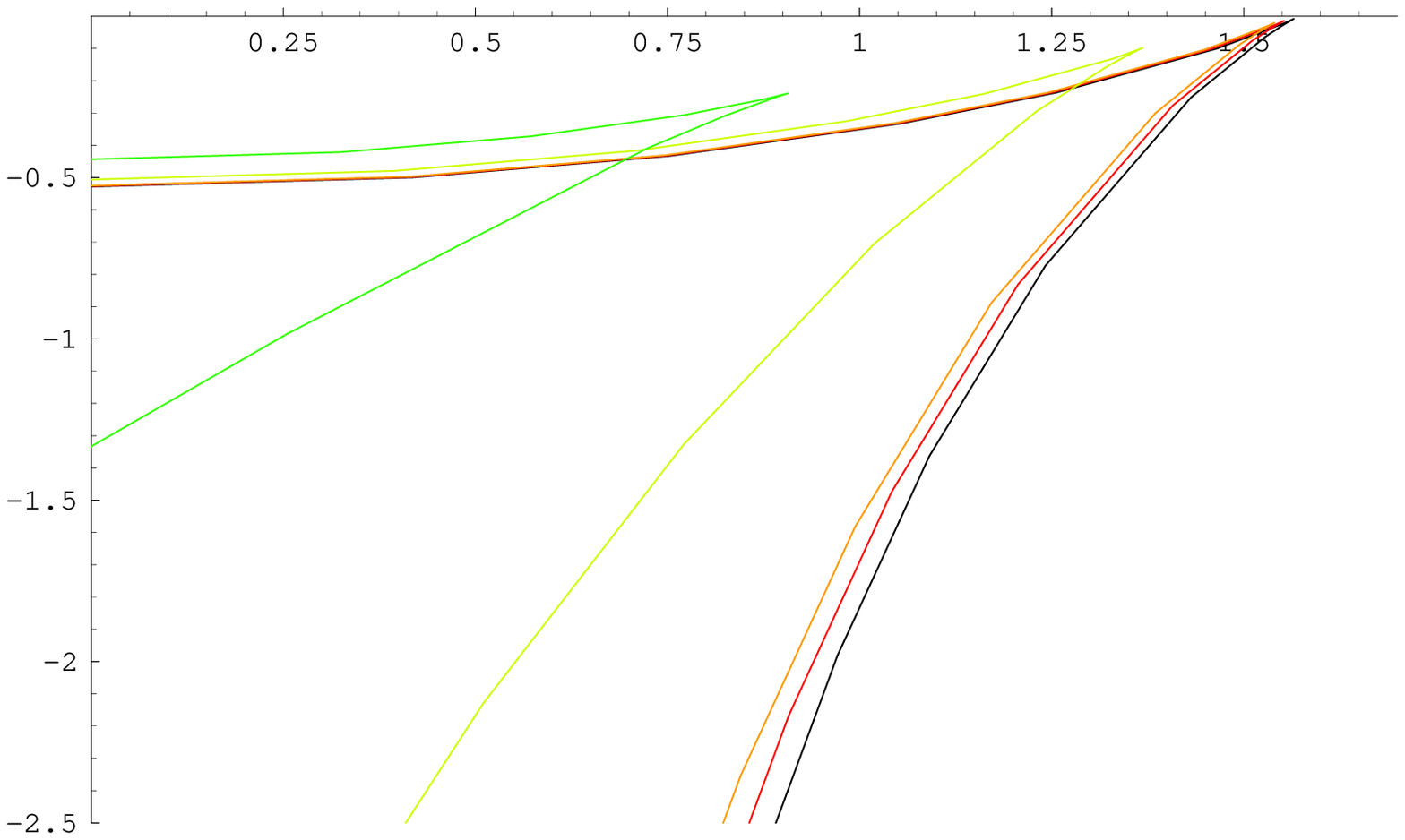}
 \begin{picture}(0,0)
   \put(0.1,3.8){$L$}
   \put(-6,4.1){$E$}
 \end{picture}\hspace{1cm}
 \includegraphics[width=6cm,height=4cm]{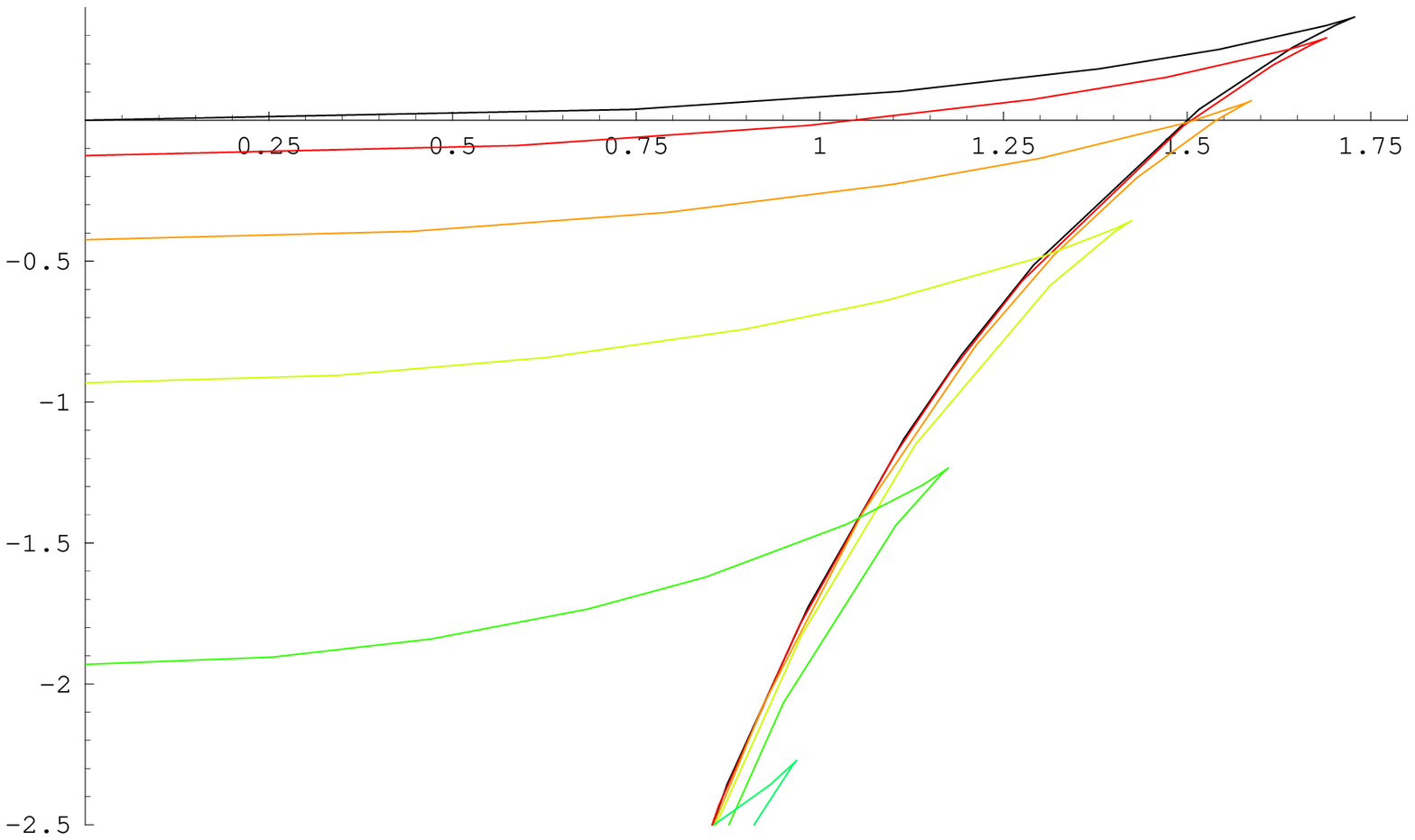}
 \begin{picture}(0,0)
   \put(0.1,3.3){$L$}
   \put(-6,4.1){$E$}
 \end{picture}
\caption{Quark-antiquark energy (in units of $T\sqrt{\lambda}/4$) as a function of
separation (in units of $1/2\pi T$), for (a) fixed $v=0.45$ and $z_m=0$ (black), 0.2 (red),
0.25 (orange), 0.5 (light green), and 0.75 (green); (b) fixed $z_m=0.2$ and $v=0$ (black),
0.2 (red), 0.4 (orange), 0.6 (light green), 0.8 (green) and 0.9 (bright green). See text for
discussion.} \label{potentialvfig}
\end{figure}

Carrying out the integrals in (\ref{L}) and (\ref{E}) numerically, one obtains plots for the
potential such as those shown in Fig.~\ref{potentialvfig}. For small separation the behavior
is again linear,
 \begin{equation}\label{Elinearplasmav}
E(L)= {\sqrt{\lambda}\over \pi}\left[-\left({1\over z_m}-{1\over z_h}\right)+{L\gamma \over
2z^2_m}\sqrt{1-\left({z_m\over z_h}\right)^4-v^2} + \mathcal{O}\left(\left({L\over
z_m}\right)^2\right)\right]~.
 \end{equation}
 The two terms in this expression have the same interpretation as was
 given for (\ref{Elinearplasma}) in the paragraph surrounding
 (\ref{Fext})-(\ref{Fcrit}).

 {}From Fig.~\ref{potentialvfig} one also sees that the value of $z_m$ has an effect on the
 location of the screening lengths $L_*$, $L_{\mbox{\scriptsize max}}$ (and the velocity beyond which
 $L_*=L_{\mbox{\scriptsize max}}$). The behavior at fixed $v$ and varying $z_m$ shown in
 Fig.~\ref{potentialvfig}a is qualitatively the same as we saw already for the static configuration in
 Fig.~\ref{potentialfig}. The dependence at fixed $z_m$ and varying $v$, which is more relevant
 for the phenomenological situation, is shown in
 Fig.~\ref{potentialvfig}b.

 \begin{figure}[htb]
 \vspace*{0.5cm}
\setlength{\unitlength}{1cm}
\includegraphics[width=6cm,height=4cm]{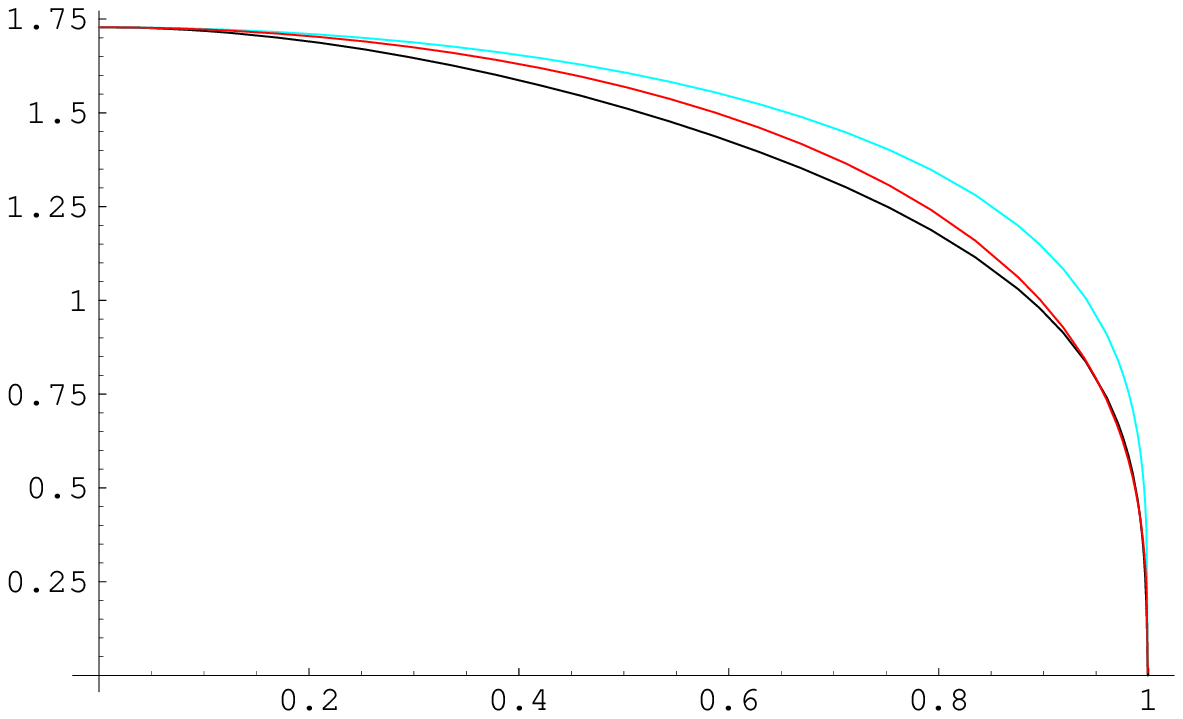}
 \begin{picture}(0,0)
   \put(0.2,0.2){$v$}
   \put(-5.6,4.1){$L_{\mbox{\scriptsize max}}$}
 \end{picture}\hspace{1cm}
 \includegraphics[width=6cm,height=4cm]{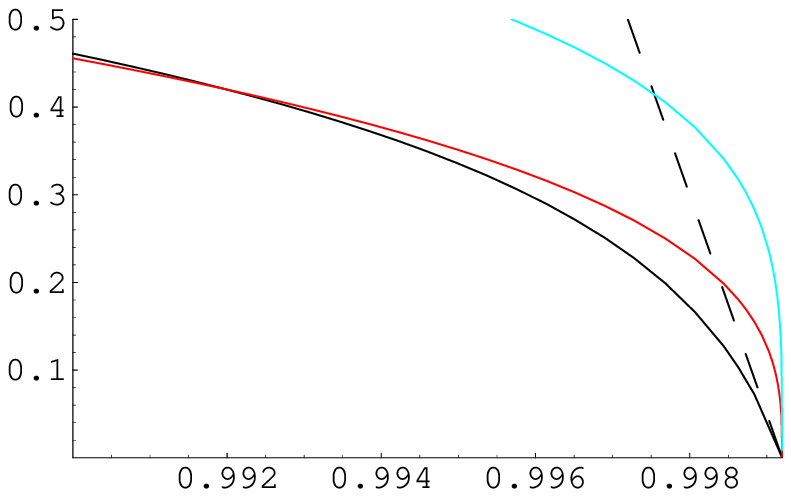}
 \begin{picture}(0,0)
   \put(0.2,0.2){$v$}
   \put(-5.6,4.1){$L_{\mbox{\scriptsize max}}$}
 \end{picture}
\caption{(a) Screening length $L_{\mbox{\scriptsize max}}$ (in
units of $2\pi T$) as a function of velocity for $z_m/z_h=0.2$ (in
black) compared against the $z_m>0$ fits (\ref{onethirdvm}) (in red)
and (\ref{onequartervm}) (in light blue). (b) Expanded version of
the same plot, showing that (\ref{onethirdvm}) gives a relatively
good approximation up to velocities very close to $v_m$, where the
asymptotic linear behavior (\ref{one}) (dotted dark blue) sets
in.} \label{lmaxfig2}
\end{figure}

\begin{figure}[thb]
 \vspace*{0.2cm}
\setlength{\unitlength}{1cm}
\includegraphics[width=6cm,height=4cm]{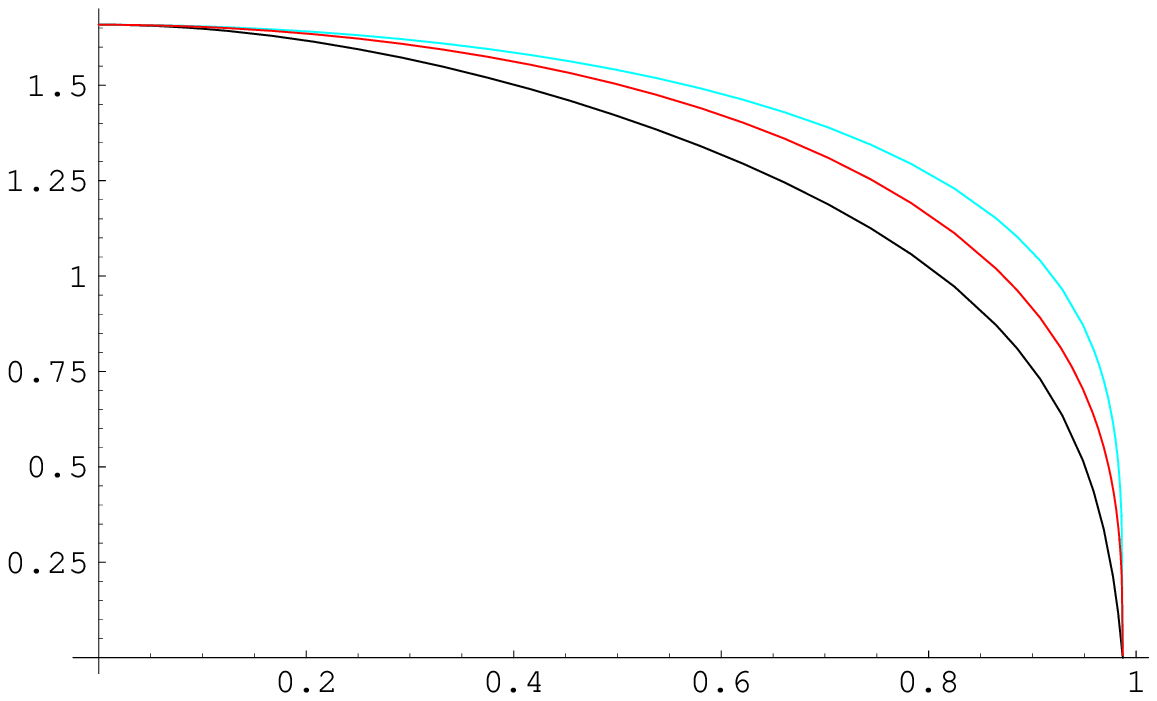}
 \begin{picture}(0,0)
   \put(0.2,0.2){$v$}
   \put(-5.6,4.1){$L_{\mbox{\scriptsize max}}$}
    \end{picture}\hspace{1cm}
 \includegraphics[width=6cm,height=4cm]{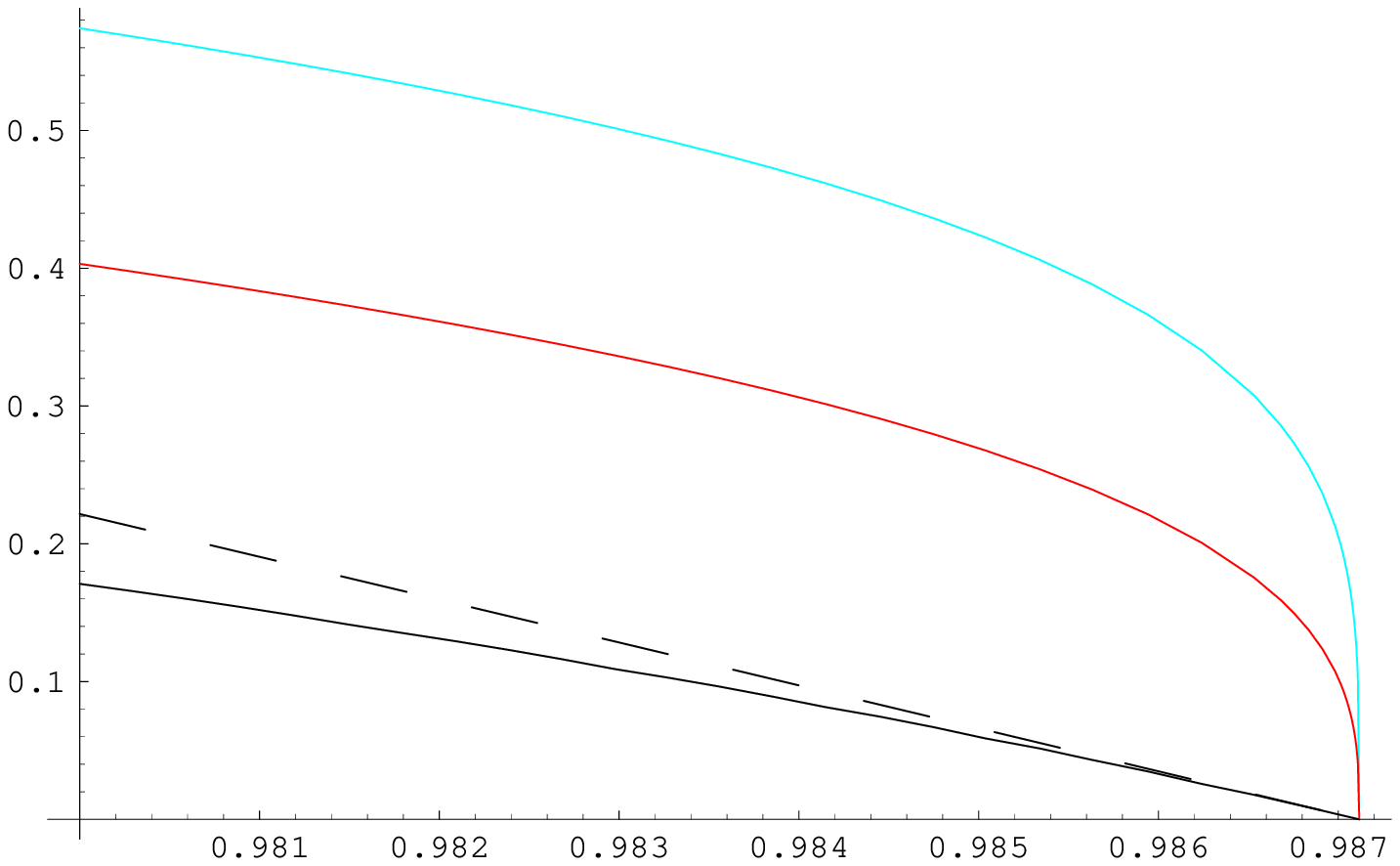}
 \begin{picture}(0,0)
   \put(0.2,0.2){$v$}
   \put(-5.8,4.1){$L_{\mbox{\scriptsize max}}$}
    \end{picture}
\caption{(a) Screening length $L_{\mbox{\scriptsize max}}$ (in
units of $2\pi T$) as a function of velocity for $z_m/z_h=0.4$ (in
black) compared against the $z_m=0$ fits (\ref{onethird}) (in red)
and (\ref{onequarter}) (in light blue). (b) Expanded version of
the same plot, showing that the region in the neighborhood of
$v_m$ where the asymptotic linear behavior (\ref{one}) (dotted
dark blue) is a significantly better approximation than
(\ref{onethird}) has grown larger in comparison to the
$z_m/z_h=0.2$ case depicted in Fig.~\ref{lmaxfig2}.}
\label{lmaxfig4}
\end{figure}

In Figs.~\ref{lmaxfig2} and
\ref{lmaxfig4}, we see that
when $z_m/z_h=0.2$, and to a lesser extent,
when $z_m/z_h=0.4$,
the screening length
$L_{\mbox{\scriptsize max}}(v)$ is still relatively well
approximated in the full range $0\le v\le v_m$ by the natural modification
of
the $z_m=0$
fit (\ref{onethird}),
\begin{equation} \label{onethirdvm}
L_{\mbox{\scriptsize max}}(v)\approx{0.865\over \pi
T v_m^{2/3}}(v_m^2-v^2)^{1/3}~.
\end{equation}
This approximation becomes worse as $z_m/z_h$
is further increased. (In all cases, the fit analogous
to (\ref{onequarter}),
\begin{equation}\label{onequartervm}
L_{\mbox{\scriptsize max}}(v)\approx
{0.865\over \pi T v_m^{1/2}}(v_m^2-v^2)^{1/4}~,
\end{equation}
 does a poorer job than
(\ref{onethirdvm}).)

 An important difference with respect to the case where the quarks are infinitely massive is
 that now the `ultra-relativistic' region would refer to the limit where the pair velocity
 approaches the limiting velocity $v_m<1$ given in (\ref{vm}) and discussed further
 in Section \ref{qqbarvmsec}. {}From this alone it is clear
 that the asymptotic formula (\ref{onequarter}) no longer holds. The behavior in this region
 can still be determined analytically, and turns out to be
 \begin{equation}\label{one}
L_{\mbox{\scriptsize max}}\to {1\over 2\pi T}{v_m^2-v^2\over
v_m(1-v_m^2)^{3/4}}={1\over 2\pi T}{z_h^3[1-(z_m/z_h)^4-v^2]\over
z_m^3\sqrt{1-(z_m/z_h)^4}}~,
 \end{equation}
 where we see that the $1/4$ exponent in (\ref{onequarteranal}) changes to $1$ at finite quark
 mass.
As shown in Figs.~\ref{lmaxfig2}b and \ref{lmaxfig4}b, for values
of $z_m/z_h$ in the neighborhood of the charm mass, the functional
form (\ref{one}) applies only for velocities extremely close to
the $v_m$ endpoint.
As $z_m/z_h$
is further increased, however,
 there is a tendency for
(\ref{one}) to apply to a larger fraction of the $0\le v\le v_m$
interval.

In \cite{liu2,liu3}, the result (\ref{onequarter}) for the
screening length in the infinitely massive case was turned into
a tentative prediction for the dissociation temperature
of charmonium or bottomonium traversing the quark-gluon plasma,
$T_{\mbox{\scriptsize diss}}\propto (1-v^2)^{1/4}$. (As
reviewed in Section \ref{qqbarreviewsec}, for $0\le v<0.991$
the  exponent
implied by the AdS/CFT calculation
is really closer to $1/3$ than to $1/4$.)
 The argument
for this velocity scaling
is heuristic, and relies on comparing the screening
length against the natural size of the bound state in question
(see, e.g., \cite{satz} and references therein).
Applying the same logic to the fits (\ref{onethirdvm}) or (\ref{one})
 for the screening length in the case of
 masses in the neighborhood of the charm quark
($z_m/z_h\sim$ 0.2-0.4, as in Figs.~\ref{lmaxfig2}-\ref{lmaxfig4}),
one would infer that  $T_{\mbox{\scriptsize diss}}\propto (v_m^2-v^2)^{n}$,
with $n\simeq 1/3$ for general $v$, and $n=1$ in the $v\to v_m$ limit
(the latter being the preferred fit only in the restricted
range $v>0.998$ or $v>0.98$,
depending on the value of $z_m$).

While this paper was in preparation, the work \cite{liu4} appeared,
which includes an explicit
 derivation of the dispersion relation for mesons,
described as open string modes on the D7-branes,
following
\cite{martinmeson}. For large meson momentum $p$,
their calculation resulted in $E=v_m p$,
with $v_m$ the limiting velocity defined in
(\ref{vm}) and discussed further in Section \ref{qqbarvmsec}
of the present paper (as well as in \cite{argyres3}, which
also appeared while our work was being written up).
 The authors of \cite{liu4} emphasized that
the $z_h$- and $z_m$-dependence of $v_m$ implies a bound on the temperature
at which a meson with velocity $v$ can exist within the plasma.
Interpreting the maximal allowed temperature
as a dissociation temperature, they found the scaling
$T_{\mbox{\scriptsize diss}}\propto (1-v^2)^{1/4}$, just
like the heuristic argument based on the potential for
infinitely-massive and ultra-relativistic quarks had suggested
\cite{liu2,liu3}.

Given the results of \cite{dragqqbar} and
the present subsection, we are not quite sure what to make of this
agreement. As explained above, for finite mass the quark-antiquark
potential does \emph{not} scale as $(1-v^2)^{1/4}$ even in the `ultra-relativistic'
region (which is now $v\to v_m$). So the fact that the
heuristic argument in \cite{liu2,liu3} based on the potential
 suggests the same scaling with velocity as the
 direct calculation in \cite{liu3}
 might be a coincidence, or it might indicate that
 there is a direct physical connection
 between the dissociation temperature
 and the potential only when the latter is computed
 for infinitely-massive quarks.
  In any case, it should be emphasized
 that the $1/4$ scaling follows not from the detailed
 computation of the dispersion relation in \cite{liu3}, but
 from the temperature-dependence of the limiting velocity (\ref{vm}).

\subsection{Screening length vs. transition distance}
\label{potentialtransitionsec}

In Fig.~\ref{xtransfig} at the end of the previous Section, we determined the location
$(x_f,v_f)$ beyond which a quark and antiquark created (with zero total momentum) within
 the plasma begin to slow down at the asymptotic rate specified by the constant
friction coefficient obtained for an isolated quark in \cite{hkkky,gubser}.
For
$q$-$\bar{q}$ separations larger than $2x_f$, then,
 the two particles are definitely oblivious to
one another and evolve independently. It is therefore interesting to compare the transition
distance $x_f$ against (half of) the screening length $L_{\mbox{\scriptsize max}}$
determined in the present section, which gives a measure of the minimal distance at which
the quark and antiquark can decouple.

\begin{figure}[htb]
 \vspace{0.2cm}
 \begin{center}
\setlength{\unitlength}{1cm}
\includegraphics[width=6cm,height=4cm]{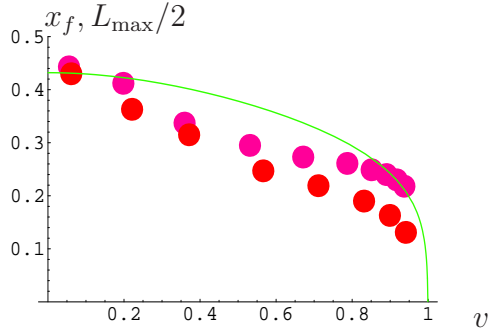}
 \begin{picture}(0,0)
   \put(0.2,0){$v$}
   \put(-5.5,4){$x_{f},L_{\mbox{\scriptsize max}}/2$}
 \end{picture}\hspace{1cm}
 \caption{(Half of) Screening length $L_{\mbox{\scriptsize max}}(v)$ (green) compared
 against the transition distance $x_f(v)$
defined at the end of Section \ref{qqbarsec}, with $z_m/z_h=0.2$, for $f=0.1$ (red) and
$f=0.05$ (magenta). The vertical axis is in units of $1/\pi T$.} \label{xlmaxfig}
\end{center}
\end{figure}

For singlet configurations, the result of this comparison is shown in Fig.~\ref{xlmaxfig},
where we see that the two separations are of comparable magnitude and scale with velocity in
a similar manner.\footnote{A similar agreement was reported very recently in
 \cite{iancu2}, which appeared while this paper was in preparation.} Notice that this is in spite of the fact that the two relevant string
configurations are quite different, with motion respectively along and perpendicular to the
plane in which the string extends. We conclude then that the transition to the
constant-drag-coefficient regime takes place immediately after the quark and antiquark lose
contact with one another. That is to say, unlike what we found for the forced isolated quark
in Section \ref{acceleratedsec}, here there is no intermediate stage where the quark and
antiquark decelerate independently from one another at a rate that differs substantially
from the asymptotic result of \cite{hkkky,gubser}.

The initial stage, where as seen
in Fig.~\ref{stage2fig} the quark and antiquark dissipate energy at a rate
\emph{larger} than (\ref{EPlossgubser}), corresponds to the period
when the string has not fallen close enough to the black hole horizon.
In other words, to a first approximation, the quark and antiquark
evolve in this region as they would in the absence of the thermal
plasma.
To the extent that this result, with all its simplifying assumptions,
might conceivably be extrapolated
to the experimental context,
this initial stage would not  differ significantly
between heavy ion and proton-proton collisions.

For adjoint configurations, the quark-antiquark potential is expected to
be suppressed by a factor of $1/N^2$ relative to the singlet case (see, e.g.,
\cite{adjointpot} and references therein). On the string theory side of the
duality, this is visible in the fact that the `line' initial conditions
 (\ref{lineic}) can be regarded as describing two disjoint strings extending
 all the way from the D7-branes to the horizon. In the
 classical limit where our calculations
 are carried out, these two halves do not influence one another, because
 signals propagating between them would have to pass through the horizon, and their
 splitting/joining/annihilating
  is suppressed by powers of the string coupling \gs, which is taken to zero.
  At this level of approximation, then,
  the quark and antiquark do not see one another,
  so the relevant `screening length' is $L_s=0$. This is consistent
  with the result $x_f=0$ we obtained for the adjoint case in Section
   \ref{qqbartransitionsec}.

The fact that in the adjoint case the quark and antiquark evolve independently from the
start makes this case directly comparable to our study of isolated quarks in Section
\ref{acceleratedsec}. Interestingly, the corresponding results seem to point in opposite
directions: whereas for an initially static quark that is externally forced we identified a
short period after release where the quark loses energy at a rate significantly
\emph{smaller} than the stationary/asymptotic rate (\ref{EPlossgubser}), for a quark that is
created with a certain initial velocity and gluonic field profile and thereafter moves only
under the influence of the plasma we found an initial rate of dissipation that is somewhat
\emph{larger} than (\ref{EPlossgubser}).

It seems, then, that the initial conditions can, to a certain extent, influence the
subsequent evolution, which does not seem altogether surprising given the nonlinear nature
of the medium. The application of an external force in Section \ref{acceleratedsec} clearly
has a significant effect on the shape of the string at the time of release, or in other
words, on the gluonic field configuration surrounding the quark. In view of our results for
quark-antiquark evolution in this and the previous section, the initial period over which
this disturbed quark behaves unlike what one would expect based on \cite{hkkky,gubser,ct}
should most likely be interpreted as reflecting the time it takes the quark to stabilize the
gluonic fields around it, or, equivalently, to move far enough away from the disturbed
region to be effectively screened from it. In any case, one should of course not lose sight
of the fact that the actual experimental situation resembles the setup of Section
\ref{qqbarsec} more closely than that of Section \ref{acceleratedsec}.

\section*{Acknowledgements}

It is a pleasure to thank Alejandro Ayala, Elena C\'aceres, Jorge Casalderrey-Solana,
Antonio Garc\'{\i}a and Guy Paic for valuable discussions. We are also grateful to Antonio
Garc\'{\i}a for collaboration in the initial stages of this project. We additionally thank
Karen Ram\'{\i}rez for help with Figs.~\ref{bhfig} and \ref{bhplasmafig}. This work was
partially supported by Mexico's National Council of Science and Technology (CONACyT) grants
50-155I and SEP-2004-C01-47211, as well as by DGAPA-UNAM grants IN116408 and IN104503.

\end{document}